\documentclass[a4paper,11pt,twoside]{StyleThese}
\RequirePackage{booktabs}
\RequirePackage[mathscr]{eucal}
\RequirePackage{tikz}
\usetikzlibrary{shapes,arrows}
\usepackage{todonotes}
\newtheorem{definition}{Definition}
\usepackage{upgreek}
\newtheorem{hypothesis}{Hypothesis}
\newcommand{\vertex}{\node[vertex]}
\tikzstyle{vertex}=[circle, draw, inner sep=0pt, minimum size=6pt]
\usetikzlibrary{plotmarks}
\usepackage{mathrsfs}
\usepackage{lmodern}
\usepackage{amssymb}
\usepackage{amsmath}
\newcommand{\NICP}{\text{CIP}}
\newcommand{\BILEVEL}{\texttt{BILEVEL}}
\newcommand{\MASTERPIECE}{\texttt{CLIQUE-INTER}}
\newcommand{\opt}{\mathrm{opt}}
\newtheorem{proposition}{Proposition}[chapter]
\newtheorem{property}{Property}[chapter]
\newtheorem{theorem}[proposition]{Theorem}
\newtheorem{lemma}[proposition]{Lemma}
\newtheorem{corollary}[proposition]{Corollary}

\usepackage{amsmath,amssymb}
\usepackage[T1]{fontenc}
\usepackage[utf8]{inputenc}
\usepackage[french]{babel}

\usepackage[left=1.5in,right=1.3in,top=1.1in,bottom=1.1in,includefoot,includehead,headheight=13.6pt]{geometry}

\usepackage{silence}

\usepackage{aecompl}
\usepackage{url}

\usepackage[printonlyused,withpage]{acronym}

\graphicspath{{.}{images/}}

\usepackage{color}
\definecolor{linkcol}{rgb}{0,0,0.4}
\definecolor{citecol}{rgb}{0.5,0,0}

\usepackage[nottoc, notlof, notlot]{tocbibind}
\usepackage{minitoc}
\usepackage{rotating}                    
\usepackage{fancyhdr}                    

\let\headruleORIG\headrule
\renewcommand{\headrule}{\color{black} \headruleORIG}

\usepackage{colortbl}
\arrayrulecolor{black}

\fancypagestyle{plain}{
  \fancyhead{}
  \fancyfoot{}
  
}

\usepackage{FrAlgorithm}
\usepackage[noend]{FrAlgorithmic}
\usepackage{scrextend}

\makeatletter

\def\cleardoublepage{\clearpage\if@twoside \ifodd\c@page\else%
  \hbox{}%
  \thispagestyle{empty}
  \newpage%
  \if@twocolumn\hbox{}\newpage\fi\fi\fi}

\makeatother

{%

\hrulefill
\vspace*{0.5cm}%
\end{minipage}
}

\let\minitocORIG\minitoc
\renewcommand{\minitoc}{\minitocORIG \vspace{1.5em}}

\usepackage{subfigure}
\usepackage{multirow}
{ \begin{list}%
	{$\bullet$}%
	{\setlength{\labelwidth}{25pt}%
	 \setlength{\leftmargin}{30pt}%
	 \setlength{\itemsep}{\parsep}}}%
{ \end{list} }

\ifpdf
  \usepackage[pagebackref,hyperindex=true]{hyperref}
\else
  \usepackage[dvipdfm,pagebackref,hyperindex=true]{hyperref}
\fi

\renewcommand*{\backref}[1]{}
\renewcommand*{\backrefalt}[4]{%
\ifcase #1 %
(Non cité.)%
\or
(Cité en page~#2.)%
\else
(Cité en pages~#2.)%
\fi}

\hypersetup
{
bookmarksopen=true,
pdftitle="Titre du manuscript",
pdfauthor="Votre nom", 
pdfsubject="Description rapide du sujet du manuscrit", 
pdfmenubar=true, 
pdfhighlight=/O, 
colorlinks=true, 
pdfpagemode=UseNone, 
pdfpagelayout=SinglePage, 
pdffitwindow=true, 
linkcolor=linkcol, 
citecolor=citecol, 
urlcolor=linkcol 
}

\begin{document}


\pagenumbering{Alph}

\begin{titlepage}
\begin{center}
\noindent {\large \textbf{UNIVERSITÉ Sorbonne Paris Nord}} \\
\vspace*{0.3cm}
\noindent \textbf{SCIENCES} \\
\vspace*{0.5cm}
\noindent \Huge \textbf{Habilitation à Diriger des Recherches } \\
\vspace*{0.3cm}
\vspace*{0.4cm}
\noindent \large {Présentée et soutenue par\\}
\noindent \LARGE Sébastien \textsc{Martin} \\
\vspace*{0.8cm}
\noindent {\Huge \textbf{Contribution to Blocker and Interdiction
optimization problems in networks.}} \\
\vspace*{0.8cm}
\noindent \large soutenue le 19 Septembre 2024 \\
\vspace*{0.5cm}
\end{center}
\noindent \large \textbf{Jury :} \\
\begin{center}
\noindent\large 
\begin{tabular}{llcllcl}
       &\textit{Referees :}\\	
    
     & François \textsc{Clautiaux}		& - & Institut de Mathématiques de Bordeaux, France\\
				& Eduardo  \textsc{Uchoa}		& - & Universidade Federal Fluminense, Brazil \\
  \vspace*{0.5cm}				& Yann  \textsc{Vaxes}		& - & Aix-Marseille Université, France\\

     & \textit{Jury Members :} \\ 
      
& Pierre \textsc{Fouilhoux}          & - & Université Sorbonne Paris Nord, France\\
      				& Imed  \textsc{Kacem}			& - & Université de Lorraine, France\\
      				& Ivana  \textsc{Ljubic}		& - & ESSEC, France\\
      				& Ridha   \textsc{Mahjoub}		& - & Université Paris-Dauphine, France\\
      				& Roberto   \textsc{Wolfler Calvo}		& - & Université Sorbonne Paris Nord, France
\end{tabular}
\end{center}
\end{titlepage}

\pagenumbering{arabic}
\sloppy

\titlepage

\pagenumbering{roman}

\setcounter{page}{0}
\cleardoublepage



\dominitoc

\section*{Abstract}
This manuscript describes the notions of blocker and interdiction applied to well-known optimization problems. The main interest of these two concepts is the capability to analyze the existence of a combinatorial
structure after some modifications. We focus on graph modification, like removing vertices or links in a network. In the interdiction version, we have a budget for modification to reduce as much as possible the size of a given combinatorial structure. Whereas, for the blocker version, we minimize the number of modifications such that the network does not contain a given combinatorial structure. Blocker and interdiction problems have some similarities and can be applied to well-known optimization problems. We consider matching, connectivity, shortest path, max flow, and clique problems. For these problems, we analyze either the blocker version or the interdiction one. Applying the concept of blocker or interdiction to well-known optimization problems can change their complexities. Some optimization problems become harder when one of these two notions is applied. For this reason, we propose some complexity analysis to show when an optimization problem, or the associated decision problem, becomes harder. Another fundamental aspect developed in the manuscript is the use of exact methods to tackle these optimization problems. The main way to solve these problems is to use integer linear programming to model them. An interesting aspect of integer linear programming is the possibility to analyze theoretically the strength of these models, using cutting planes. For most of the problems studied in this manuscript, a polyhedral analysis is performed to prove the strength of inequalities or describe new families of inequalities. The exact algorithms proposed are based on Branch-and-Cut or Branch-and-Price algorithm, where dedicated separation and pricing algorithms are proposed.


\tableofcontents

\mainmatter

\chapter*{Introduction}

Combinatorial optimization is a field of mathematics and computer science that deals with finding the best solution among a finite set of possible solutions to a problem. It involves the study of algorithms and mathematical models for solving optimization problems that arise in various applications, such as scheduling, logistics, network design, and resource allocation. The goal of combinatorial optimization is to find an optimal solution that satisfies certain constraints and criteria, such as minimizing cost, maximizing efficiency, or optimizing performance. This field has significant practical applications in industries such as transportation, manufacturing, and telecommunications, and has been instrumental in the development of modern optimization techniques and algorithms. 
During the last decades, powerful methods based on mathematical programming have been developed to face these problems and the scalability issue. Branch-and-Cut and Branch-and-Price algorithms have shown their efficiencies in solving large-scale instances. 

The first well-known example is the dedicated Branch-and-Cut algorithm able to solve large networks for the traveling salesman problem \cite{TSP}. 
To develop an efficient Branch-and-Cut algorithm, the key ingredient is the separation algorithm that allows adding dynamically the most interesting constraints. Two elements are primordial to designing an efficient separation algorithm. The first one is the analysis of the strength of the constraint, given by the polyhedral analysis. The second one is the design of an algorithm with low complexity. Furthermore, it is important to propose new valid inequalities to strengthen the linear relaxation and thus reduce the number of nodes in the branching tree to speed up the global computational time.

A second well-known example is the coloration problem where a dedicated Branch-and-Price algorithm is able to solve more significant instances thanks to the decomposition in smaller sub-problems \cite{COLOR}. The key ingredients to developing an efficient Branch-and-Price algorithm are the choice of the decomposition, the meaning of variables, and the design of the branching rules. 
Some recent works \cite{MAGNOUCHE202186} proposed efficient methods based on Branch-and-Cut-and-Price algorithms to solve the separator problem. These algorithms add constraints and columns dynamically. 

This manuscript focuses on optimization methods based on Branch-and-Cut and Branch-and-Price algorithms. For some studied problems, a complexity analysis is given. The complexity of an optimization problem is important to design efficient and adapted methods. 

The problems studied, in this document, are related to the blocker and interdiction notions. These two notions can be applied to several combinatorial optimization problems. Considering interdiction or blocker notion allows analyzing the strength of a particular property to face anomalies. These can represent the absenteeism of people or link/node failures in a network. In some cases, the blocker and interdiction concepts allow knowing where monitors must be installed to check information flow or to detect clusters during a pandemic. 
Blocker or interdiction problems are bi-level problems where: an actor, the outer problem, destroys the structure; whereas a second actor, the inner problem, optimizes after destruction. This perspective of two levels can imply harder problems. As shown in the next chapters, some well-known combinatorial problems can be seen as bi-level problems like blocker or interdiction problems. 

In this manuscript, we consider several combinatorial optimization problems. The first one is the matching problem. The blocker version of the matching problem is considered. This problem is related to the problem studied during my Ph.D. Indeed, we worked on a particular case of a matching problem.
The application associated with this problem is to detect if an algebraic-differential system is well described under some conditional constraints. This problem can be seen as a blocker problem in a bipartite graph where one side corresponds to the conditional constraints and the other side to the variables. The goal is to detect a combination of constraints, where the associated conditions are fixed, such as the induced graph does not have a perfect matching. This problem can be seen as a blocker problem where all perfect matching must be deleted to find if there exists a solution without perfect matching. Chapter one presents the bipartite complete matching blocker problem which is closely related to the problem studied during my Ph.D. This problem is based on the bipartite complete matching problem which can be solved in polynomial time. First, we show that the blocker version of this problem stays solvable in polynomial time and we provide the associated polynomial algorithm. This allows modeling the strength of a matching. The application consists in analyzing the crew assignment against absenteeism.  Second, we analyze the problem where several time slots are considered. This problem is modeled like several bipartite graphs where some nodes must be assigned only once on a bipartite graph to ensure that a complete matching exists in each graph. We show that this problem is NP-hard and we propose an efficient Branch-and-Cut algorithm and we analyze the strength of the separated inequalities.  
In the second chapter, we consider the vertex $k$-cut problem that can be seen as a vertex $k$-connected-component blocker problem. This partitioning problem allows us to analyze the connectivity. Another application is related to matrix decomposition by reordering the matrix into a bordered diagonal matrix. We analyze the complexity and provide an efficient Branch-and-Price algorithm to solve this problem. 
The third chapter is dedicated to path and flow problems. The shortest path blocker problem has many applications in the resilience of path computation and in the monitoring of paths. For this problem, we show the NP-hardness and provide an integer linear model 
with an NP-Hard separation problem. We derive an efficient Branch-and-Cut algorithm. We also propose a model to solve the maximum flow blocker problem based on the min-cut problem which is the dual of the maximum flow problem. This transformation allows solving efficiently the problem with a nice compact model. The main application of the maximum flow blocker problem is the monitoring of flow.
Chapter four provides results on the $m$-clique free interval sub-graph problem. This problem consists in ensuring that an induced graph does not contain some structure. Indeed, in scheduling problems where several machines are considered then an induced graph of a solution must be an interval graph without clique of size $m$ where $m-1$ is the number of machines. We propose several inequalities to remove forbidden graph structures and  show the efficiency of this approach on the Unrelated Parallel Machines Scheduling Problem With Job Disjunctives Graph.
In chapter five, we propose a model to solve the clique interdiction problem. We analyze the facial aspect of the constraints, we propose a lifting procedure and a combinatorial lower bound. With these two key ingredients, we derived an efficient Branch-and-Cut algorithm. The main application of this problem is the detection of influencers  in social networks.

Along this manuscript, we present some contributions on matching blocker \cite{MBCMVBP2020,MBCMVBP2014-v2,MBCMVBP2014,MBCMVBP2017}, on vertex cut related to the $k$-connected-component blocker \cite{cornaz2014mathematical,cornaz2019vertex,magnouche2016multi2,magnouche2016multi}, on flow blocker \cite{magnouche2020most, isma},  on clique and interval subgraph \cite{hassan2018m,furini2019maximum}.  
We show the link between these problems. In the last chapter, we provide some contributions to telecommunication network problems \cite{detnetOur1,detnetOur2,detnetOur3,detnetOur4,detnetOur5,slicingOur,slicingOur1,patentMTR,patent1Det,patent2Det,patentVIGP}.

\newpage


\chapter{Bipartite Complete Matching blocker}

\newcommand{\uip}{{\tilde{u}}}

\newcommand{\mjob}{\mathcal{U}}
\newcommand{\kup}{\upkappa}

The matching problem is a well-known combinatorial optimization problem. This problem has many applications, for instance, in scheduling or crew assignments.  The goal is to find a one-to-one assignment. We focus on the complete matching in a bipartite graph. A complete matching in a bipartite graph is a set of links that covers all nodes of the smallest set of the bipartition, and two links of this set do not cover the same node. Formally, given a bipartite graph $G=(U\cup V,E)$ with $|U|\leq |V|$, a complete matching is a matching covering all vertices of $U$. In the rest of this chapter, $|U|\leq |V|$ and thus a complete matching is always considered on $U$.
This problem can be solved in polynomial time using a combinatorial algorithm or mathematical programming \cite{lovasz2009matching}.

In this chapter, we present some results from original research published in \cite{MBCMVBP2020} and extension of these results can be found in \cite{MBCMVBP2017},\cite{MBCMVBP2014} and \cite{MBCMVBP2014-v2}. 
We analyze the complexity and propose mathematical models to solve the Bipartite Complete Matching Blocker Problem (BCMBP) and the extension to the Multiple Bipartite Complete Matching Blocker Problem (MBCMP). 

\vspace{-0.5cm}
\paragraph{Applications} One application of these two problems is to measure the impact of absenteeism on the assignment of nurses to services. Another application is the determination of the maximum possible perturbation (failure or attack) such that beyond that point, no solution exists (scheduling on machine or truck/network assignment), where the problem is reduced to the complete matching problem.

\section{Bipartite Complete Matching Blocker Problem}

In this section, we present the Bipartite Complete Matching Blocker Problem and show that this problem can be solved in polynomial time using a mathematical model. One interesting result comes from the fact that the blocker version of this problem is still polynomial. This property is uncommon and, to the best of our knowledge, it is the first operations research problem where the blocker version is still polynomial.

We first introduce the notion of $k$-CM to help the definition of the BCMBP. Let $G[U \cup V']$ be the induced bipartite graph where $V'\subseteq V$.

\begin{definition} [$k$-CM] Let $G=(U\cup V,E)$ be a bipartite graph. We say that $G$ is k-Complete Matching ($k$-CM) on $U$ if, for all $V'\subseteq V$ with $|V'|=|V| - k$, $G[U \cup V']$ has a complete matching on $U$. For short, we denote by $k$-CM a $k$-Complete Matching on $U$.
\end{definition}

We can deduce the following definition.

\begin{definition} [BCMBP] Let us denote by $\kappa(G)$ the maximum number $k$ such that $G$ is $k$-CM. The Bipartite Complete Matching Blocker Problem (BCMBP) consists in finding $\kappa(G)$ for a given bipartite graph $G$.
\end{definition}

First, let us introduce Hall theorem. 
\begin{theorem}\cite{hall} Hall theorem:
Let $G=(U\cup V,E)$ be a bipartite graph. $G$ has a complete matching on $U$, if and only if, for all subset $U'\subseteq U$, $|N_G(U')| \geq |U'|$.
\end{theorem}

The first main result in this research project is the extension of  Hall theorem \cite{hall}.
\begin{theorem}\cite{MBCMVBP2020}
Let $G=(U\cup V,E)$ be a bipartite graph. $G$ is $k$-CM if and only if, for all non-empty subset $U'\subseteq U$, $|N_G(U')| \geq |U'|+k$.
\end{theorem}
The proof is based on the Hall theorem where we consider any possible $k$ vertices removed from the set $V$.

We can deduce the following corollary.
\begin{corollary}\cite{MBCMVBP2020}\label{c1}
Let $G=(U\cup V, E)$ a bipartite graph. Then,
$$\kup(G) = \min_{U'\subseteq U,~ U'\neq\emptyset} \Big(|N_G(U')| - |U'| \Big).$$
\end{corollary}

This theorem and the induced corollary have a strong impact on the complexity and mathematical formulation of the BCMBP.

\subsection{Mathematical formulation and complexity}

For $U'\subseteq U$ an optimal solution of the BCMBP, let $x\in \{0,1\}^{|U \cup V | }$ be given by

\begin{quote}
$\qquad x_u =  \left \{ \begin{array}{ll} 1 & \mbox{ if }  u\in U',\\0 &\mbox{ otherwise, }  \end{array} \right. \quad \forall u\in U,
$\bigskip

\qquad $ x_v =  \left \{ \begin{array}{ll} 1 & \mbox{ if }  v\in N_G(U'),\\0 &\mbox{ otherwise, }  \end{array} \right. \quad \forall v\in V .
$\bigskip
\end{quote}

The BCMBP is equivalent to the following integer program:

\begin{alignat}{8}
\label{fo} & & 	&\min \left(\sum_{v \in V } x_v -  \sum_{u \in U} x_u\right) \\
\label{ctn1:-1}	& 	&\quad 	&\sum_{u \in U} x_u\geq 1, & & \\
\label{ctn1:0}	& (IP) 	\qquad &\quad 	& x_u-x_v \leq 0, & &
\forall uv\in E, \ u\in U, \ v\in V ,\\
\label{ctn1:1}  &  &\quad & x_w \leq 1, &  &  \forall w\in U \cup V ,\\
\label{ctn1:4}  & &\quad & -x_w \leq 0, &  &  \forall w\in U \cup V ,\\
\label{ctn1:5} &  &\quad & x_w  \in \{0, 1\}, &  &  \forall w\in U \cup V .
\end{alignat}
\\
The objective function expresses  Corollary~\ref{c1}.
Inequality \eqref{ctn1:-1} guarantees that the set $U'$ is not empty.
Inequalities \eqref{ctn1:0} ensure that the variables $x_v$, associated with the neighborhood of $U'$, will be set to 1.

By considering one mathematical model for each $u\in U$ where the inequality \eqref{ctn1:-1} is replaced by $x_u=1$ we can deduce that the matrix induced by the constraints is totally unimodular and thus the following theorem holds.

\begin{theorem}\cite{MBCMVBP2020}\label{th:polynomial}
The BCMBP can be solved in polynomial time.
\end{theorem}

\section{Multi Bipartite Complete Matching Blocker Problem}

The MBCMBP is a generalization of the BCMBP for the case where the set $U$ is a set of disjoint subsets forming a partition of $U$.

An application of the MBCMBP is the assignment of people to a job by considering different time slots. 

We will discuss the complexity of this problem.

\begin{definition}[$k$-MCM]
Let $G=(U\cup V, E)$ be a bipartite graph, $\mathcal{U}=\{U_1,...,U_m\}$  a partition of $U$ and $\mathcal{V}=\{V_1,...,V_m\}$ a partition of $V$.
For all $ i\in M= \{1,...,m\}$, denote by $H_i$ the subgraph $G[U_i\cup V_i]$. We say that $(G, \mathcal{U}, \mathcal{V})$ is $k$-\emph{Multiple Complete Matching on $\mathcal{U}$} if, for all $ i\in M$,  $H_i$  is $k$-CM on $U_i$. For short, we denote by $k$-MCM a $k$-Multiple Complete Matching on $\mathcal{U}$.
\end{definition}

\begin{definition}[MBCMBP]
Let $G=(U\cup V, E)$ be a bipartite graph and  $\mathcal{U}=\{U_1,...,U_m\}$ a partition of $U$. The \emph{Multiple Bipartite Complete Matching Vertex Interdiction Problem (MBCMBP)} consists in finding a partition $\mathcal{V}=\{V_1,...,V_m\}$  of $V$ such that $ \min\big\{\kup(H_i)\ | \ i\in M\big\}$ is maximum. We denote this maximum number by $\kup(G, \mathcal{U}, \mathcal{V})$. We call this problem $m$-MBCMBP when the cardinality of the partition $\mathcal{U}$ is $m$.
\end{definition}

We proved that the MBCMBP is NP-hard even with 2 sets of $U$. We propose a reduction from the stable set problem. The vertices, corresponding to a stable set solution from $U$, are assigned to one set of $\mathcal{V}$.

\begin{theorem}\cite{MBCMVBP2020}\label{exthall}
The MBCMBP is strongly NP-hard even when $|\mathcal{U}|=2$.
\end{theorem}

\subsection{Natural formulation}
 
 In \cite{MBCMVBP2020} we also propose a natural formulation based on the extension of Hall theorem. 
Our goal is to find a partition $\mathcal{V}= \{V_1,...,V_m\}$ of $V$ such that $(G, \mathcal{U}, \mathcal{V})$   is $\kup(G, \mathcal{U}, \mathcal{V})$-MCM.

We now present the integer linear program formulated to find $\mathcal{V}$.
Let $ x_v^i $, for $v\in V, \forall i\in M$, be a binary variable equals to 1 if $v\in V_i$ and 0 otherwise, and $z\in \mathbb{N}_+$ the value of $\kup(G, \mathcal{U}, \mathcal{V})$.

The MBCMBP is equivalent to the following integer program:
\begin{alignat}{6}
\nonumber & & \quad && \quad\max z \\
\label{ctn2:0} & &  \quad &&  \sum_{i\in M } x_v^i  =1, &\quad  \mbox{ for all } v\in V ,\\
\label{ctn2:1}  &(P') &\quad && \sum_{v\in N_G(U')} x_v^i -z \geq |U'|, & \quad \mbox{ for all } i\in M \mbox{ and for all } U'\subseteq U_i,\\
\label{ctn2:3}  & &\quad && x_v^i \in \{0,1\}, &  \quad \mbox{ for all } v\in V  \mbox{ and for all } i\in M,\\[1ex]
\label{ctn2:4}  & &\quad && z \in \mathbb{N}_+
\end{alignat}
where  inequalities \eqref{ctn2:0} ensure that each vertex $v\in V $ belongs to only one set of the partition $\mathcal{V}$ and  inequalities \eqref{ctn2:1}
express the condition of   Theorem \ref{exthall} on each $H_i$, $i\in M$.

A facial study and a valid inequality are given in \cite{MBCMVBP2020}, and can be summarized in the following results.

To simplify the polyhedral study without modifying the optimal solution, we relax \eqref{ctn2:0} by the following inequalities:

\begin{alignat}{6}
\label{ctn2:-1} & &\quad & \sum_{i\in M} x_v^i  \leq1 & \!\!\! & \quad \mbox{ for all } v\in V.
\end{alignat}

Let $P(G,\mathcal{U})$ be the convex hull of the solution of program $(P')$, that is,
$$P(G,\mathcal{U})=conv(\{x\in \{0,1\}^{|V|\times m},z \quad | \quad x, z \text{ satisfy } \eqref{ctn2:1}, \eqref{ctn2:-1}\}).$$

\begin{hypothesis}\label{1sol}
We consider  that, for all $v\in V$,  there exists a matching in $G\setminus \{v\}$.
\end{hypothesis}
We can verify this hypothesis in polynomial time. If it is not verified, then $z=0$.

\begin{proposition}\cite{MBCMVBP2020}\label{prop:fulldim}
The polytope $P(G,\mathcal{U})$ is full dimensional if and only if  a 1-MCM solution exists.
\end{proposition}

All facet proofs associated with the propositions given in this section are done by maximality. 

\begin{proposition}\cite{MBCMVBP2020}
If   a 1-MCM solution does not exist, then $dim(P(G,\mathcal{U}))=|V|\times m$.
\end{proposition}

\begin{proposition}\cite{MBCMVBP2020}
The inequality \eqref{ctn2:-1}, associated with $v\in V$, defines a facet of $P(G,\mathcal{U})$.
\end{proposition}

\begin{proposition}\cite{MBCMVBP2020}\label{prop:facet}
The inequality \eqref{ctn2:1}, associated with $i\in M$ and $U'\subseteq U_i$, defines a facet of $P(G,\mathcal{U})$
if and only if
\begin{enumerate}
\item if $U' \subset U_i$, then there does not exist $U''\subseteq U_i$ where $U'\subset U''$ and $|N_G(U')|-|U'|\geq |N_G(U'')|-|U''|$,
\item there does not exist $U''\subseteq U_i$ where $|U''|=|U'|$ and $N_G(U'')\subset N_G(U')$,
\item $G[U'\cup N_G(U')]$ is connected.
\item there exists a partition $\breve{\mathcal{V}}=\{\breve{V}_0,...,\breve{V}_m\}$ and $k>0$ such that $(G,\mathcal{U}, \breve{\mathcal{V}})$ is $k$-MCM, satisfying  $|\breve{V}_i\cap N_G(U')|=|U'|+k$.
\end{enumerate}
\end{proposition}

Recall that inequalities (\ref{ctn2:1}) give an upper bound on $z$ considering each  subset $U_i \in \mathcal{U}=\{U_1,...,U_m\}$ separately. Now, we extend this family of inequalities by considering these subsets two by two, simultaneously. In the following example, we can see the interest in this approach.
Let $U_s, U_t \in \mathcal{U}$, $s \neq t$,  suppose that $|U_s|=|U_t|=2$ and  $N_G(U_s) = N_G(U_t)$ with cardinality 4. Using inequalities  (\ref{ctn2:1}), $z$ is bounded by 2. However, it is straightforward to see that $z = 0$. \medskip

For any subsets $U'_s \subseteq U_s$ and   $U'_t \subseteq U_t$, we denote by $N_G(U'_s U'_t) =N_G(U'_s) \cap N_G(U'_t)$ the set of common neighbors  of $U'_s$ and  $U'_t$.
Let us consider $\tilde{N}_G(U'_s) = N_G(U'_s) \setminus N_G(U'_s U'_t)$  the set of vertices  belonging exclusively to the neighborhood of $U'_s$. In the same way,
we define  $\tilde{N}_G(U'_t) = N_G(U'_t) \setminus N_G(U'_s U'_t)$.

Let us denote by \medskip

$\tilde{k}_{\max} = \max \Big\{\; |\tilde{N}_G(U'_s)| - |U'_s|,  \ |\tilde{N}_G(U'_t)| - |U'_t|\;\Big\}$, \\$\tilde{k}_{\min} = \min \Big\{\;  |\tilde{N}_G(U'_s)| - |U'_s|, \ |\tilde{N}_G(U'_t)| - |U'_t|\;\Big\}$,

the upper bounds on the optimal solutions for $U_s$ and $U_t$, respectively, when considering only the exclusive neighborhoods. Then, it is clear that the upper bound of the optimal solutions are less than or equal to $ \tilde{k}_{\min}+ |N_G(U'_s U'_t)|$.

By considering the common neighborhood $N_G(U'_s U'_t)$, two cases have to be studied in order to know how vertices of this set can complete $\tilde{N}_G(U'_t)$ and $\tilde{N}_G(U'_s)$ to give a better upper bound $k_{\sup}$ for both sets.

\begin{enumerate}
	\item  $\tilde{k}_{\max} - \tilde{k}_{\min}  \geq |N_G(U'_s U'_t)|$:  only one of the two subsets, $\tilde{N}_G(U'_s)$ or $\tilde{N}_G(U'_t)$,  can be completed with the common vertices in such a way that $k_{\sup} = \tilde{k}_{\min}+ |N_G(U'_s U'_t)|$.   \medskip
	\item $\tilde{k}_{\max} - \tilde{k}_{\min}  <  |N_G(U'_s U'_t)|$: first, one of the two subsets, $\tilde{N}_G(U'_s)$ or $\tilde{N}_G(U'_t)$,  can be completed with $\tilde{k}_{\max}- \tilde{k}_{\min}$ common vertices; the remaining common vertices can complete both $\tilde{N}_G(U'_s)$   and  $\tilde{N}_G(U'_t)$ equally. We deduce that $k_{\sup} = \tilde{k}_{\max}+ \Big( |N_G(U'_s U'_t)| - (\tilde{k}_{\max} - \tilde{k}_{\min})\Big)/2 $.
\end{enumerate}

Let $\ell = s$ if $|\tilde{N}_G(U'_s)| - |U'_s| <  |\tilde{N}_G(U'_t)| - |U'_t|$ and  $\ell = t$ otherwise.

Remark that, if in a solution, a common vertex $v \in N_G(U'_s U'_t)$  is assigned to a subset $V_i$ where $i \in M \setminus  \{s, t\}$, this vertex cannot be used to complete  $\tilde{N}_G(U'_s)$  or  $\tilde{N}_G(U'_t)$. Then, $k_{\sup}$ will be decreased by at least $1/2$.

Furthermore, if in a solution a  vertex $v \in \tilde{N}_G(U'_\ell )$  is assigned to a subset $V_i$ where $i \in M \setminus  \{s, t\}$, this vertex cannot be used to complete  $\tilde{N}_G(U'_s)$  or  $\tilde{N}_G(U'_t)$. Then, $k_{\sup}$ will be decreased by  $1$.
Thus the following inequalities are valid for all $U'_s$ and  $U'_t$:

\begin{alignat}{8}
\label{ctn:new} & & 	& z +\frac{1}{2}\sum_{v\in N_G(U'_s U'_t)}\sum_{i\in M\setminus \{s, t\}} x_v^i  + \sum_{v\in N_G(U'_\ell)}\sum_{i\in M\setminus \{s, t\}} x_v^i\leq k_{sup}.
\end{alignat}

\subsection{Separation algorithm}

Given a solution $(x^*, z^*)\in \mathbb{R}_+^{|V |\times m} \times \mathbb{R}_+$, the \emph{separation problem} for a family of inequalities consists in determining
whether $(x^*, z^*)$ satisfies these inequalities and, if not, in finding one of these inequalities which is violated by $(x^*, z^*)$. An algorithm that solves this problem is called a separation algorithm associated with these inequalities.

\subsubsection{Separation of inequality \eqref{ctn2:1}} 
The number of inequalities \eqref{ctn2:1} in $(P')$ is exponential. Hence, a polynomial time separation algorithm is necessary to allow using these inequalities inside a cutting plane algorithm.
The equivalence between separation and optimization in combinatorial optimization \cite{G81} implies that the linear relaxation of problem $(P')$ can be solved in polynomial time. Let $(x^*, z^*)\in \mathbb{R}_+^{|V |\times m} \times \mathbb{R}_+$ be a solution of the linear relaxation.
For all $i\in \{1,...,m\}$, we want to find a set $U_i'\in U_i$ such that
$\sum_{v\in N_G(U_i')} x_v^{i*} -z^* < |U_i'|$.

We separate the problem in $m$ subproblems, each problem is associated with a single set $U_i \subseteq U$. In the following, we consider the graph $G[U_i\cup V]$. Let $w: V\rightarrow \mathbb{R}_+ $ be the weight function where $w(v)=x_v^{i*}$ for all $v\in V$.
The separation problem consists in solving the integer linear program $(P')$ with the objective function
$z'=\min \left( \sum_{v\in V } w(v) x_v-\sum_{u\in U_i} x_u \right)$.
If $z'< z^*$ then the inequality $\sum_{v\in N_G(U_i')} x_v^i -z \geq |U_i'|$, where $U_i'$ is given by the optimal solution of $(P')$, is violated when respecting $(x^*, z^*)$. In this case, we add the inequality associated with $i$ and $U_i'$. Otherwise, all inequalities \eqref{ctn2:1} are verified.

It is possible that the added constraint does not define a facet.
For this reason, we propose some improvements according to Conditions 1  -- 3 of Proposition \ref{prop:facet}.

\begin{itemize}
\item[{Imp 1.}] If the resulting $U_i'$ does not respect Condition 1, then there must exist a set $\tilde{U}_i'$ where $\sum_{v\in N_G(\tilde{U}_i') \setminus N_G(U_i')} w(v) = |\tilde{U}_i'|$ and $N_G(\tilde{U}_i') \setminus N_G(U_i') = |\tilde{U}_i'|$, which implies that  $w(v) = 1$, for all $v\in N_G(\tilde{U}_i') \setminus N_G(U_i')$. Thus, we aim to maximize $|U_i'|$. By subtracting $\sum_{u\in U} \epsilon x_u$ from the objective function, where $\epsilon$ is sufficiently small, we ensure the maximality of $U_i'$.
\item[{Imp 2.}] If the resulting $U_i'$ does not respect Condition 2,
then there must exist a set $U''\subseteq U_i$ with $|U_i''|=|U_i'|$ such that $N_G(U_i'')\subset N_G(U_i')$. Then, $\sum_{v\in N_G(U_i') \setminus N_G(U_i'')} w(v) = 0$.
Thus, we aim to minimize $|N_G(U_i')|$. By adding $\sum_{v\in V} \epsilon x_v$ to the objective function, where $\epsilon$ is sufficiently small, we ensure the minimality of $N_G(U_i')$.
\item[{Imp 3.}] If $G[U_i' \cup N_G(U_i')]$  is not connected then, for each connected component, if the induced inequality is violated, then we add it.
\end{itemize}

\paragraph{Separation of inequalities \eqref{ctn:new} }
We propose a heuristic for solving the separation problem for inequalities. Let $M'\subseteq M$ and  $\mathcal{U'} = \{U_i', i \in M'\}$ the set of sets found by the previous algorithm.   This heuristic is based on  the set $\mathcal{U'}$. Indeed, for all couples $(U_i', U_j')\in \mathcal{U'}^2$, if the inequality \eqref{ctn:new} induced by this couple is violated, then we add it.

\subsection{Branch-and-Cut Algorithm and experimental results}
In this section, we present a Branch-and-Cut algorithm for solving $(P')$.
Our aim is to address the algorithmic applications of the model and the theoretical
results presented in the previous sections. To start the optimization, we
consider the linear program $$\max\{z|x\in [0,1]^{m\times |V|}, z\in \mathbb{N}_+,\ x\mbox{ satisfies } \eqref{ctn2:0}\}.$$

An important task in the Branch-and-Cut algorithm is to determine whether
or not an optimal solution of the linear relaxation of the MBCMVIP is feasible.
An optimal solution $x^*$ of the linear relaxation is feasible for the MBCMVIP if $x^*$ is
integer and $x^*$ satisfies inequalities \eqref{ctn2:1}. Thus, whether or not $x^*$ is feasible for MBCMVIP can be verified in polynomial time. If not, the Branch-and-Cut algorithm
uses the inequalities \eqref{ctn2:1} and \eqref{ctn:new}, and their separations are successively performed. We remark that all inequalities are global
(\emph{i.e.} valid in all the Branch-and-Cut tree) and several inequalities may be added
at each iteration.

Each line of Table \ref{tabResPrimalVSDual} represents five randomly generated instances, sharing the same parameters. The benchmark used is restricted to the smallest instances, allowing to solve at least one instance over five,  in less than one hour.

The three first columns identify the instances. In column 4 (resp. column 5) is indicated the average computing time in seconds over the five instances of the dual formulation (resp. the second formulation). In columns 6 and 7 are given the average numbers of nodes of the branching tree for both variants. The last two columns present the ratios of instances solved in less than one hour.
\begin{table}[h!]
\centering
\begin{tabular}{|c|c|c|c|c|c|c|c|c|}
\hline
$|U|$	&	$|V|$	&	density	&	CPU 1	&	CPU 2	&	No 1	&	No 2	&	opt 1	&	opt 2	\\ \hline	\hline
50	&	62	&	10	&	14	&	0	&	494.6	&	0	&	5/5	&	5/5	\\ \hline	
50	&	62	&	20	&	429.8	&	1.4	&	254.4	&	9.8	&	5/5	&	5/5	\\ \hline	
50	&	62	&	40	&	1428	&	0	&	566.4	&	0	&	5/5	&	5/5	\\ \hline	
50	&	75	&	10	&	2.8	&	0.6	&	0	&	0	&	5/5	&	5/5	\\ \hline	
50	&	75	&	20	&	123.8	&	2.6	&	553.2	&	4.6	&	5/5	&	5/5	\\ \hline	
50	&	75	&	40	&	1926.6	&	1.6	&	708.8	&	2.6	&	4/5	&	5/5	\\ \hline	
50	&	87	&	10	&	3.6	&	0.6	&	18.6	&	0	&	5/5	&	5/5	\\ \hline	
50	&	87	&	20	&	379	&	3	&	330.4	&	11.8	&	5/5	&	5/5	\\ \hline	
50	&	87	&	40	&	2298.4	&	4	&	462.8	&	1562.6	&	2/5	&	5/5	\\ \hline	
100	&	125	&	10	&	1241.8	&	4	&	44.2	&	6	&	4/5	&	5/5	\\ \hline	
100	&	125	&	20	&	3234.2	&	19.6	&	69.4	&	11480.6	&	2/5	&	5/5	\\ \hline	
\end{tabular}
\caption{Comparison of the two formulations}
\label{tabResPrimalVSDual}
\end{table}

The second formulation allows solving all instances in less than 50 seconds, whereas the dual formulation only solves 48 instances over 55. Furthermore, the computing time of the dual formulation is far greater. The standard deviation of the numbers of nodes in the branching tree for the dual formulation is smaller than that in the second formulation. As the second formulation leads to models considering less variables, the computing time and memory amount needed to solve the problem are better than the ones in the dual formulation.

Unfortunately, the improvement proposed in the previous section was not able to reduce the computational time significantly. 

\section{Conclusion and open questions}

In this chapter, we have presented several results on the Bipartite Complete Matching Blocker Problem and the extension to the Multiple Bipartite Complete Matching Blocker Problem. We have discussed the complexity of these problems. For the Bipartite Complete Matching Blocker Problem, we have designed a polynomial time algorithm thanks to the integrity of the associated polytope. For the Multiple Bipartite Complete Matching Blocker Problem, we have designed two mathematical formulations, one based on the duality and the second on the natural variables with an exponential number of constraints. Thanks to a well-designed Branch-and-Cut algorithm based on the natural formulation, we have solved large instances in comparison with the formulation based on the duality. In   \cite{MBCMVBP2017},\cite{MBCMVBP2014} \cite{MBCMVBP2014-v2} we have discussed more about the application and how to consider new constraints like incompatibility constraints between employees.

For the Bipartite Complete Matching Blocker Problem, we have proved that the blocker problem is still polynomially solvable as the classical problem. One interesting opening question is "Does there exist a property ensuring that if a problem can also be solved in polynomial time then the blocker or interdiction version can be solved in polynomial time ?". Furthermore, the weighted version of the blocker problem has not been explored for the moment. Also, the interdiction version of this problem has not been extensively studied.

For the Multiple Bipartite Complete Matching Blocker Problem, we have proved that the problem is NP-hard to solve and provided a natural formulation to solve efficiently this problem. An interesting research direction is to provide an efficient separation algorithm or other valid inequalities to strengthen the natural formulation and solve harder instances. We can also consider other well-known techniques to solve the problem as Benders decomposition.

\chapter{Vertex $k$-cut}

In this chapter, we focus on graph connectivity to analyze, for instance, the strength of graphs or how to decompose them. 

One of the common definitions of connectivity in a graph is that for every two vertices, there exists at least one path between them. A graph $G=(V,E)$ is said to be $k$-edge-connected (resp. $k$-vertex-connected) if after removing any subset of $k-1$ edges (resp. vertices) or less, the graph is still connected. In this case, the blocker of the $k$-edge-connected (resp. $k$-vertex-connected) problem consists in finding the smaller subset of edges (resp. vertices) to remove so that the graph is not $k$-connected. In this chapter, we do not focus on this problem but on the vertex $k$-connected-component blocker problem which consists in finding the minimum number of vertices such that their removal yields a graph with more than $k$ connected components. 

More formally, the vertex $k$-connected-component blocker problem consists, given a graph $G=(V, E)$ and an integer $k$, in finding a set of vertices $V_0$ such that removing these vertices provides at least $k$ connected components in the remaining graph.   

This problem is equivalent to the vertex $k$-cut problem, where the goal is to find a minimum number of vertices to remove in such a way that the resulting graph consists of $k$ disjoint subgraphs.  

\begin{corollary}
The vertex $k$-cut problem is equivalent to the vertex $k$-connected-component blocker problem
\end{corollary}

In this chapter, we present original research from \cite{cornaz2019vertex}. We analyze the complexity and provide mathematical models to solve efficiently the vertex $k$-cut problem.
In \cite{magnouche2016multi}, \cite{magnouche2016multi2} \cite{CORNAZ201911} and \cite{MAGNOUCHE202186} we propose Branch-and-cut and Branch-and-price algorithms based on polyhedral investigation and advanced pricing approaches for solving the variant of the problem where terminals are considered. We also study in \cite{cornaz2014mathematical} the variant where the sizes of the components are balanced. A compact model with valid inequalities and an extended model are given. More recently, in \cite{minich1,minich2}, we propose a fully polynomial time approximation scheme algorithm for a balanced variant where the graph is a tree. In \cite{grange}, we propose heuristic algorithms to solve a more generic problem where the vertex separator is considered in a hyper-graph. Another variant is considered in \cite{HEALY2023} where the goal is to separate the graph into exactly $k$ connected components. For this variant, we consider the edge cut and not the vertex cut to separate the graph.  

\vspace{-0.5cm}
\paragraph{Applications}The vertex $k$-cut problem has many applications in decomposition problems like automatic Dantzig-Wolfe decomposition and automatic code decomposition to parallelize the execution of code. The vertex cut represents the elements shared by or copied on each set of these decompositions. Another application is in scheduling where some jobs can share resources (the vertex cut) and copy them on each machine.

\vspace{-0.5cm}
\paragraph{Complexity}
In \cite{cornaz2019vertex} we show that the problem of the vertex $k$-cut problem is NP-hard if $k\geq 3$. The proof is done by reduction from the stable set problem in a tripartite graph. For the case $k=2$, using a graph transformation it is possible to solve in polynomial time the problem by solving the $s-t$ min-cut problem for each pair of vertices \cite{didi2011exact}.

\section{Mathematical formulations}

\subsection{Compact formulation}

In this section, we show that the vertex $k$-cut problem can be reformulated as a maximum stable set problem on a specific $k$-partite graph with additional requirements. We also derive a compact integer linear program based on this reformulation.

\medskip

Let $G=(V,E)$ and $k \geq 2$ be an instance of the vertex $k$-cut problem. As previously noted, a subset $V_0 \subset V$ is a vertex $k$-cut of $G$ if and only if $V\setminus V_0$ can be partitioned into $k$ nonempty pairwise disconnected sets. Hence, the vertex $k$-cut problem is equivalent to finding $k$ nonempty disjoint sets $V_1,\ldots,V_k$ of $V$ which are pairwise disconnected such that $|\bigcup_{i \in K} V_i|$ is maximum. Let $K=\{1,...,k\}$.
In the following, we consider that the graph $G$ contains a stable set of size $k$, otherwise no solution exists. 

We construct a $k$-partite graph $G'=(V',E')$ so that the vertex $k$-cut problem on $G$ reduces to the maximum stable set problem on $G'$. 
Formally the construction of $G'$ is as follows. The set $V'$ is obtained by considering $k$ copies $v^1,\ldots,v^k$ of every vertex $v \in V$. We define the $k$-partition of $V'$ as $\pi=\{V'_1,\ldots,V'_k\}$ with
 $V'_{i} = \{v^i: v \in V\}$ for all $i=1,\ldots,k$. In other words, $V'_i$ corresponds to a copy of $V$. 
 The edge set $E'$ is the union of two sets $E'_{\alpha}$ and $E'_{\beta}$. $E'_{\alpha} = \{v^iv^j: i \neq j \in K\}$ is the edge set obtained by considering a clique between all the copies of the same vertex  $v\in V$. For $E'_{\beta}$, we consider for each $uv \in E$ an edge between every copy of $u$ and every copy of $v$. Hence, $E'_{\beta} = \{u^iv^j: uv \in E, i \neq j \in K\}$.
 There is a 1-to-1 correspondence between nonempty pairwise disconnected disjoint sets $V_1,\ldots,V_k$ of $V$ and stable sets of $G'$ intersecting each $V'_i$, $i \in K$.  Indeed, let $V_1,\ldots,V_k$ satisfy the aforementioned requirements. Let $S \subseteq V'$ be the set obtained by taking in $V'_i$ the copies of the vertices in $V_i$ for all $i \in K$. $S$ is a stable set because no edge exists between $V_i$ and $V_j$ and $V_i \cap V_j = \emptyset$ for $i \neq j \in K$. Moreover, $S$ intersects every $V'_i$, $i \in K$, since $V_1,\ldots,V_k$ are nonempty. Finally $S = |\bigcup_{i \in K} V_i|$. The converse also holds and the result follows.

	We now give a formulation of the vertex $k$-cut problem in terms of an integer linear program. By the previous reformulation, we look for a stable set $S$ of $G'$ intersecting every $V'_i$ of the $k$-partition. For each vertex $v \in V$ and each integer $i\in K$, let us associate a binary variable $x_v^i$ such that:

\begin{equation*}
 x_{v}^{i}=
\begin{cases}
 1 &\mbox{if copy $v^i \in V'_i$ of vertex } v \in V \mbox{ belongs to } S \\
 0 & \mbox{otherwise} \\
\end{cases}  \qquad \text{for all } i\in K, v\in V.
\end{equation*}

\noindent
The first natural compact ILP formulation (called ILP\textsubscript{C}) reads as follows:

\begin{align}
\label{VKSP_CF_OBJ} \text{(ILP\textsubscript{C})} & & \max \sum_{i\in K}\sum_{v\in V} x_v^i \\
\label{VKSP_CF_1}  & &  \sum_{i \in K} x_v^i &\le 1 & v\in V,\\
\label{VKSP_CF_2_bis}& & x_u^i + \sum_{j\in K \setminus \{i\}} x_v^j & \leq 1 & i\in K,  uv \in E,\\
\label{VKSP_CF_3}   & &  \sum_{v \in V}x_v^i & \ge 1 & i \in K,\\
\label{VKSP_CF_var} & & x_v^i &\in \{0,1\} & i\in K,v \in V.
\end{align}

The objective function maximizes the size of $S$.
Constraints~\eqref{VKSP_CF_1} and \eqref{VKSP_CF_2_bis} are the clique constraints associated with cliques of $E'_{\alpha}$ and edges of $E'_{\beta}$, respectively.
Constraints~\eqref{VKSP_CF_3} impose that $S$ intersects each $V'_{i}$ for $i \in K$.

\subsection{Extended formulation}

We derive an alternative formulation for the vertex $k$-cut problem having an exponential number of variables with respect to the input size. Let
$\mathcal{S}= \{S \subseteq V, S \neq  \emptyset \}$
be the family of all non-empty subsets of vertices of $V$. 

For a subset $S \in \mathcal{S}$, let us associate a binary variable $ \xi_{S}$ such that:
\begin{equation*}
 \xi_{S}=
\begin{cases}
 1 &\mbox{if $S$ corresponds to one of the $k$ disconnected subsets of $G$}  \\
 0 & \mbox{otherwise} \\
\end{cases}  \qquad S\in \mathcal{S}.
\end{equation*}
 
The vertices that do not appear in any selected subset are assigned to the vertex cut. In the following, let $\mathcal{C}$ be  an edge-covering family of cliques of $G$, that is, a family of cliques so that for each edge $uv \in E$, there is at least one clique $C \in \mathcal{C}$ containing both  $u,v \in C$. The exponential-size ILP formulation for the vertex $k$-cut problem reads as follows

\begin{align}
 \text{(ILP\textsubscript{E})} & & \max \sum_{S \in \mathcal{S}}  |S| \xi_{S}    &\label{eq:EF_obj}\\
 & &  \sum_{S \in \mathcal{S}:v\in S} \xi_{S} & \le 1 & v \in V,   \label{eq:EF_1}\\
& & \sum_{S \in \mathcal{S}: C \cap S \neq \emptyset}  \xi_{S} & \le 1 & C \in \mathcal{C}, \label{eq:EF_2}\\
& &  \sum_{S \in \mathcal{S}} \xi_{S} & = k \label{eq:EF_3}\\
& & \xi_{S} & \in \{0,1\} & S \in \mathcal{S}. \label{eq:EF_var}
\end{align}

The objective function \eqref{eq:EF_obj} maximizes the sum of the cardinalities of the selected subsets $S$ of vertices, which is equivalent to minimizing the cardinality of the vertex cut. Constraints \eqref{eq:EF_1} impose that each vertex $i \in V$ does not appear in more than one of the selected subsets. Constraints \eqref{eq:EF_2} impose that, for each clique, $C \in \mathcal{C}$, at most one subset containing any vertex of the clique can be selected. Constraint \eqref{eq:EF_3} imposes that exactly $k$ subsets are selected. Constraints \eqref{eq:EF_var} impose the variables to be binary. Finally, by relaxing the integrality of constraints \eqref{eq:EF_var} to
\begin{equation}
    \xi_{S} \ge 0 \quad \quad S \in \mathcal{S}, \label{eq:dom2}
\end{equation}
we obtain the Linear Programming relaxation of ILP\textsubscript{E}, that is denoted as LP\textsubscript{E} in what follows.

\vspace{-0.5cm}
\subsubsection{Subproblem Matching blocker}
The \emph{master problem} (MP) can be initialized with the $n$ subsets of $V$ containing a single vertex. Since we assumed that $G$ contains a stable set of cardinality $k$, this initialization ensures the existence of a feasible solution to start the column generation. 
Additional variables, needed to optimally solve the MP, are then generated by separating the associated dual constraints.
The \emph{pricing problem} (PP) (see, e.g., \cite{desaulniers2006column} for definition and more details on column generation)  can be solved efficiently as described in the following.\\

At each column generation step, the optimal values $\lambda^*\in \mathbb R_+^V$, $\pi^*\in \mathbb R_+^{\mathcal{C}}$, $\gamma^*\in \mathbb R$ (respectively)  of the dual variables associated with constraints (\ref{eq:EF_1}), (\ref{eq:EF_2}), (\ref{eq:EF_3}) (respectively) are given. The separation of the family of dual constraint is equivalent  to finding a non-empty subset $S^* \in \mathcal{S}$ such that

\begin{equation}
	\sum_{v \in S^*} \lambda_v^* + \sum_{ C \in \mathcal{C}: C \cap S^* \neq \emptyset} \pi_{C}^* + \gamma^*  < |S^*| \label{sep} \nonumber
\end{equation}which can be reformulated as
\begin{equation}\label{eq:pricing}
	\sum_{v \in S^*} \nu_v^* - \sum_{ C \in \mathcal{C}: C \cap S^* \neq \emptyset}  \pi_{C}^* >  \gamma^*,
\end{equation}where $\nu_v^* = 1 - \lambda_v^*$.

If such a subset exists, the corresponding variable $\xi_{S^*}$ is added to the MP, and the procedure is iterated; otherwise, the MP is solved to prove optimality. 
Hence PP amounts to find a $S^*$ maximizing the left-hand side in \eqref{eq:pricing} and to check whether or not it is bigger than the right-hand side.  It can be modeled as a Binary Linear Program using variables $x_v$ ($ v \in V$), which represent $S^*$, and variables $y_C$ ($C \in \mathcal{C}$), each of which takes value 1 if clique $C$ intersects set $S^*$. This binary linear program can be stated as follows:

\begin{align}
&&  \max \  \sum_{v \in V}   \nu_v^* x_v - \sum_{C \in \mathcal{C}} \pi_{C}^* y_{C} \label{eq:obj-sub} \\
&&y_{C}&\ge x_v & v \in C\in \mathcal{C},\label{eq:clique-sub}\\
&& \sum_{v \in V}   x_v &\ge 1\label{eq:nonempty-sub}\\
&& x_v &\in \{0,1\} & v \in V, \label{eq:intx-sub}\\
&& y_{C} &\in \{0,1\} & C \in \mathcal{C}. \label{eq:inty-sub}
\end{align}

Constraints \eqref{eq:clique-sub} impose $y_C=1$ ($C \in \mathcal{C} $) if at least a vertex $v$ of a clique $C$ belongs to $S^*$; while constraints \eqref{eq:nonempty-sub} impose $S^*$ to be not empty.
If the value of the optimal solution of the PP is larger than $\gamma^*$, $S^*=\{v \in V, x_v^* =1\}$, and the associated variable $z_{S^*}$ is added to the MP.
Note that, since $\pi_{C} \ge 0$ $(C \in \mathcal{C})$ and variables $x_v$ $(v \in V)$ are binary, we can relax constraints (\ref{eq:inty-sub}) to $y_{C} \ge 0$  $(C \in \mathcal{C})$.

We notice that the model \eqref{eq:obj-sub}-\eqref{eq:inty-sub} is equivalent to the model \eqref{fo}-\eqref{ctn1:5} in the previous chapter, where weights are added in the objective function. We shown that the associated polytope is integer, this implies that the pricing problem is polynomial. 

\vspace{-0.5cm}
\paragraph{Branching Scheme:}
Let $\xi^*$  be the current (fractional) solution of the MP. A two-level branching scheme has to be considered.
The first level branching imposes that, for each vertex $v \in V$, either $v$ is in the vertex $k$-cut $V_0$ or it belongs to the vertex-set $S$ of some component of the subgraph of $G$ induced by $V \setminus V0$.
In the second level branching, for two vertices $u$ and $v$ outside $V_0$, we impose that either $u$ and $v$ are in the same component, or they belong to different
ones.
    
\begin{proposition}\cite{cornaz2019vertex}
The two branching rules, applied in sequence, provide a complete branching scheme for model $(ILP_E)$.
\end{proposition}

\section{Experimental results}
The resulting model is then solved using the MIP solver of {\tt Cplex 12.6.0} in single-thread mode and default parameter setting. The resulting solution method is denoted as Cplex in what follows.

The extended formulation ILP\textsubscript{E} is solved via the Branch-and-Price algorithm, initialized with $n$ variables $\xi_S$, where $S=\{v\},~v \in V$. At each column-generation iteration, linear programs are solved with {\tt Cplex 12.6.0}. The pricing subproblem, formulated as a max flow-min cut problem, is solved by means of the pre-flow algorithm of Goldberg and Tarjan \cite{GT88}. We very rarely observed a branching requiring to solve the subproblem as a MIP (i.e., introducing incompatibility constraints between vertices). The exploration of the branching tree is performed in a depth-first fashion.

The experiments have been performed on a computer with a 3.40 Ghz 8-core Intel Core i7-3770    
processor and 16Gb RAM, running a 64 bits Linux operating system. Both exact approaches were tested with a time limit of 3600 seconds of computing time. 

 In the computational experiments, we considered one class of classical graph instances from DIMACS challenges. All considered instances have up to 150 vertices. We only consider graphs for which the size of the largest stable $\alpha(G)$ is at least 5. The results are reported in Table \ref{tab:istFeature}. In the table, after the instance name, we report the number of vertices $n$, the number of edges $m$, the density $d$, and the size of the largest stable set in the graph $\alpha(G)$. This last parameter determines whether a graph instance is feasible for a given value of $k$, i.e., $\alpha(G)\ge k$; and the corresponding stable set provides a feasible solution. We use Cplex to find the maximum stable set in $G$.

 \begin{table}[]

 \centering
 \scriptsize
 \setlength{\tabcolsep}{9pt}
 \renewcommand \arraystretch{1.1}
 \begin{tabular}{lrrrrrlrrrrr}
  \toprule

			&	$n$	&	$m$	&	$d$	&	$\alpha(G)$	&	&				&	$n$	&	$m$	&	$d$	&	$\alpha(G)$	\\
\cmidrule(lr){2-6}\cmidrule(lr){8-11}					\cmidrule(lr){1-1}\cmidrule(lr){7-7}							

{\tt	karate}	&	34	&	78	&	13.90	&	20	&	&		{\tt	polbooks}	&	105	&	441	&	8.08	&	43		\\
{\tt	chesapeake}	&	39	&	170	&	22.94	&	17	&	&	{\tt	adjnoun}	&	112	&	425	&	6.84	&	53			\\
{\tt	dolphins}	&	62	&	159	&	8.41	&	28	&	&	{\tt	football}	&	115	&	613	&	9.35	&	21		\\
{\tt	lesmis}	&	77	&	254	&	8.68	&	35	&	&				&		&						\\[2 ex]

      \bottomrule
\end{tabular}
 \caption{Instance Features} \label{tab:istFeature}
\end{table}

\begin{table}[]
\tiny
\centering
 \renewcommand \arraystretch{1.02}
 \begin{tabular}{lrrrrrrrrrrrrrrrrrrr}
  \toprule
  
&	\multicolumn{ 5}{c}{$k=5$} 								&	&	\multicolumn{ 5}{c}{$k=10$} 								\\

\cmidrule(lr){2-6}\cmidrule(lr){8-12}

&	\multicolumn{ 2}{c}{Branch and Price} 			&	&	\multicolumn{ 2}{c}{Cplex} 			&	&	\multicolumn{ 2}{c}{Branch and Price} 			&	&	\multicolumn{ 2}{c}{Cplex} 			\\

\cmidrule(lr){2-3}\cmidrule(lr){5-6}\cmidrule(lr){8-9}\cmidrule(lr){11-12}

 			&	time	&	nodes	&	&	time	&	nodes	&	&	time	&	nodes	&	&	time	&	nodes	\\

\cmidrule(lr){2-3}\cmidrule(lr){5-6}\cmidrule(lr){8-9}\cmidrule(lr){11-12}
{\tt	karate}	&	 	0.11	 	&	13	&	&	{\bf0.03}	&	0	&	&	{\bf0.03}	&	4	&	&	 	0.06	 	&	0	\\
{\tt	chesapeake}	&	 	1.28	 	&	86	&	&	{\bf0.80}	&	793	&	&	{\bf0.10}	&	15	&	&	 	11.20	 	&	11755	\\
{\tt	dolphins}	&	 	0.72	 	&	1	&	&	{\bf0.29}	&	30	&	&	{\bf0.07}	&	4	&	&	 	8.91	 	&	2887	\\
{\tt	lesmis}	&	 	17.53	 	&	11	&	&	{\bf0.14}	&	0	&	&	 	1.01	 	&	2	&	&	{\bf0.42}	&	8	\\
{\tt	polbooks}	&	 	2036.05	 	&	359	&	&	{\bf58.83}	&	20170	&	&	{\bf330.90}	&	288	&	&	 	$tl$	 	&	631201	\\
{\tt	adjnoun}	&	 	$tl$	 	&	1	&	&	{\bf1.88}	&	40	&	&	 	$tl$	 	&	10	&	&	{\bf139.90}	&	7030	\\
{\tt	football}	&	 	$tl$	 	&	35	&	&	 	$tl$	 	&	615634	&	&	 	$tl$	 	&	149	&	&	 	$tl$	 	&	189697	\\
\cmidrule(lr){2-3}\cmidrule(lr){5-6}\cmidrule(lr){8-9}\cmidrule(lr){11-12}

&	\multicolumn{ 5}{c}{$k=15$} 								&	&	\multicolumn{ 5}{c}{$k=20$} 								\\
\cmidrule(lr){2-6}\cmidrule(lr){8-12}

&	\multicolumn{ 2}{c}{Branch and Price} 			&	&	\multicolumn{ 2}{c}{Cplex} 			&	&	\multicolumn{ 2}{c}{Branch and Price} 			&	&	\multicolumn{ 2}{c}{Cplex} 			\\

\cmidrule(lr){2-3}\cmidrule(lr){5-6}\cmidrule(lr){8-9}\cmidrule(lr){11-12}

 			&	time	&	nodes	&	&	time	&	nodes	&	&	time	&	nodes	&	&	time	&	nodes	\\

\cmidrule(lr){2-3}\cmidrule(lr){5-6}\cmidrule(lr){8-9}\cmidrule(lr){11-12}
{\tt	karate}	&	{\bf0.02}	&	4	&	&	 	0.06	 	&	0	&	&	{\bf0.01}	&	3	&	&	 	0.08	 	&	100	\\
{\tt	chesapeake}	&	{\bf0.06}	&	9	&	&	 	8.46	 	&	4903	&	&	 		 	&		&	&	 		 	&		\\
{\tt	dolphins}	&	{\bf0.16}	&	8	&	&	 	316.13	 	&	195342	&	&	{\bf0.10}	&	4	&	&	 	$tl$	 	&	1688935	\\
{\tt	lesmis}	&	{\bf0.58}	&	6	&	&	 	1.23	 	&	804	&	&	{\bf0.41}	&	4	&	&	 	7.44	 	&	2226	\\
{\tt	polbooks}	&	{\bf25.94}	&	44	&	&	 	$tl$	 	&	279125	&	&	{\bf3.13}	&	11	&	&	 	$tl$	 	&	267568	\\
{\tt	adjnoun}	&	 	$tl$	 	&	12	&	&	{\bf2337.24}	&	157113	&	&	 	$tl$	 	&	28	&	&	 	$tl$	 	&	90669	\\
{\tt	football}	&	 	$tl$	 	&	1544	&	&	 	$tl$	 	&	117103	&	&	 	$tl$	 	&	9228	&	&	 	$tl$	 	&	57614	\\
\cmidrule(lr){2-3}\cmidrule(lr){5-6}\cmidrule(lr){8-9}\cmidrule(lr){11-12}

    \bottomrule
\end{tabular}
 \caption{Formulation performance comparison  \textcolor{blue}{on the DIMACS instances} ($k=5$ and $k=10$)} \label{tab:comparison1}
\end{table}

In Table \ref{tab:comparison1} we consider values of $k=5,10,15,20$, and report, for Branch-and-Price and Cplex, the CPU time is seconds ($tl$ for time limit) and the number of explored nodes in the Branch-and-Bound tree. For each instance and for each value of $k$, we report in bold the fastest method. The missing lines correspond to infeasible instances. At the end of each block, we report the number of instances solved to optimality by each method.

From these results, we can conclude that Cplex has an average good performance for $k=5$, and has increasing difficulties for larger values of $k$. This can be explained by the fact that when $k$ increases, the number of variables also increases, and the algorithm becomes slower. A partial explanation can be found in the increase in the number of variables ($n$ more variables for each incremental value of $k$). For $k=5$, Cplex outperforms Branch-and-Price. For Branch-and-Price, an opposite behavior is experienced when increasing the value of $k$. In this case, the performance of the method is improved. For example, instance {\tt polbooks} needed 2036.05 seconds for $k=5$, while 330.90, 25.94, and 3.13 seconds were needed for $k=10,15$ and 20, respectively. For $k=10,15$ and $20$, Branch-and-Price outperforms then  Cplex.

\section{Conclusion and open questions}

In this chapter, we have proved that the vertex $k$-cut problem is NP-hard if $k\geq3$. We have proposed two mathematical formulations. The first one is the compact formulation that can be solved using a commercial solver like Cplex. The second formulation proposed, called extended formulation, has an exponential number of variables. To solve this model, we have proposed an efficient Branch-and-Price algorithm thanks to the two following key ingredients, column generation to add on the fly the variables and a complete branching scheme. We have showed that the first model is the better one if the value of $k$ is small, less than 10, otherwise the second model is the best one. 

In other papers, we have considered other variants of the vertex $k$-cut problem.
The first variant consists in ensuring that sets of the partition have a balanced size. In \cite{cornaz2014mathematical}, we formalize the problem by describing one metric of "balanced size" and propose mathematical models to solve this problem. In \cite{grange2021approximate}, we show the relationship between this problem and scheduling problems.  
The second ones have in addition terminals. In this case, we consider $k$ vertices called terminals where the goal is to find the minimum number of vertices to remove in order to disconnect the terminals. This implies that no path exists in the remaining graph between any pair of terminals. In \cite{magnouche2016multi} we propose an efficient model based on path deletion and strengthen the model with valid cuts. In \cite{magnouche2016multi2} and \cite{MAGNOUCHE202186} we propose an extended model to solve the multi-terminal vertex separator. 

The vertex $k$-cut problems appear in several real problems of partitioning/clustering/decomposition. For each problem, additional constraints are considered. The most common constraint is the notion of "balanced size" as the fairness between each set of the partition obtained after removing the vertex cut. 
\chapter{Flow Blocker Problems}

In this chapter, we present original research from \cite{magnouche2020most},\cite{isma}. These two papers focus on the blocker variant associated with the shortest path and the maximum flow problems. Some extensions or variants can be found in \cite{ismaPhD}.
We analyze the complexity, and some structural properties, and provide mathematical models to solve efficiently these two problems.

\vspace{-0.5cm}
\paragraph{Applications:} In telecommunication networks, blocker problems allow monitoring the traffic to deduce traffic fluctuation and improve load balancing and the Quality of Service. Indeed, the goal is to cover paths or flows with a given property, like a small delay or flow with a big size. Another application, in telecommunication networks, is the analysis of the strength of the network against simultaneous failures.

\vspace{-0.5cm}
\section{Shortest path blocker problem}

The shortest path problem is a well-known problem. In this chapter, we consider the most vital vertices and the blocker point of view. The two problems are very close.

In \cite{khachiyan2008short}, Khachiyan et al. introduce the Minimum Vertex Blocker to Shortest Path problem (MVBSP)
as follows:

let $D = (V\cup\{s,t\},A)$ be a digraph, where $V \cup \{s,t\}$ is a set of nodes with two distinguished nodes $s$ and $t$, and $A$ is a set of arcs. Given a nonnegative length $l(a)$ associated with each arc $a\in A$ and a threshold $k \in \mathbb Z^+$, a \textit{vertex blocker} is a set of vertices whose removal increases the $s$-$t$-distance to at least $k$. The objective is to find the smallest vertex blocker i.e., $ \min \{|U|~| ~ d_{D[V\setminus U]}(s,t) \geq k, U \subseteq V\setminus \{s,t\}\}$, where $d_{D}(s,t)$ is the $s$-$t$-distance in $D$ (the length of the shortest path between $s$ and $t$ in $D$).

The most vital vertices for the shortest $s$-$t$ path problem (MVVSP) are defined as follows:
given a digraph $D=(V\cup\{s,t\},A)$ with a nonnegative length $l(a)$ associated with every arc $a\in A$ and a threshold $d>0$, a set of vertices is {\em vital} if its removal ensures that it does not exist a $s$-$t$ path of length less or equal than $d$. The aim is to find the smallest vital set.

In \cite{magnouche2020most} we have shown that the MVVSP is NP-hard on a reduction from the covering set problem. 

For some optimization problems, the most vital notion and the blocker notion are the same. 
\vspace{-0.5cm}

\subsection{Terminal connectivity and formulation}

In this section, we present a mathematical model based on the model given in \cite{cornaz2019vertex} for solving the terminal vertex 2 cut problem.  

Let $x\in \{0,1\}^{|V|}$ defined as
\begin{quote}
$\qquad x_v =  \left \{ \begin{array}{ll} 1 & \mbox{ if }  v \mbox{ is a vital vertex},\\0 &\mbox{ otherwise, }  \end{array} \right. \quad \forall v\in V.$
\end{quote}

The MVVSP is equivalent to the following ($P'$):
\begin{alignat}{6}
\nonumber & & \quad && \quad\min \sum_{v\in V} x_v\\
\label{cnt:1}  &(P') &\quad && \sum_{v\in P} x_v \geq 1, & \quad \mbox{ for all } P\in \mathcal{P},\\
\label{cnt:3}  & &\quad && x_v \in \{0,1\}, &  \quad \mbox{ for all } v\in V,
\end{alignat}

where $\mathcal{P}$ is the set of all $s$-$t$ paths of length less or equal to $d$. The inequalities \eqref{cnt:1} ensure that at least one node must be interdicted in any path of length less or equal to $d$. 
Remark that, the length of $s$-$t$ paths does not appear in the model. Indeed, the length is in the definition of $\mathcal{P}$.
\\

Let $P(D,d)= conv(x \in \{0,1\}^{|V|} |\textit{ } x \textit{ satisfies } (\ref{cnt:1}))$ be the polytope of vital vertices sets for a length $d$ in graph $D=(V,A)$. \\

Given an inequality $ax \leq b$, where $a \in \mathbb R^V$, the support graph of $ax \leq b$ is the subgraph induced by the vertices corresponding to variables having a non-zero coefficient in the inequality.

\begin{theorem}
$P(D,d)$ is full-dimensional.
\end{theorem}
\begin{theorem}
For a path $P\in \mathcal P$, inequality (\ref{cnt:1}) defines a facet of $P(D,d)$ if and only if
$P$ is minimal.
\end{theorem}

Inequalities \eqref{cnt:1} are in exponential numbers. In order to solve ($P'$) using a Branch-and-Cut approach, one needs an efficient algorithm for the separating inequalities \eqref{cnt:1}.
\vspace{-0.5cm}
\paragraph{Separation algorithm}

The separation problem for inequalities \eqref{cnt:1} consists, given a solution $x^* \in \mathbb{R}^{|V|}$, in determining whether $x^*$ satisfies inequalities \eqref{cnt:1}, and if not in finding an inequality violated by  $x^*$. 

The separation problem consists in finding a $s$-$t$ path $P$ such that $l(P)\leq d$  and minimizing the cost function $\sum_{v\in P} x^*_v$. We denote by $A(P)$ the sequence of arcs associated with the path $P$.
In the following, we prove that the separation problem is NP-hard using a polynomial reduction from the shortest weight-constrained $s$-$t$ path problem \cite{garey1979guide}, known to be NP-hard. Given a directed graph $D=(V,A)$, a cost function $c: A\Rightarrow \mathbb Z^+$, a weight function $w: A\Rightarrow \mathbb Z^+$, a threshold $W\in \mathbb N$ and two terminal vertices $s,t\in V$, the shortest weight-constrained $s$-$t$ path problem consists in finding a $s$-$t$ path $P$ with minimum cost $\sum \limits _{a\in A(P)}c(a)$ such that $\sum \limits _{a\in A(P)}w(a)\leq W$.
Let $D'=(V', A')$ be the graph obtained from $D$ by adding, for each arc $a=(u,v)\in A$, a vertex $z^a$ to $V'$ with a cost $c(a)$  and replacing arc $a$ by two arcs $(u, z^a)$ and $(z^a, v)$, each with a weight $\dfrac{w(a)}{2}$. The costs of $u$ and $v$ are 0.
Clearly, the optimal solution given by solving the separation problem on $D'$ can be transformed to an optimal solution of the shortest weight-constrained $s$-$t$ path problem on $D$, in a polynomial time. Therefore the separation problem is NP-Hard.\\
However, if $x^* \in \mathbb{N}^{|V|}$, then the separation becomes polynomial.
This case consists in finding a $s$-$t$ path $P$ such that $l(P)\leq d$ and $\sum_{v\in P} x^*_v < 1$. Since the vector $x^*$ is integer, we search a shortest $s$-$t$ path without any vertex $v\in V$ satisfying $x^*_v=1$.

\vspace{-0.5cm}
\subsection{Branch-and-Cut algorithm and experimental results}
We have developed a Branch-and-Cut algorithm to solve the MVVSP. As mentioned in the previous sections, the integer linear program has an exponential number of inequalities \eqref{cnt:1}. In our Branch-and-Cut, we only separate the integer vector $x^*$ by separating inequalities \eqref{cnt:1} using Dijkstra algorithm for solving the shortest $s$-$t$ path problem. 

We now describe the framework of our algorithm. To start the optimization, we consider the linear program  with only trivial inequalities. The current optimal solution $x^*\in \mathbb R^V$ of the linear relaxation is feasible for the problem if $x^*$ is an integer vector that satisfies all inequalities \eqref{cnt:1}. Usually, the solution $x^*$ is either fractional or not feasible for the MVVSP. In each iteration of the Branch-and-Cut algorithm, if $x^*$ is fractional, one has to branch on a fractional variable $x_i$ by generating two child nodes, one with an additional constraint $x_i = \lfloor x_i \rfloor$ and the other one with $x_i = \lceil x_i \rceil$. However, when $x^*$ is integer but not feasible, it is necessary to generate further inequalities \eqref{cnt:1} violated by $x^*$. For this, one has to solve the separation problem. 
The Branch-and-Cut algorithm only uses inequalities \eqref{cnt:1}. In our implementation, the solver Cplex is used to handle the branching tree and solve the linear programs.

The computational results are obtained using Cplex 12.6 and Lemon 1.3.1. 

The required CPU time is measured in seconds. We limit to 3600 seconds the running time for each instance using at most 8 GB of RAM and a processor Intel Core i5-3340M CPU of 2.70GHz $\times$ 4.\newline
The integer linear program $(P')$ is tested on random instances. The graph density is equal to 2, 10, 25, 50 and 75 percent. For each instance, we consider all different values of $d$, between $sp+1$ and $disc$ where $sp$ is equal to the length of the shortest $s$-$t$ path and $disc$ is the minimum value for $d$ for which the smallest vital set disconnects $s$ and $t$.
The next tables provide the following information: 
\begin{itemize}
\item $|V|$, the number of vertices;
\item density, the density of $D$; 
\item $d$, the bound of remaining shortest $s$-$t$ paths (all shortest $s$-$t$ path must be of length strictly greater than $d$);
\item Nodes, the number of nodes in the branching tree;
\item CPU, Computational time (limited to 1 hour);
\item $\#$cuts, the number of inequalities \eqref{cnt:1} added in the Branch-and-Cut algorithm;
\item Size Blocker, the size of the blocker for this instance.
\end{itemize}

\begin{table}[h]
 \centering
 \caption{Average of CPU time.}
 \label{tab:ins}
 \begin{tabular}{l c c c r r r  }
  \toprule
$|V|$  & density & $d$ & Nodes & CPU & $\#$cuts & Size Blocker \\
  \toprule 
16000 &   2 & 3 & 1 & 9.61 & 7 & 7\\
16000 &  2 & 4 & 1 & 1136.42 & 857 & 334\\
16000 &   2 & 5 & 1 & 1267.29 & 871 & 335\\
19000 &   2 & 3 & 1 & 16.35 & 7 & 7\\
19000 &   2 & 4 & 1 & 2482.30 & 1110 & 360\\  
6000 &  10 & 3 & 1 & 64.24 & 77 & 77\\ 
8000 &  10 & 3 & 1 & 138.77 & 83 & 83\\  
8000 &  10 & 4 & 1 &  Time Limit & 2759 & 780\\  
\toprule 
1500 &  25 & 3 & 1 & 15.19 & 87 & 87\\ 
1500 &  25 & 4 & 1 & 175.03 & 1293 & 365\\ 
2000 &  25 & 3 & 1 & 44.44 & 130 & 130\\ 
2000 &  25 & 4 & 1 & 490.33 & 1816 & 515\\ 
2500 &  25 & 3 & 1 & 50.73 & 148 & 148\\ 
2500 &  25 & 4 & 1 & 955.44 & 2333 & 610\\  
1500 &  50 & 3 & 1 & 110.14 & 376 & 376\\ 
1500 &  50 & 4 & 1 & 297.69 & 1672 & 753\\ 
2000 &  50 & 3 & 1 & 161.38 & 468 & 468\\ 
2000 &  50 & 4 & 1 & 521.51 & 2406 & 967\\ 
2500 &  50 & 3 & 1 & 351.03 & 648 & 648\\ 
2500 &  50 & 4 & 1 & 1765.47 & 2722 & 1258\\  
1500 &  75 & 3 & 1 & 169.87 & 844 & 844\\ 
1500 &  75 & 4 & 1 & 218.69 & 1812 & 1111\\ 
2000 &  75 & 3 & 1 & 444.06 & 1144 & 1144\\ 
2000 &  75 & 4 & 1 & 817.83 & 2470 & 1496\\ 
2500 &  75 & 3 & 1 & 1451.76 & 1389 & 1389\\ 
2500 &  75 & 4 & 1 & 1567.47 & 2622 & 1826 \\
\bottomrule
 \end{tabular}
\end{table}
In Table \ref{tab:ins}, we consider random instances. Note that all these instances are solved in the root node even for the large graphs. Furthermore, we solved all the instances in less than 1 hour. Notice that, when $d$ is equal to $sp+1$ then the number of cuts and the size blocker are equal. This is not true when $d$ is equal to $disc$. Our Branch-and-Cut algorithm can solve instances with 2500 vertices in less than 30 minutes. We remark that the density and the bound $d$ impact the efficiency of our algorithm. In most of instances we note that $sp+1= disc-1$, except when the density is $2\%$, there is an instance with $sp+1 = disc-2$.

\section{Maximum flow blocker problem}

The maximum flow problem is a well-studied problem with many applications in telecommunication networks and vehicle routing. This problem can be solved in polynomial time and in this section, we consider the maximum flow blocker problem. We show NP-completeness and provide a strong relationship between the maximum flow blocker and the maximum flow interdiction problems. To the best of our knowledge, this kind of relationship between interdiction and blocker is new.  

Let $G=(V,A)$ be a directed graph with \textit{m} $= |A|$ arcs and \textit{n} $= |$\textit{V}$|$ vertices that contains two special vertices: the source $s \in V$ and the destination $t \in V$. Each arc $a \in A$ is given a \textit{capacity} $c_a \in \mathbb{Z}_+$. 
The classical \textit{maximum flow problem} (MFP) asks to determine the maximum value of a flow from the source to the destination respecting all arc capacities. The blocker variant of the MFP, called the \textit{maximum flow blocker problem} (MFBP), in which each arc $a \in A$ is also given an interdiction cost $r_a \in \mathbb{Z}_+$, consists in finding a minimum-cost subset of arcs to be removed, i.e., interdicted from the graph, in such a way that the maximum flow value between $s$ and $t$ in the remaining graph is no larger than a given threshold. The threshold is called the \textit{target flow value} and it is denoted by $\Phi$.

The main theorem of article \cite{isma} gives a relationship between the maximum flow interdiction problem and the maximum flow blocker problem. To introduce this let us first give some notations.
We denote by ${\bf c} \in \mathbb{Z}_+^m$ the arc-capacity vector, by ${\bf q} \in \mathbb{Z}_+^m$ the interdiction cost vector and by ${\bf r} \in \mathbb{Z}_+^m$ the blocker cost vector. Accordingly, the tuple $(G, {\bf c}, {\bf q}, \Psi)$ represents the associated MFIP instance and the tuple $(G, {\bf c}, {\bf r}, \Phi)$ represents the associated MFBP instance. The following theorem allows to obtain an optimal MFBP solution starting from an optimal MFIP solution of an instance where the interdiction costs are set to the capacities of the arcs, the capacities of the arcs are set to the blocker costs of the arcs and the interdiction budget is set to the target flow.

\begin{theorem}
\label{PROP_EQUIVALENCE_RELATION}

Given an optimal solution $({\bf w}, {\boldsymbol \beta}, {\boldsymbol \alpha})$ for the MFIP instance $(G,{\bf {\bf r}}, {\bf {\bf c}},\Phi)$, an optimal solution for the MFBP instance $(G,{\bf {\bf c}}, {\bf {\bf r}},\Phi)$ is $x_a = \beta_a$ for all $a \in A$.
\end{theorem}

This theorem is based on the max-flow/min-cut relationship. By proving that a solution of the MFBP is included in an edge-cut of the graph, we can deduce the strong relationship described in Theorem \ref{PROP_EQUIVALENCE_RELATION}.

\subsection{ILP formulations for the MFBP}

\label{SECTION_COMPACT_MFBP}

By using the vector of binary variables ${\bf x} \in \{0,1\}^m$ representing the blocked arcs and two additional vectors of binary variables ${\boldsymbol \omega} \in \{0,1\}^m$ and ${\boldsymbol \gamma \in \{0,1\}^n}$ representing the vectors of binary variables ${\bf w}$ and $\boldsymbol \alpha$ of model given in \cite{Wood11} to solve the interdiction problem, respectively, Theorem \ref{PROP_EQUIVALENCE_RELATION} allows us to derive the following ILP formulation for the MFBP:
\begin{subequations}
\label{COMPACT_MFBP}
\begin{align}
&& \zeta({\rm MFBP}) \;=\; \min_{{\bf x}, {\boldsymbol \omega} \,\in\, \{0,1\}^m, {\boldsymbol \gamma} \,\in\, \{0,1\}^n} ~~~~ {\sum_{a \in A} r_{a} \; x_{a}} \label{COMPACT_MFBP_OBJ} \\[1.5ex]
&& \; x_{uv} + \omega_{uv} + \gamma_v - \gamma_u &\geq 0, & \forall \; (u,v) \in {A}, \label{COMPACT_MFBP_C1}\\[1.5 ex]
&& \; \gamma_{s} - \gamma_t &\geq 1, \label{COMPACT_MFBP_C2}\\[1.5ex]
&& \; \sum_{a \in A} c_a \; \omega_a &\leq \Phi.\label{COMPACT_MFBP_C3}
\end{align}
\end{subequations}

It is worth noticing that constraints \eqref{COMPACT_MFBP_C1} and \eqref{COMPACT_MFBP_C2} are related to ensuring that the solution is in a cut.  For this reason, we observe that in any optimal solution of Model \eqref{COMPACT_MFBP}, for a given arc $a \in A$, we can have either $\omega_a=1$ or $x_a=1$ but not both. Accordingly, an optimal solution of \eqref{COMPACT_MFBP} is also a cut $\delta(U^{G}({\bf x}))$ in the graph $G$, which depends on an optimal blocker policy ${\bf x}$. This cut $\delta(U^{G}({\bf x}))$ is given by the arcs $a \in A$ where $\omega_a = 1$ or $x_a = 1$ and it is the union of the set of non-blocked arcs such that $x_a = 0$ and $\omega_a = 1$ and the set of blocked arcs such that $x_a = 1$ and $\omega_a = 0$. If a variable $\gamma_u$ is equal to $1$, it indicates that vertex $u$ is in the subset $U^{G}({\bf x})$ containing the source $s$, and if it is equal to $0$, it indicates that vertex $u$ is in the subset $V \setminus U^{G}({\bf x})$ containing the destination $t$.  In addition, as for the MFIP, any optimal solution $({\bf x}, {\boldsymbol \omega}, {\boldsymbol \gamma})$ of Model \eqref{COMPACT_MFBP} contains the minimum cut $\delta(U^{\mathcal{G}_{NB}}({\bf x}))$ in the non-blocked graph $\mathcal{G}_{NB}({\bf x})$ given by the arcs $a \in \mathcal{A}_{NB}({\boldsymbol \omega})$ such that $\omega_a = 1$, and $U^{\mathcal{G}_{NB}}({\bf x})$ is a set of vertices containing the source $s$.

The objective function \eqref{COMPACT_MFBP_OBJ} minimizes the total cost of blocked arcs. Constraint \eqref{COMPACT_MFBP_C3},  called \textit{target flow constraint}, imposes that the capacity remaining in the cut $\delta(U^{G}({\bf x}))$, i.e., the maximum flow in the non-blocked graph, is less than or equal to the target flow $\Phi$. Constraint \eqref{COMPACT_MFBP_C1} enforces that an arc $(u,v)$ must be in the cut $\delta(U^{G}({\bf x}))$ if $\gamma_u = 1$ and $\gamma_v = 0$, which implies that either $\omega_{(u,v)} = 1$ or $x_{(u,v)} = 1$.  On the other hand, constraint \eqref{COMPACT_MFBP_C2} imposes that the source $s$ belongs to the set $U^{G}({\bf x})$, i.e., $\gamma_s = 1$ and the destination $t$ belongs to the set $V \setminus U^{G}({\bf x})$, i.e, $\gamma_t = 0$.

\subsubsection{Benders decomposition}

The Benders decomposition consists in dividing the variables of the original problem into two subsets so that a first-stage master problem is solved over the first set of variables, and the values for the second set of variables are determined in a second-stage subproblem for a given first-stage solution \cite{benders}. 
By applying a Benders-like decomposition to the bilevel Model, where variables $x$ are considered in the first-stage master problem. The  polytope of the feasible solutions for the follower subproblem, which does not depend on the leader variables, can be defined as follows: 
\begin{equation}
\label{KKKK}
 \mathcal{P}_{f} = \bigg\{~ {{\bf y} \,\in\,\mathbb{Q}^m_+}: \sum_{a \,\in\, \delta^+(s)}  y_a \ge \Phi+1,~~ \text{ flow conservation constraints },~  y_a   \leq c_a,~ \forall \; a \in A~~\bigg\}.
 \end{equation}
 
 We obtain the following single-level ILP reformulation for the MFBP: 
\begin{subequations}
\label{MODEL_NAT_Benders_MFBP}
\begin{align}
&& \zeta({\rm MFBP}) \;=\; \min_{{\bf x} \,\in\, \{0,1\}^m} \quad \sum_{a \, \in \, A} r_a \; x_a \\[2 ex]
     &&\sum_{a \, \in \, \delta^+(s)} y_a    -  \sum_{a \in A} x_a \; y_a &\leq \Phi,  & \forall \; {\bf y} \in ext(\mathcal{P}_{f}).
\label{BendersCuts} 
\end{align}
\end{subequations}

where constraints \eqref{BendersCuts}, {called \textit{Benders cuts}}, is an exponential-size family of constraints one for each extreme point of  $\mathcal{P}_{f}$. This ILP model is called \textit{natural formulation} since it features only the natural variables associated with the arcs. It is denoted  by n-ILP in the remainder of the article.


In order to solve n-ILP, we develop a Branch-and-Benders-Cut approach, i.e., a Branch-and-Cut algorithm where Benders cuts \eqref{BendersCuts} are separated in the nodes of the branching tree for integer and fractional solutions. This exact algorithm requires defining a \textit{relaxed master problem} (RMP) where the binary variables are replaced with continuous variables taking values between 0 and 1.  Only a subset of constraints are included in the RMP in the initialization phase.

\paragraph{Target-flow inequalities}
For a given vector ${\bf y} \in ext(\mathcal{P}_{f})$, we define the subset of arcs $\mathcal{A}_S({\bf y)} \subseteq A$ routing a strictly positive flow in the extreme point ${\bf y}$ as follows: $$\mathcal{A}_S({\bf y)} = \big \{a \in A : y_a > 0 \big \}.$$   

These arcs induce the \textit{support graph} $\mathcal{G}_S({\bf y}) = (V, \mathcal{A}_S({\bf y)})$ in which, by construction, the maximum flow value $\psi(\mathcal{G}_S(\bf y))$ is larger than or equal to $\Phi + 1$. The following set of constraints, called \textit{target-flow inequalities}, are valid for the natural formulation n-ILP:

\begin{align}
     &&\sum_{a \, \in \mathcal{A}_S({\bf y})}  x_a  &\geq 1,  & \forall \; {\bf y} \in ext(\mathcal{P}_{f}).
     \label{COVER} 
\end{align}
For any vector ${\bf y} \in ext(\mathcal{P}_{f})$, the associated constraint \eqref{COVER} is valid since it imposes to block at least one arc in the subset $\mathcal{A}_S({\bf y}) \subset A$. In other words, the constraint prevents having a flow of value strictly larger than $\Phi$ in the support graph $\mathcal{G}_S({\bf y})$. Clearly, it is an exponential family of constraints, one for each extreme point ${\bf y} \in ext(\mathcal{P}_{f})$.

In \cite{isma} the separation algorithms and Branch-and-Benders-Cut algorithm are described in detail. 

\vspace{-0.5cm}
\subsection{Comparison of the effectiveness of the natural and the compact formulations}

\label{SEC_COMPARE_NAT_COMP}
In this section, we compare the computational performance of our best Branch-and-Cut algorithm for the natural formulation n-ILP, i.e., {\tt BENDERS\_TF}, against the compact formulation c-ILP solved by {\tt CPLEX} MIP solver. This study is performed on instances from the {\tt SYNTHETIC} class, including graphs with $50, 100$, and $300$ vertices, with the same density values as mentioned earlier, i.e., $d(G) \in \{0.2, 0.4, 0.6, 0.8\}$. Additionally, instances from the {\tt GRID} class and the {\tt REAL-NETWORKS} class are also examined. We consider three values of the target flow defined by $\lambda \in \{0.2, 0.6, 0.9\}$. Results are reported in Table \ref{Table_RANDOM_CF_NF}. For each formulation, we report the number of instances solved to optimality, the average and maximum computing time, and the average number of nodes explored in the branching tree (nodes). 

Table \ref{Table_RANDOM_CF_NF} directly shows that c-ILP outperforms n-ILP. For the smallest graphs ($n=50$), c-ILP solves all instances in record time; less than $0.1$ seconds for all densities. For the same group of instances, n-ILP manages to solve all instances in a reasonable time, i.e., at most $27.88$ seconds for a density of $0.8$. However, for larger graphs, the performance spread between the two formulations is  exacerbated. More precisely, we observe that for graphs of $100$ vertices with a density of $0.4$, n-ILP fails to solve three instances to optimality. This number increases to seven for a density $d(G)$ equal to $0.8$. Subsequently, for graphs of $300$ vertices and a density larger than $0.2$, none of the instances were solved to optimality within the time limit. In addition, for the natural formulation n-ILP, the number of nodes explored in the branching tree tends to increase as the size of the graph grows. In particular, for graphs with $50$ vertices and a density of $0.8$, the maximum number of explored nodes reaches an approximate value of $190$ while for graphs with $100$ vertices, this number averages around $530$.
For larger instances, such as graphs of $300$ vertices with a density greater than or equal to $0.6$, no nodes are explored within the time limit. This observation highlights that the linear relaxation of n-ILP requires a substantial amount of time. Furthermore, based on the results from previous instances, it can be inferred that the time invested in the linear relaxation does not contribute to an overall reduction in the computing time of the Branch-and-Cut algorithm for n-ILP. In contrast to n-ILP, c-ILP successfully solves all instances with $100$ and $300$ vertices at the root node, achieving a maximum computing time of $2.13$ seconds. It is worth noticing that the high performance of c-ILP comes essentially from enhancements made by the solver. Indeed, due to the combinatorial nature of integer programs, recent versions of {\tt CPLEX} incorporate specific operations to improve the branching algorithm,  which are applied automatically in the default setting. However, {\tt CPLEX} offers many parameters that allow users to customize the problem-solving approach. After conducting a series of experiments, we have identified two settings that significantly impact the performance of c-ILP. The first setting decides whether {\tt CPLEX} applies presolve during preprocessing, which involves performing various reductions to eliminate variables and consequently reduce the problem size. The second setting involves the incorporation of additional cuts into the model, such as Gomory cuts, clique cuts, and various other types of cuts. Our experiments have shown that performing presolve and adding cuts to the model can help enhance the efficiency of the solver, especially when dealing with large graphs. This can be explained by the structure of the compact formulation c-ILP. More precisely, it has been shown that the solution of the MFBP is contained in a cut of the graph and constraints \eqref{COMPACT_MFBP_C1} and\eqref{COMPACT_MFBP_C2} of c-ILP are constraints defining a cut. Furthermore, constraint \eqref{COMPACT_MFBP_C3} belongs to a well-known optimization problem, namely the knapsack problem. 
Hence, constraints of c-ILP are familiar to {\tt CPLEX}, that will be able to perform good preprocessing and general improvements.

\begin{center}
\begin{table}[h!]
\centering
\tiny
\tabcolsep 7pt
\renewcommand
\arraystretch{1.2}
\begin{tabular}{lrrrrrrrrrrrrrrrrrrrrrrrrrrrrrrrrr}
\toprule

	&	&	&	&  	\multicolumn{4}{c}{n-ILP } 		&	&	\multicolumn{4}{c}{c-ILP}	\\
\cline{5-8}
\cline{10-13}
	&		&		&&		&	\multicolumn{2}{c}{time}		& nodes		&&	&	\multicolumn{2}{c}{time}			& nodes		\\
\cline{6-7}
\cline{11-12}

$n$	&	$d(G)$	&	 $\#$	&&	 $\#$ opt	&	avg.	&	max	&	avg.	&&	$\#$ opt	&	avg.	&	max	&	avg.	\\

\cline{1-3}
\cline{5-8}
\cline{10-13}

50	&	0.2	&	15	&&	15	&	0.02	&	0.07	&	7	&&	15	&	0.01	&	0.01	&	0	\\
	&	0.4	&	15	&&	15	&	0.18	&	0.61	&	37.53	&&	15	&	0.03	&	0.06	&	0	\\
	 &	0.6	&	15	&&	15	&	0.7	&	2.65	&	71.33	&&	15	&	0.04	&	0.05	&	0	\\
	&	0.8	&	15	&&	15	&	3.41	&	27.88	&	189.67	&&	15	&	0.05	&	0.08	&	0	\\[2ex]
100	&	0.2	&	15	&&	14	&	77.8	&	t.l.	&	214	&&	15	&	0.05	&	0.08	&	0	\\
	&	0.4	&	15	&&	11	&	184.2	&	t.l.	&	462.87	&&	15	&	0.06	&	0.08	&	0	\\
	&	0.6	&	15	&&	8	&	321.1	&	t.l.	&	263.73	&&	15	&	0.08	&	0.09	&	0	\\
	&	0.8	&	15	&&	7	&	351.3	&	t.l.	&	530.87 	&&	15	&	0.11	&	0.21	&	0	\\[2ex]
300	&	0.2	&	15	&&	5	&	469.73	&	t.l.	&	251.6	&&	15	&	0.26	&	0.41	&	0	\\
	&	0.4	&	15	&&	0	&	t.l.	&	t.l.	&	2.27	&&	15	&	0.53	&	0.92	&	0	\\
	&	0.6	&	15	&&	0	&	t.l.	&	t.l.	&	0	&&	15	&	0.94	&	1.51	&	0	\\
	&	0.8	&	15	&&	0	&	t.l.	&	t.l.	&	0	&&	15	&	1.48	&	2.13	&	0	\\[2ex]
\cline{1-3}
\cline{5-8}
\cline{10-13}
Total	&		&	180	&&	105	&		&		&		&& 180	&		&	&		\\
\bottomrule
\end{tabular}

\caption{Performance comparison between c-ILP and n-ILP on {\tt SYNTHETIC} instances}
\label{Table_RANDOM_CF_NF}
\end{table}
\end{center}

\section{Conclusion and open questions}

In this chapter, we have provided two strong results. In the first studied problem, the shortest path blocker problem, we proved that the problem is NP-hard. 
We have derived an efficient Branch-and-Cut algorithm able to solve large-scale instances with a small number of nodes in the branching tree. The second problem studied is the maximum flow blocker problem, where we have proved that any algorithm able to solve the interdiction version can be directly used to solve the blocker version of the maximum flow problem.

A further a natural question that can be considered is: Would it be possible to extend the results of the paper to more general flow blocker problems? A first generalization may concern the problem with multiple sources and destinations. This problem, also known as the multi-commodity flow problem, aims at maximizing the flow between a set of source-destination pairs.
Another extension would consider the fractional blocker, where instead of blocking arcs in the graph, we only block a fraction of the capacity of the arc.

\chapter{The $m$-clique free interval subgraph problem}

In this chapter, we present original research from \cite{hassan2018m}. 
We focus on interval subgraph and graphs without cliques of size $m$+1. The $m$-clique free interval subgraph problem can be seen as a blocker problem where the goal is to ensure that several graph structures do not exist: bipartite claw, umbrella, $n$-net, $n$-tent, a hole of a given size, and a clique of a given size (Figure \ref{chap4:fig:fsg}).

\vspace{-0.5cm}
\paragraph{Applications:}  The property of interval subgraph and $m$-clique free is a core structure of scheduling problems when several machines are considered. Furthermore, some forbidden structures appear in other problems. For instance, the $m$-clique free structure has an application in the clique blocker problem. 

Algorithmic aspects of
interval graphs have been the subject of ongoing research for several decades,
stimulated by their numerous applications; see e.g. \cite{gonzalez1976open}. In some
applications, interval representations with special properties are
required.\newline Numerous applications of interval graphs have appeared in
the literature including applications to genetic structure, sequential storage, and scheduling (see \cite{gacias2010parallel}). An application of interval graphs arises
in the context of scheduling jobs in cloud computing. Here we do
not only have to determine how many, but also which jobs should be allocated
to a virtual machine. In scheduling, for example, jobs can have a certain
duration that should be reflected by the lengths of their intervals and two
consecutive jobs can require a certain handover period that is determined by
how much their intervals should intersect.\newline

 Let $G=(V,E)$ be
a graph. An undirected graph $G$ is called an \textit{interval graph} if its vertices
can be put into a one-to-one correspondence with a set of intervals $I$ of a
linearly ordered set (like the real line) such that two vertices are connected
by an edge of $G$ if their corresponding intervals have nonempty
intersection.\newline An interval graph is a graph showing intersecting
intervals on a line. Thus, we associate a set of intervals $I=\{I_{1}%
,...,I_{n}\}$ on a line with the interval graph $G=(V,E)$, where
$V=\{1,...,n\}$ and two vertices, $u$ and $v$, are linked by an edge if and
only if $I_{u}\cap I_{v}\neq\emptyset$.

In parallel machines, scheduling some jobs in different machines can share any
time units, i.e. running at the same time.  The jobs can be represented by nodes and edges indicating
there is a shared time units between jobs. Normally, the solution is
mathematically formalized as a graph $G=(V,E)$ where $V$ is associated with jobs and $E$ represents the intersections of jobs. However, when we assign jobs to
parallel machines, the solution is valid for two types of graphs (i.e.,
interval graph and $m$-clique free, where $m$ is the number of machines).

\section{Interval graph analysis}
\begin{figure}[th]
\centering
\begin{subfigure}[b]
\centering
\begin{tikzpicture}[line width=1pt]
	\vertex (2) at (0,.5) [label=right:$2$] {};
	\vertex (1) at (0,0) [label=right:$1$] {};
	\vertex (3) at (-.5,-.5) [label=left:$3$] {};
	\vertex (4) at (.5,-.5) [label=right:$4$] {};
	\vertex (5) at (0,1) [label=right:$5$] {};
	\vertex (6) at (-1,-1) [label=left:$6$] {};
	\vertex (7) at (1,-1) [label=right:$7$] {};
	\path
		(1) edge (2)
		(1) edge (3)
		(1) edge (4)
		(5) edge (2)
		(3) edge (6)
		(4) edge (7)
	 ; 
\end{tikzpicture}
\caption{{Bipartite Claw}}
\label{chap4:fig:bc}
\end{subfigure}
\qquad\begin{subfigure}[b]
\centering
\begin{tikzpicture}[line width=1pt]
	\vertex (2) at (-1,0) [label=below:$2$] {};
	\vertex (1) at (0,1) [label=right:$1$] {};
	\vertex (3) at (-.5,0) [label=below:$3$] {};
	\vertex (4) at (0,0) [label=below:$4$] {};
	\vertex (5) at (.5,0) [label=below:$5$] {};
	\vertex (6) at (1,0) [label=below:$6$] {};
	\vertex (7) at (0,-1) [label=right:$7$] {};
	\path
		(1) edge (2)
		(1) edge (3)
		(1) edge (4)
		(1) edge (5)
		(1) edge (6)
		(2) edge (3)
		(3) edge (4)
		(4) edge (5)
		(5) edge (6)
		(4) edge (7)
	 ; 
\end{tikzpicture}
\caption{Umbrella}
\label{chap4:fig:umb}
\end{subfigure}
\qquad\begin{subfigure}[b]
\centering
\begin{tikzpicture}[line width=1pt]
	\vertex (2) at (0,0.5) [label=left:$b$] {};
	\vertex (1) at (0,1)   [label=right:$a$] {};
	\vertex (3) at (-.9,0) [label=below:$1$] {};
	\vertex (4) at (-.4,0) [label=below:$2$] {};
	\vertex (5) at (.4,0)  [label=below:$3$] {};
	\vertex (6) at (.9,0)  [label=below:$..n$] {};
	\vertex (7) at (-1.5,-.3) [label=below:$c$] {};
	\vertex (8) at (1.5,-.3)  [label=below:$d$] {};
	\path
		(1) edge (2)
		(2) edge (3)
		(2) edge (4)
		(2) edge (5)
		(2) edge (6)
		(3) edge (7)
		(6) edge (8)
		(3) edge (4)
		(4) edge (5)
		(5) edge (6)
	 ;  
\end{tikzpicture}
\caption{$n$-net, $n\geq2$}
\label{chap4:fig:nnet}
\end{subfigure}
\begin{subfigure}[b]
\centering
\begin{tikzpicture}[line width=1pt]

	\vertex (1) at (0,1) [label=right:$a$] {};
	\vertex (2) at (-1,0) [label=left:$b$] {};
	\vertex (3) at (1,0) [label=right:$c$] {};
	\vertex (4) at (-2,-1) [label=below:$1$] {};
	\vertex (5) at (-1,-1) [label=below:$2$] {};
	\vertex (6) at (0,-1) [label=below:$3$] {};
	\vertex (7) at (1,-1) [label=below:$4$] {};
	\vertex (8) at (2,-1) [label=below:$...n$] {};
	\path
		(1) edge (2)
		(1) edge (3)
		(2) edge (3)
		(2) edge  (4)
		(2) edge (5)
		(2) edge (6)
		(2) edge (7)
		(3) edge (5)
		(3) edge (6)
		(3) edge (7)
		(3) edge (8)
		(4) edge (5)
		(5) edge (6)
		(6) edge (7)
		(7) edge (8)
	 ;   
\end{tikzpicture}
\caption{$n$-tent, $n\geq3$}
\label{chap4:fig:ntent}
\end{subfigure}
\qquad\qquad\qquad\qquad\begin{subfigure}[b]
\centering
\begin{tikzpicture}[line width=1pt]

	\vertex (2) at (-1,0) [label=left:$1$] {};
	\vertex (3) at (1,0) [label=right:$2$] {};
	\vertex (5) at (-1,-1) [label=below:$3$] {};
	\vertex (7) at (1,-1) [label=below:$...n$] {};
	\path

		(2) edge (3)
		(2) edge (5)
		(5) edge (7)
		(7) edge (3)
	 ;   
\end{tikzpicture}
\caption{Hole, $n\geq4$}
\label{chap4:fig:c}
\end{subfigure}
\caption{Forbidden Subgraphs Characterization}%
\label{chap4:fig:fsg}%
\end{figure}

Let $\mathcal{I}:=\{I\subseteq E|G[I] \text{ induces an $m$-clique free interval graph}\}$. The vector $z^I$ is called the incidence vector associated with $I$, i.e., $z^I=(z^I_e)_{e\in E}$, where $z^I_e=1$ if $e\in I$ and $z^I_e=0$ otherwise. We define the $m$-Clique Free Interval Subgraph Polytope as follows: 
$P_{\mathcal{I}}(G,m):= \text{conv}\{z^I \in \{0,1\}^{|E|} | I \in \mathcal{I}\}$

In the following subsection, we propose some valid inequalities 
associated with all forbidden subgraphs and prove that these inequalities define facets for
$P_{\mathcal{I}}(G,m)$.

Figure \ref{chap4:cbc} shows these subsets.

\begin{figure}[th]
\centering
\begin{subfigure}[b]
\centering
\begin{tikzpicture}[line width=1pt]

	\vertex (2) at (0,.5) [label=right:$2$] {};
	\vertex (1) at (0,0) [label=right:$1$] {};
	\vertex (3) at (-.5,-.5) [label=left:$3$] {};
	\vertex (4) at (.5,-.5) [label=left:$4$] {};
	\vertex (5) at (0,1) [label=right:$5$] {};
	\vertex (6) at (-1,-1) [label=left:$6$] {};
	\vertex (7) at (1,-1) [label=right:$7$] {};
	\path
	  
		(1) edge [bend left](5)
		(1) edge [bend left](6)
		(1) edge [bend left](7)

	 ; 
\end{tikzpicture}
\caption{Subset $\overline{BC}_i$}
\label{chap4:fig:bca}
\end{subfigure}
\qquad\begin{subfigure}[b]
\centering
\begin{tikzpicture}[line width=1pt]

	\vertex (2) at (0,.5) [label=right:$2$] {};
	\vertex (1) at (0,0) [label=below:$1$] {};
	\vertex (3) at (-.5,-.5) [label=left:$3$] {};
	\vertex (4) at (.5,-.5) [label=right:$4$] {};
	\vertex (5) at (0,1) [label=right:$5$] {};
	\vertex (6) at (-1,-1) [label=left:$6$] {};
	\vertex (7) at (1,-1) [label=right:$7$] {};
	\path

		(4) edge (5)
		(7) edge(2)
		(3) edge(7)
		(4) edge(6)
		(3) edge (5)
		(2) edge (6)
		
	 ;   
\end{tikzpicture}
\caption{{Subset $\overline{BC}_h^4$}}
\label{chap4:fig:bcb}
\end{subfigure}
\qquad\begin{subfigure}[b]
\centering
\begin{tikzpicture}[line width=1pt]

	\vertex (2) at (0,.5) [label=right:$2$] {};
	\vertex (1) at (0,0) [label=below:$1$] {};
	\vertex (3) at (-.5,-.5) [label=below:$3$] {};
	\vertex (4) at (.5,-.5) [label=below:$4$] {};
	\vertex (5) at (0,1) [label=right:$5$] {};
	\vertex (6) at (-1,-1) [label=left:$6$] {};
	\vertex (7) at (1,-1) [label=right:$7$] {};
	\path

		(6) edge (5)
		(6) edge (7)
		(5) edge (7)
		
	 ;   
\end{tikzpicture}
\caption{Subset {$\overline{BC}_h^5$}}
\label{chap4:fig:bcc}
\end{subfigure}
\qquad\begin{subfigure}[b]
\centering
\begin{tikzpicture}[line width=1pt]

	\vertex (2) at (0,.5) [label=right:$2$] {};
	\vertex (1) at (0,0) [label=below:$1$] {};
	\vertex (3) at (-.5,-.5) [label=left:$3$] {};
	\vertex (4) at (.5,-.5) [label=right:$4$] {};
	\vertex (5) at (0,1) [label=right:$5$] {};
	\vertex (6) at (-1,-1) [label=left:$6$] {};
	\vertex (7) at (1,-1) [label=right:$7$] {};
	\path

		(3) edge (2)
		(4) edge(3)
		(2) edge(4)

	 ;   
\end{tikzpicture}
\caption{Subset $\overline{BC}_\triangle$}
\label{chap4:fig:bcd}
\end{subfigure}
\caption{Subsets of the complementary \textit{Bipartite Claw}}%
\label{chap4:cbc}%
\end{figure}

\vspace{-0.5cm}
\paragraph{Bipartite Claw}
\label{chap4:sssec:bc}

In this subsection, we propose inequalities to avoid the \textit{bipartite claw}
forbidden subgraph. An example is given in Figure \ref{chap4:fig:bca}. \newline We
 give some notations to help in analyzing the \textit{bipartite claw}
free subgraphs.\newline Let us consider the complete graph $K_{7}$ with
seven nodes. We partition this graph into $BC$ and $\overline{BC}$, where $BC$
is the set of all edges that form the \textit{bipartite claw} as in Figure
\ref{chap4:fig:bc} and $\overline{BC}$ is the set of edges in the associated
complementary graph of $BC$. Moreover, $\overline{BC}$ is partitioned as
follows:

\begin{itemize}
\item[$\circ$] Subset $\overline{BC}_{h}^{4}$ contains all the edges such that
each of them enables to form a hole of size $4$ in a bipartite claw.

\item[$\circ$] Subset $\overline{BC}_{\triangle}$ contains three edges such
that when we add one of them to $BC$, then we obtain a central triangle.

\item[$\circ$] Subset $\overline{BC}_{i}$ contains the edges that are able to
form a triangle with the inner vertex.

\item[$\circ$] Subset $\overline{BC}_{h}^{5}$ is composed of all edges such
that each one enables forming a hole of size 5 when added to the \textit{bipartite claw}. 
\end{itemize}
As a consequence, the previous definitions lead explicitly to the following subsets:

\begin{itemize}
\item $BC=\{(1,2),(1,3),(1,4),(2,5),(4,7),(3,6)\}$.

\item $\overline{BC}=\{$ $(7,1)$, $(7,2)$, $(7,3)$, $(7,5)$, $(7,6)$, $(6,1)$,
$(6,2)$, $(6,4)$, $(6,5)$, $(5,1)$, $(5,3)$, $(5,4)$, $(4,2)$, $(4,3)$,
$(3,2)$ $\}$.

\item $\overline{BC}_{h}^{4}=\{(3,5)$,$(2,6)$,$(5,4)$,$(2,7)$,$(3,7)$%
,$(4,6)\}$.

\item $\overline{BC}_{\triangle}=\{(2,3)$,$(2,4)$,$(3,4)\}$,

\item $\overline{BC}_{i}=\{(1,5),(1,6),(1,7)\}$.

\item $\overline{BC}_{h}^{5}=\{(5,6),(5,7),(6,7)\}$.
\end{itemize}

We consider two cases, when $m=2$, and when $m\geq3$.\newline If $m=2$ then
the following inequality is valid:
\begin{alignat}{10}
\label{chap4:ctn2:3} & \sum_{{e}\in BC}z_{e}\leq 5.
\end{alignat}
Indeed, when we add an edge from $\overline{BC}_{\triangle}$
in Figure \ref{chap4:fig:bc}, by definition, the resulting subgraph will contain a
clique of size 3, which is not $m$-clique free in this case (as well
it is $2-net$). Moreover, if we add an edge $e\in\overline{BC}_{h}%
$, then we obtain a $hole$. If we add another edge to break
this hole, then we obtain a clique of size 3.\newline

\begin{proposition}\label{BCF2}
The inequality \eqref{chap4:ctn2:3} defines a facet of $P_{\mathcal{I}}(G,m)$ when $m=2$.
\end{proposition}

Now, if $m\geq3$ then the following inequality is valid.
\begin{alignat}{10}
\label{chap4:ctn2:6} & \sum_{{e}\in BC}2z_{e}-\sum_{{e}\in\overline{BC}_{h}^{4}\cup\overline
{BC}_{\triangle}}z_{e}-2\sum_{{e}\in\overline{BC}_{i}}z_{e}\leq10
\end{alignat}
This inequality is valid, if we add one edge of $\overline{BC}_{\triangle}$ to
the bipartite claw, then the resulting subgraph contains $2-net$. If we add
one edge of $\overline{BC}_{h}^{5}$ or $\overline{BC}_{h}^{4}$, then we obtain
a hole of size 5 respectively 4. It is clear that when we add one, two, or
three edges of $\overline{BC}_{i}$, then the resulting graph becomes interval
and $m$-clique free.\newline

\begin{proposition}\label{BCF3}
Inequality \eqref{chap4:ctn2:6} defines a facet of $P_{\mathcal{I}}(G,m)$ when $m\geq3$.
\end{proposition}

\vspace{-0.5cm}
\paragraph{Umbrella Inequalities}

For the umbrella subgraph as shown in Figure \ref{chap4:fig:umb}, let $G_{u}%
=(U_{u},E_{u})$ be a graph that formulates the umbrella and let $\overline
{E_{u}}$ be a set of the complementary edges for $G_{u}$. In the following, we
will present a family of valid inequalities that delete the umbrella
subgraphs. To analyze this forbidden subgraph we need the following
notations:\newline Let $E_{u}^{i}\subset E_{u}$ be the set of the inner three
edges in the umbrella subgraph. Let $E_{u}^{t}\subset\overline{E_{u}}$ be the
set of the edges such that when we add one of these edges to the umbrella we
create a new triangle. Finally, $E_{u}^{a}\subset E_{u}$ is the set of the
around edges, and $E_{u}^{h}\subset\overline{E_{u}}$ is the set of edges such
that if they are connected, then they will form a hole of size 4 or of size 5.
Let
\begin{itemize}
\item $E_{u}^{i}=\{$ $(1,3)$, $(1,4)$, $(1,5)$ $\}$.

\item $E_{u}^{t}=\{$ $(1,7)$, $(3,7)$, $(5,7)$ $\}$.

\item $E_{u}^{c}=\{(2,4),(3,5),(4,6)\}$.

\item $E_{u}^{a}=\{(1,2)$, $(2,3)$, $(3,4)$, $(4,5)$, $(5,6)$, $(1,6)$,
$(4,7)$ $\}$.

\item $E_{u}^{h}= \{$ $(2,7)$, $(6,7)$, $(2,5)$, $(2,6)$, $(3,6)$ $\}$.
\end{itemize}

Remark that the graph induced by $H^{u}=\{E_{u}^{i}\cup E_{u}^{t}\cup
E_{u}^{c}\cup E_{u}^{a}\cup E_{u}^{h}\}$ is a complete graph.\newline When
$m=2$, the triangle becomes a forbidden subgraph (and then it is not possible
to find an umbrella). For this forbidden subgraph, we focus on instances where
$m\geq3$.

When $m=3$, in order to keep all edges of $E_{u}$ it is necessary to add at
least one edge of $E_{u}^{t}$. Moreover, when we add an edge from $E_{u}^{c}$
in this case, the subgraph contains a clique of size 4. If we add an edge from
$E_{u}^{h}$, then the induced subgraph will contain a hole.\newline Thus, the
valid inequalities when $m=3$ will be:
\begin{alignat}{10}
\label{chap4:ctn2:8} & \sum_{{e}\in E_{u}^{a}\setminus\{(4,7)\}}z_{e}+z_{(2,6)}+z_{(2,5)}%
+z_{(3,6)}\leq 5.
\end{alignat}

\begin{proposition}\label{UMBF3}
 Inequality \eqref{chap4:ctn2:8} defines a facet of $P_{\mathcal{I}}(G,m)$ if $m=3$.
\end{proposition}

When $m\geq4$, to find a valid solution we can add also the edges from
$E_{u}^{c}$. Then, the valid inequalities when $m \geq4$ will be:
\begin{alignat}{10}
\label{chap4:ctn2:9} & \sum_{{e}\in E_{u}^{a}} z_{e} - \sum_{{e}\in E_{u}^{t} \cup
E_{u}^{c}} z_{e} \leq6.
\end{alignat}

\begin{proposition}\label{UMBF4}
Inequality \eqref{chap4:ctn2:9} defines a facet of $P_{\mathcal{I}}(G,m)$ if $m\geq4$.
\end{proposition}

\vspace{-0.5cm}
\paragraph{$n$-net Inequalities}

The $n-net$ forbidden subgraph is shown in Figure \ref{chap4:fig:nnet}. We will give
some notations to help in analyzing the $n-net$ forbidden subgraph.\newline
Let $G_{net}=(U_{net},E_{net})$ be the graph that forms a $net$ of size $n$
(i.e., $n-net$) and $\overline{E_{net}}$ a set of complementary edges of
$G_{net}$. To avoid having a subgraph that represents an $n-net$, where
$n\geq2$ we need either to eliminate an edge from the $n-net$ without having a
hole denoted by $E_{net}^{h}$, or to add an edge that does not construct a
hole denoted by $E_{net}^{\bar{h}}$.\newline To analyze this forbidden
subgraph we will use the following notations. From Figure \ref{chap4:fig:nnet} let us consider:

- $E^{\bar{h}}_{net}=\{(a,c),(a,d)\}\cup\{(c,3),(c,4),...,(c,n)\}\cup
\{(d,1),(d,2),...,(d,n-2)\}\cup\{(c,d)\}$.\newline- $E^{h}_{net}%
=\{(b,2),...,(b,n-1)\}$.\newline

We propose valid inequalities that delete the $n-net$ forbidden subgraphs.
\begin{alignat}{10}
\label{chap4:ctn2:9nt} & \sum_{{e}\in E_{net}\setminus E^{h}_{net}} z_{e} -
\sum_{{e}\in\overline{E_{net}} \setminus E^{\bar{h}}_{net}} z_{e} \leq\vert
E_{net}\setminus E^{h}_{net} \vert- 1.
\end{alignat}

\begin{proposition}\label{nnetF}
Inequality \eqref{chap4:ctn2:9nt} defines a facet of $P_{\mathcal{I}}(G,m)$.
\end{proposition}

\vspace{-0.5cm}
\paragraph{\textit{n}-tent Inequalities}
Figure \ref{chap4:fig:ntent} shows the $n-tent$ forbidden subgraph, the graph is
non-interval if it contains an $n-tent$ forbidden subgraph $G_{tent}$.
Let the graph $G_{tent}=(U_{tent},E_{tent})$ be a graph that formulates
$n-tent$ for all $n\geq3$, and $\overline{E_{tent}}$ be the set of
complementary edges.\newline From Figure \ref{chap4:fig:nnet} \newline

- $E^{h}_{tent}= \{(b,c),(c,4),(b,2)\}$,\newline- $E^{\bar{h}}_{tent}=
\{(1,4),(2,5),...,(n,n+3)\}$.\newline

Remark that each $n$-tent where $n\geq5$ contains a clique of size 5. Then the
clique inequality dominates inequalities induced by  $n$-tent subgraphs if $m\leq4$. It is the same
idea for all $n$-tents where $n=4$ (resp. $n=3$). Then the clique inequality dominates inequalities induced by $n$-tent subgraphs if $m=3$ (resp. $m=2$).

In the following, we will describe valid inequalities that delete the $n$-tent
forbidden subgraphs.
\begin{alignat}{10}
\label{chap4:ctn2:9tnt} & \sum_{{e}\in E_{tent}\setminus E^{h}_{tent}} z_{e} -
\sum_{{e}\in\overline{E_{tent}}\setminus E^{\bar{h}}_{tent}} z_{e} \leq\vert
E_{tent}\setminus E^{h}_{tent} \vert-1.
\end{alignat}

\begin{proposition}\label{tentF}
Inequality \eqref{chap4:ctn2:9tnt} defines a facet of $P_{\mathcal{I}}(G,m)$, when $m\geq5$ or ($n=4$ and
$m=3$) or ($n=3$ and $m=2$).
\end{proposition}

\vspace{-0.5cm}
\paragraph{Hole inequalities}

Here, it is convenient to define a hole as an induced subgraph of $G$
isomorphic to $C_{k}$ for some $k\geq4$ \cite{schrijver2003combinatorial}. The hole $C$ is a
forbidden subgraph as depicted in Figure \ref{chap4:fig:c}.

Let $C$ be the set of edges that construct the hole, i.e., $C=\{$
$(u_{1},u_{2})$,$(u_{2},u_{3})$,...,$(u_{|C|-1},u_{|C|})$,$(u_{|C|},u_{1})$
$\}$. If $(i+k)>|C|$, then $u_{i+k}=u_{i^{\prime}}$, $i^{\prime}=(i+k)-|C|$.
Let $\overline{C}$ denote the set of all chords of hole $C$.\newline Suppose
we have a hole of size $4$, this graph is a non-interval graph. The induced
subgraph by a hole is an interval graph only if we add to it at least one chord.

\begin{proposition}\label{Hole}
For a hole $C$, the minimum number of necessary chords that should be added to
the hole to be an interval graph is $|C|-3$, when $|C|\geq4$.
\end{proposition}

In the following, we will present valid inequalities for the hole forbidden
subgraph.\newline If $m=2$, then\ the inequality \eqref{chap4:ctn2:12} is valid.

\begin{alignat}{10}
\label{chap4:ctn2:12}&\sum_{{e}\in C}z_{e}+\sum_{{e}\in\overline{C}}z_{e}\leq|C|-1.
\end{alignat}
Indeed, if we add one chord to $C\subset\{e\}$, then we will obtain a
triangle or another cycle not valid for $m=2$. Remark that this
inequality for $m=2$ is equivalent to the clique inequalities described in the
next subsection.

If $m\geq3$, then inequality \eqref{chap4:ctn2:10} is valid.
\begin{alignat}{10}
\label{chap4:ctn2:10} & \sum_{{e}\in C} (\vert C \vert-3)z_{e} - \sum_{{e}%
\in\overline{C}} z_{e} \leq(\vert C\vert-1)(\vert C \vert-3).
\end{alignat}

\begin{proposition}\label{HF}
Let $C$ be a hole of size greater than 3, then inequality \eqref{chap4:ctn2:10}
induced by cycle $C$ defines a facet of $P_{\mathcal{I}}(G,m)$ if $m\geq3$.
\end{proposition}
\vspace{-0.5cm}
\section{Clique analysis}

\paragraph{Clique inequalities}

In this subsection, we will study the clique subgraph, describe valid
inequalities and characterize facets.

\begin{proposition}\label{VFK1}
Let $K$ be a clique and let $V(K)$ be its set of vertices. If $m=2$, then the
inequality

\begin{alignat}{10}
\label{ctplus}& \sum_{{e}\in E(K)}z_{e}\leq|V(K)|-1.
\end{alignat}
is valid and defines a facet of $\mathrm{P}_{\mathcal{I}}(G,m)$.
\end{proposition}

\begin{proposition}\label{FK1}
Let $K$ be a clique of size $m+1$. Then the inequality
\begin{alignat}{10}
\label{chap4:ctn2:11} &  \sum_{{e}\in E(K)} z_{e} \leq|E(K)|-1
\end{alignat}
defines a facet of $\mathrm{P}_{\mathcal{I}} (G,m)$.
\end{proposition}

Let $f(K,m)$ be a function giving the minimum number of edges necessary to be
removed from $E(K)$ such that the resulting graph $G^{\prime}(K)$ is
$m$-clique free. Let $\alpha=\lceil\frac{|V(K)|}{m}\rceil$, $n_{\alpha
-1}=m\alpha-|V(K)|$ and $n_{\alpha}=\frac{|V(K)|-(n_{\alpha-1})(\alpha
-1)}{\alpha}$

\begin{proposition}\label{FKM}
$f(K,m) = n_{\alpha-1} \frac{(\alpha-1)(\alpha-2)}{2}+n_{\alpha} \frac
{(\alpha)(\alpha-1)}{2}$
\end{proposition}

\begin{proposition}\label{KV}
Let $K$ be a clique. Then the inequality
\begin{alignat}{10}
\label{chap4:ctn2:11m} &  \sum_{{e}\in E(K)} z_{e} \leq|E(K)|- f(K,m)
\end{alignat}
is valid  for $\mathrm{P}_{\mathcal{I}} (G,m)$.
\end{proposition}
\vspace{-0.5cm}
\paragraph{Clique-Hole inequalities}

Remark that if we remove $m$ disjoint cliques in $G(K)$, then we obtain a
complete bipartite subgraph between all pairs of two cliques.

Let $H_{ij}=(K_{i},K_{j},E_{ij})$ be a complete bipartite graph. Remark that
$H$ contains a hole if $|K_{i}|\geq2$ and $|K_{j}|\geq2$. To remove every hole
in $H_{ij}$, the minimum number of edges $E^{\prime}$ necessary to be removed
to obtain a hole-free graph is equal to $\max(|K_{i}|,|K_{j}|)-1,$ otherwise
we can always take 2 nodes in $K_{i}$ or $K_{j}$ such that these two nodes are
not covered by $E^{\prime}$. Note that $E^{\prime}$, with size $\max
(|K_{i}|,|K_{j}|)-1$, covers the maximum nodes of $K_{i}$ and $K_{j}$. We can
strengthen inequality \eqref{chap4:ctn2:11m} by the following inequality. Let
$\alpha=\sum_{i\in m}(\max\{|K_{i}|-1,0\})-\max_{i\in m}|K_{i}|-1$.

\begin{proposition}\label{KVF}
Let $K$ be a clique. Then the inequality
\begin{alignat}{10}
\label{chap4:ctn2:11mp} &  \sum_{{e}\in E(K)} z_{e} \leq|E(K)|- (f(K,m)+\alpha)
\end{alignat}
is valid and defines a facet of $\mathrm{P}_{\mathcal{I}} (G,m)$.
\end{proposition}

\section{Application to the generalized open shop problem with disjunctive constraints problem}

The Generalized Open Shop with Disjunctive Constraints (GOSDC) can be
formulated as follows. Let $M$ be a set of machines. For every $i\in M$ we
consider the set of jobs $J_{i}$ to be performed on machine $i$, and 
denote by $\mathcal{J}=\{J_{1},...,J_{m}\}$ the set of all these sets and by
$J=\bigcup_{i\in M}J_{i}$ the union of these sets. We denote by $p_{ij}$ the
processing time of job $j$ on  machine $i$. We consider an
incompatibility graph $G_{I}=(V_{I},E_{I})$ such that for each job $%
j\in J$  a vertex $v_{j}\in V_{I}$ is considered and there exists an edge
between $v_{j_{1}}$ and $v_{j_{2}}$ if $j_{1}$ and $j_{2}$ cannot run at the
same time. Remark that it is necessary to consider a linear ordering on
each machine. The GOSDC problem consists in assigning all jobs to machines such that the maximum completion time is minimum. 

\subsection{Integer linear programming formulation}

In this subsection we present an integer linear programming model for solving the GOSDC problem. For this, we first describe
the variables used in the model:\newline
$\bar{z}_{j_{1},j_{2}}=\left\{
\begin{array}{l}
1\text{ if job }j_{1}\text{ runs before job }j_{2} \\
0\text{ otherwise}%
\end{array}%
\right. \forall j_{1},j_{2}\in J.$\newline
\newline

\noindent $z_{j_{1},j_{2}}=\left\{
\begin{array}{l}
1\text{ if }j_{1}\text{ and }j_{2}\text{ run at the same time } \\
0\text{ otherwise}%
\end{array}%
\right. \forall j_{1}\in J_{i},j_{2}\in J_{i^{\prime }}|i\neq i^{\prime }\in
M.$\newline
\newline
For every $j\in J$ we consider the variable $y_{j}\in \mathbb{N}^{+}$
representing the starting time of job $j$.\newline
$C_{\text{max}}\in \mathbb{N}^{+}$ is the maximum completion time.\newline
The GOSDC problem is equivalent to the following ILP, denoted by $(P_{GOS})$:

\begin{alignat}{10}
\nonumber & \min C_{max}   \\
\label{ctngos:1}&y_{j} + p_{ij}\leq C_{max},& &\quad \forall i\in M\qquad \forall j \in J_i,\\
\label{ctngos:2}&y_{j_1}+ p_{ij_1} \leq y_{j_2}+C \bar{z}_{j_2,{j_1}},& &\quad  \forall i\in M\qquad \forall j_1\in J_i \text{ and } j_2 \in J,\\
\label{ctngos:3}&\bar{z}_{j_1,{j_2}}+\bar{z}_{j_2,j_1}=1,& &\quad \forall i\in M\qquad \forall j_1,j_2 \in J_i,\\
\label{ctngos:4}&\bar{z}_{j_1,{j_2}}+\bar{z}_{j_2,j_1}=1,& &\quad  \forall (v_{j_1},v_{j_2}) \in E_I,\\
\label{ctngos:5}&\bar{z}_{j_1,{j_2}}+\bar{z}_{j_2,j_1}+z_{j_2,j_1}=1,& &\quad  \forall (v_{j_1},v_{j_2}) \notin E_I,
\end{alignat}
\begin{alignat}{10}
\label{ctngos:6}&\sum_{{(j_1,j_2)}\in E(\bar{I})} z_{{j_1},{j_2}}- \sum_{{(j_1,j_2)}\in  E\setminus E(\bar{I})} z_{{j_1},{j_2}} \leq \vert E(\bar{I})\vert -1,& &\quad  \forall  \overline{I} \subseteq \mathcal{\overline{I}},\\
\label{ctngos:7}&\sum_{{(j_1,j_2)}\in E(K)} z_{{j_1},{j_2}}\leq |E(K)|-1 , & &\quad\forall K \subseteq \mathcal{K},
\end{alignat}

The objective function is to minimize the makespan. Inequalities %
\eqref{ctngos:1} ensure that the starting time for each job plus its
processing time is less than or equal to the total completion time.
Inequalities \eqref{ctngos:2} and inequalities \eqref{ctngos:3} guarantee
that there is no two jobs running on the same machine at the same time and
control the linear ordering. Inequalities \eqref{ctngos:4} ensure that if
two jobs are linked by an edge in the compatibility graph, then they do not
run at the same time. Indeed, for these two jobs $j_{1}$ and $j_{2}$ either $j_{1}$ is
before $j_{2}$ or $j_{2}$ is before $j_{1}$. Inequalities \eqref{ctngos:5}
ensure the three possibilities: $j_{1}$ before $j_{2}$, $j_{2}$ before $%
j_{1}$ and $j_{1}j_{2}$ they run at the same time. Inequalities \eqref{ctngos:6} and %
\eqref{ctngos:7} guarantee that the induced subgraphs are interval and $m$%
-clique free subgraphs. The number of inequalities may be exponential and
thus we will use the separating algorithm presented in this chapter.

\section{Experimental results}

In order to evaluate the efficiency of the inequalities mentioned in this chapter, we developed the mentioned exact and heuristic separations. All the
computational results are obtained using Cplex 12.6 and JAVA for
implementing exact and heuristic algorithms. The ILP with the valid
inequalities is tested on the following proposed benchmark of
instances.\newline The processing times are uniformly distributed between 50
and 150 as it is common in the literature.
The graph density (GD) is equal to 0.5 and
calculated as follows: $GD=\frac{|E|}{|V|(|V|-1)}$ where $E$ is the set of
edges associated with the precedence constraints between jobs, and $V$ is the set
of vertices associated with jobs.
The results are given for 4 families of instances. Each family contains 5 instances with the same parameter.

The required CPU time is measured in seconds. We limit to 3600 seconds the
algorithm running time for each instance, by using 4.0 GB of RAM.\newline

The next tables provide the following information:

\begin{itemize}
\item $|J_i|$: Number of jobs per machine.

\item $m$: Number of machines.

\item Method:
\begin{itemize}
\item 0: Basic model;
\item 1: Bipartite claw inequalities (H1BC-Sep),
\item 2: Umbrella Inequalities (H1U-sep),
\item 3: Hole Inequalities,
\item 4: Clique-Hole Inequalities,
\item 5: $n$-net Inequalities,
\item 6: $n$-tent Inequalities,
\item 7: All inequalities of methods 1 to 6,
\end{itemize}
\item Nodes: The number of nodes in the branching tree.

\item Gap: The gap between the lower bounds and the upper bounds ($%
100\times \frac{UB-LB}{LB}$),

\item CPU: Computational time (limited to 1 hour).

\item o/p: The number of solved instances (5 instances over 5 or 0 over 5)
\end{itemize}

\begin{table}[h!]
  \centering
  \caption{Average CPU time.}
  \label{tab:ins}
  \begin{tabular}{c c c c c c c c c c c c c c c c c c}
   \toprule
$|J_i|$	&	m	&	Method	&	CPU	&	Nodes	&	Gap	&	o/p	\\
5	&	2	&	0	&	0	&	172,8	&	0	&	5/5	\\
5	&	2	&	1	&	0	&	119	&	0	&	5/5	\\
5	&	2	&	2	&	0	&	200	&	0	&	5/5	\\
5	&	2	&	3	&	0	&	58	&	0	&	5/5	\\
5	&	2	&	4	&	0	&	172,8	&	0	&	5/5	\\
5	&	2	&	5	&	0	&	172,8	&	0	&	5/5	\\
5	&	2	&	6	&	0	&	133,2	&	0	&	5/5	\\
5	&	2	&	7	&	0	&	51	&	0	&	5/5	\\
5	&	4	&	0	&	$>1h$	&	307851	&	0,5950	&	0/5	\\
5	&	4	&	1	&	$>1h$	&	141328,6	&	0,4239	&	0/5	\\
5	&	4	&	2	&	$>1h$	&	125544,6	&	0,5249	&	0/5	\\
5	&	4	&	3	&	$>1h$	&	120168,8	&	0,2015	&	0/5	\\
5	&	4	&	4	&	$>1h$	&	290959	&	0,5975	&	0/5	\\
5	&	4	&	5	&	$>1h$	&	187125	&	0,5934	&	0/5	\\
5	&	4	&	6	&	$>1h$	&	138112	&	0,5519	&	0/5	\\
5	&	4	&	7	&	$>1h$	&	144547,8	&	0,1065	&	0/5	\\
5	&	6	&	0	&	$>1h$	&	114346,6	&	0,7728	&	0/5	\\
5	&	6	&	1	&	$>1h$	&	15098	&	0,7537	&	0/5	\\
5	&	6	&	2	&	$>1h$	&	18630,8	&	0,7507	&	0/5	\\
5	&	6	&	3	&	$>1h$	&	41487,4	&	0,7576	&	0/5	\\
5	&	6	&	4	&	$>1h$	&	112547,8	&	0,7666	&	0/5	\\
5	&	6	&	5	&	$>1h$	&	91569	&	0,7691	&	0/5	\\
5	&	6	&	6	&	$>1h$	&	35554,4	&	0,7678	&	0/5	\\
5	&	6	&	7	&	$>1h$	&	10257	&	0,7628	&	0/5	\\
10	&	2	&	0	&	148	&	214562	&	0	&	5/5	\\
10	&	2	&	1	&	887,5	&	170593,4	&	0	&	5/5	\\
10	&	2	&	2	&	83,25	&	330710	&	0	&	5/5	\\
10	&	2	&	3	&	39,25	&	296249,6	&	0	&	5/5	\\
10	&	2	&	4	&	148,5	&	214562	&	0	&	5/5	\\
10	&	2	&	5	&	149,75	&	214562	&	0	&	5/5	\\
10	&	2	&	6	&	683,25	&	441429	&	0	&	5/5	\\
10	&	2	&	7	&	98	&	218567,2	&	0	&	5/5	\\
    \bottomrule
  \end{tabular}
\end{table}

We present the results where we already select the best separation algorithm for each family of inequalities.
Table \ref{tab:ins} presents the results for seven different methods. First, we observe that for all instances on 2 machines we can solve all of them to the optimality. Furthermore, with 5 jobs per machine the number of nodes decreases when we add inequalities. In this case, we notice that adding all the inequalities is the best option in order to reduce the number of generated nodes. With 10 jobs per machine, we notice that even if we generate more nodes, we reduce the CPU Time. In this case, we remark that the hole inequalities are the most efficient for solving this family of instances.
Moreover, if we increase the number of machines, then we cannot solve instances with 5 jobs per machine. However, we can reduce the gap by using our valid inequalities. For 5 jobs per machine and 4 machines, the gap is divided by 5  using all the inequalities (almost 6), whereas with 6 machines, the best method is to use only the umbrella inequalities.

\section{Conclusion and Perspectives}
\vspace{-0.2cm}

\label{cp}
In this chapter, we have presented a polyhedral study for the interval and m-clique free graphs problem.
We have also applied the obtained results to the problem of unrelated parallel machines with disjunctive constraints, and designed and implemented Branch-and-Cut algorithms based on families of strong valid inequalities presented in this chapter. We separate some famillies of inequalities associated with the forbidden subgraphs. Computational experiments on a set of instances have shown that the algorithms are capable to solve all instances to optimality within reasonable CPU time.

The presented inequalities are able to block some specific subgraphs. Deep analysis has been done to show the theoretical strength of each class of inequalities. 
From the provided results, we can derive several mathematical formulations to solve the interdiction/blocker problem. For instance, we can deduce a mathematical model to solve the edge clique blocker problem where the goal is to minimize the number of edges to delete so that the remaining maximum clique in the graph is of size less than a given threshold. Indeed, a mathematical model with only inequalities \eqref{chap4:ctn2:11m} for all cliques of the graph allows solving the edge clique blocker problem.

\chapter{Clique interdiction}

In this chapter, we present original research from \cite{furini2020integer}. 
we focus on the well-known clique interdiction problem. 
The maximum vertex clique interdiction problem (CIP) consists in finding a subset of at most $k$ vertices to be removed from $G$ so that the size of the maximum clique in the remaining graph is minimum.

\vspace{-0.5cm}
\paragraph{Applications} The CIP has many applications. The first example is in social networks to analyze the most influential people. The second example is during a pandemic to determine in priority people to vaccine or to test for detecting the virus. Indeed, the interdiction problem consists of a budget for vaccines or tests and tries to cover the biggest clique community within the available budget.

\section{Bi-level formulation and complexity}

The following binary decision variables are needed to model the problem as a bilevel integer linear program:
\begin{align*}
w_u &= \begin{cases}  1, & \text{ if vertex $u$ is interdicted by the leader}, \\
		    0, & \text{ otherwise}
      \end{cases} && \quad  u \in V\\
x_u &= \begin{cases}  1, & \text{ if vertex $u$ is used in the maximum clique of the follower}, \\
		    0, & \text{ otherwise}
      \end{cases} && \quad  u \in V
\end{align*}

Let $\mathcal{W}$ be the set of all feasible interdiction policies of the leader, i.e.:
\begin{equation}\label{eq:C}
\mathcal{W} = \left\{ w \in \{0,1\}^n : \sum_{u \in V} w_u  \leq k \right\}.
\end{equation}
Similarly, let $\mathcal{K}$ represent the set of incidence vectors of all cliques in the graph $G$, i.e.:
\begin{equation}\label{eq:K}
\mathcal{K}
= \left\{ x \in \{0,1\}^n : x_u + x_v  \leq 1, uv \in \overline E \right\},
\end{equation}
where the constraints $x_u + x_v  \leq 1$ ensure that two vertices cannot be part of a clique if there is no edge connecting them.
With a slight abuse of notation, we will use both notations $K \in \mathcal{K}$ and $x \in \mathcal{K}$, where $x$ is the incident vector of $K$.
Given an interdiction strategy $w^* \in \mathcal{W}$, let $V_{w^*}$ be the associated set of interdicted (deleted) vertices.

The \NICP\ can be formulated as follows:
\begin{equation}\label{eq:NICP}
\min_{ w \in \mathcal W
}
\max_{K \in \mathcal{K} } \left\{ |K| - \sum_{u \in K} w_u \right\}.
\end{equation}

In the objective function of \eqref{eq:NICP} we express the min-max nature of the problem: 

the leader controls (with the variables $w$)  the outer minimization problem by interdicting at most $k$ vertices of the graph, while 
the inner  problem consists of 
calculating the clique number  in the graph induced after the removal of the interdicted vertices.  We call this inner problem the \emph{follower's subproblem}.

Observe that for the follower, the size of each clique $K$ in $G$ reduces by the number of interdicted vertices from $K$, which follows from the hereditary property of the clique (i.e., every vertex-induced subgraph of $K$ is a clique itself). Therefore, the value $|K| - \sum_{u \in K} w_u $ denotes the size of the clique $K$  after applying the interdiction strategy defined by $w \in \mathcal{W}$. Hence, the formulation \eqref{eq:NICP} states that the leader chooses a set of vertices to interdict, so that among all possible cliques $K \in \mathcal{K}$, the size of the maximum remaining clique is the smallest possible.

The \NICP\ can be equivalently stated as:

\begin{equation}
\min_{ w \in \mathcal W} ~ \max_{x \in \mathcal{K} } \left \{ \sum_{u \in V} (1 - w_u)x_u  \right \}
\label{eq:NICP01}
\end{equation}
Indeed, we can rewrite the objective function of the follower in \eqref{eq:NICP01} as $\sum_{u\in V}(1-w_u)x_u=|K|-\sum_{u\in K} w_ux_u$.  The latter sum corresponds to the objective function in \eqref{eq:NICP} if and only if 
$w_ux_u=w_u$ if $u\in K$ and $w_ux_u=0$, otherwise. The latter is always true, as the vector $x$ encodes the clique $K$, and hence $x_u=1$ if $u \in K$, and $x_u=0$, otherwise. 

Finally, the \NICP\ can also be reformulated as the following bilevel integer linear program:
\begin{equation}
\min_{ w \in \mathcal W} ~ \max_{x \in \mathcal{K} } \left \{ \sum_{u \in V} x_u ~ : ~ x_u  \leq 1-w_u, ~u \in V  \right \}.
\label{eq:NICP1}
\end{equation}
Indeed, given an interdiction strategy $w^* \in \mathcal{W}$, the optimal solution of the follower's subproblem \eqref{eq:NICP} and the optimal solution of the follower's subproblem \eqref{eq:NICP1} are the same. Inequalities $x_u  \leq 1-w_u$ ($u \in V$) are the linking interdiction constraints, making sure that the follower cannot choose a vertex $u$ if it has been interdicted by the leader. They ensure that for a given vector $w^*$, the follower searches for the maximum clique in the support graph in which the vertices $v$ such that $w^*_v=1$ have been removed.  

The bilevel ILP formulation \eqref{eq:NICP1} can be solved by using a Benders-like decomposition approach \cite{benders}. 

\begin{proposition}
CIP is $\Sigma_2^p$-complete.
\end{proposition}

\section{Combinatorial bounds, tighter gaps and preprocessing} \label{sec:LB}
We propose combinatorial algorithms for calculating tight global lower and upper bounds. Besides being useful for exact Branch-and-Bound-based approaches, we also demonstrate how these tight bounds help in reducing the size of the input graph. In the following, let  $\ell_{\min}$, $\ell_{\max}$ denote the global lower and upper bounds on the solution value, respectively, and let $\ell_{\opt}$ be the optimal solution value (i.e., the size of the maximum clique after applying an optimal interdiction policy).

\subsection{Computing the global lower bound $\ell_{\min}$} \label{sec:lmin}
With the following result, we show how removing edges from $G$ allows us to compute a global lower bound on the \NICP\ value.

\bigskip

\begin{proposition} \label{prop:LB}
Given a subgraph $G'=(V,E')$ with $E' \subset E$,
an optimal \NICP\ solution on $G'$ provides a valid lower bound for the optimal \NICP\ solution on $G$.
\end{proposition}

The result of Proposition \ref{prop:LB} is rather counter-intuitive, as it states that by reducing the input graph, instead of obtaining a valid upper bound for a minimization problem, we obtain a valid lower bound. The key issue here is that we are not reducing the feasibility space of the leader (as the set of vertices in $G'$ remains the same as in $G$), but only the feasibility space of the follower (which is reduced from all cliques in $E$ to all cliques in $E'$). Observe, furthermore, that in terms of the problem complexity, the result of Proposition \ref{prop:LB} does not help us much in solving the problem, since we still have to solve the same \NICP\ problem, only on a smaller graph. However, this result can be particularly useful for special classes of graphs on which solving the \NICP\ on $G'$ is easier than on $G$. 

\bigskip 

\begin{corollary} \label{cor:lmin}
  Given a set $\mathcal{Q}_{p+1}= (K_1, \dots, K_{p+1})$ of vertex-disjoint cliques of $G$, such that $|K_1| \ge \dots \ge |K_{p+1}|$, a valid lower bound $\ell_{\min}$ for the \NICP\ can be obtained by computing
\begin{equation}
\ell_{\min} = 
\begin{cases}
 \max \left\{ |K_{{p+1}}|, |K_{{p}}|-1-\left\lfloor\frac{k-k(\mathcal{Q}_{p})}{{p}}\right \rfloor \right\}, & \text{ if $k < k(\mathcal{Q}_{p+1})$ } \\
|K_{p+1}|-1-\left\lfloor\frac{k-k(\mathcal{Q}_{p+1})}{p+1}\right\rfloor, & \text{ otherwise, } 
 \end{cases} 
 \label{eq:lmin}
  \end{equation}
\end{corollary}
where $k(\mathcal{Q}_{p+1})$ denotes the size of an optimal interdiction strategy necessary to reduce the size of all cliques in $\mathcal{Q}_{p+1}$ to the size of the largest clique of $\mathcal{Q}_{p+1}$ minus 1. 
The quality of this lower bound depends on the number of cliques $p+1$ and their sizes.

\subsection{Reducing the input graph}\label{sec:reduction}

We start by describing another important solution property that is exploited in our study.
Given a clique $K$ and an interdiction policy $w^* \in \mathcal{W}$, we say that $K$ is \emph{covered} by $V_{w^*}$ if and only if at least one vertex from $K$ is interdicted by $w^*$, i.e.,  $K \cap V_{w^*} \neq \emptyset$.  

\bigskip 

\begin{property}\label{prop:covering}
If there exists a feasible interdiction policy $w^* \in \mathcal{W}$, such that all cliques of size $\ell_{\max} +1$ in $G$ are covered by $V_{w^*}$ and there exists at least one clique $K^*$ in $G$, $|K^*| = \ell_{\max}$ which is not covered by $V_{w^*}$, then $\ell_{\max}$ is a valid (non-trivial) upper bound on the CIP.
\end{property}

The latter property suggests another way of seeing the CIP: find a feasible interdiction policy that minimizes the value of $\ell_{\max}$ while making sure all cliques of size $\ell_{\max} + 1$ are covered.

We say that an interdiction policy $w \in \mathcal{W}$ is \emph{minimal}, if for the associated set of interdicted vertices $V_w$, we have:
\[ \omega (G[V\setminus V_w]) < \omega(G[(V \setminus V_w) \cup \{v\}]), \quad \forall v \in V_w. \]
The following result identifies redundant vertices in the input graph $G$.

\bigskip

\begin{proposition} \label{prop:preprocess}
Let $v$ be an arbitrary vertex from $V$.
If
$\omega_G(v) \le \ell_{\opt}$, then $v$ cannot be part of a minimal optimal interdiction policy. 
\end{proposition}

Hence, by focusing on the minimal interdiction policies, which can be done without loss of generality,
one can preprocess the graph $G$ and remove redundant vertices  from it.
To properly exploit Proposition \ref{prop:preprocess}, instead of using the (unknown) value of $\ell_{\opt}$ for removing the redundant vertices, one can employ a tight lower bound $\ell_{\min}$.
The following result gives a connection between the optimal solution value and the solution value found on the preprocessed graph from which redundant vertices are removed.

\bigskip 

\begin{proposition}
Let      $V_{prep} = \{ v \in V : \omega_G(v) \le \ell_{\min} \}$
be the set of vertices $v$ satisfying the (weakened) property of Proposition \ref{prop:preprocess}.
Let $\tilde V= V \setminus V_{prep}$ and $\tilde G = G[\tilde V]$ and let $\tilde \ell_{\opt}$ denote the optimal \NICP\ solution value on $\tilde G$. Then
\[ \ell_{\opt} = \max\{ \tilde \ell_{\opt}, \ell_{\min} \}. \]
Furthermore, the optimal interdiction policy $\tilde w$ on $\tilde G$, is also optimal for $G$.
\end{proposition}

Observe that the latter result allows us not only to fix the binary decision variables $w_v$ to zero, but also to 
modify the input graph $G$ by removing its vertices, resulting into a smaller input graph for solving the follower's subproblem. This reduction  preserves optimality, since the removal of a vertex $v \in V_{prep}$ only reduces the size of maximal cliques $K$ with $|K| \le \ell_{\min}$. These are never going to constitute an optimal follower's response to an optimal interdiction policy.

In a standard implementation,  one would start with an arbitrary vertex $v$, and solve the MCP by fixing this vertex to one.
If the size of the obtained clique $K$ is smaller than $\ell_{\min}$, vertex $v$ is removed from $G$ and the process is repeated for the remaining vertices.

This procedure can be time-consuming, as the maximum clique algorithm has to be called per each vertex.  
To overcome this drawback, the degree of a vertex can be used instead, as a trivial upper bound on the clique number of a vertex $v$.

\subsection{Computing the global upper bound $\ell_{\max}$}
\label{sec:lmax}

We have implemented several heuristics in order to calculate a tight upper bound $\ell_{\max}$ and create a pool of  initial feasible solutions that are  given later to the MIP solver.
In all heuristics, if a vertex has been fixed to zero (e.g., removed by the preprocessing described above), it is not considered as a candidate for being interdicted.
Our heuristics work in a greedy fashion, based on four different criteria: vertices are interdicted one-by-one until the interdiction budget is exhausted. The chosen criteria are as follows:
\begin{itemize}
\item Vertex-degree: At the beginning, the vertices are sorted in non-increasing order according to their degrees. We start by interdicting the vertices with the highest degree first, and  stop once the interdiction budged is exhausted. 
\item Updated vertex-degree: The major difference to the ``vertex-degree'' heuristic is in the fact  
that now we recompute the vertex degrees, each time a vertex is interdicted.
\item Vertex-coreness-number: at the beginning, the vertices are sorted in non-decreasing order according to their coreness number. 

The intuition behind this approach is that a vertex with a high coreness value is likely to belong to large cliques and shall be interdicted first. As mentioned above, $core(v) + 1$ is a rough upper bound on the size of the maximum clique containing this vertex, $\omega_G(v)$.
 \item Vertex-color-number:  This heuristic exploits the well-known result that the size of any feasible coloring gives the upper bound on the clique number of a graph \cite{balas1986finding}.
We first apply the greedy sequential coloring heuristic based on independent sets, where color classes are obtained incrementally using bitmasks, see \cite{san2011exact,san2013improved}. This procedure runs in $O(|V|^2)$ in the worst case and returns a feasible coloring. 
We then assign a label (corresponding to the color number) to each vertex. In this sequential greedy coloring heuristic, the higher the color associated to a vertex is, the higher are the chances that this vertex belongs to a large clique. 
Therefore, we interdict the vertices in non-increasing order with respect to their color numbers.
Notice that the reordering of vertices, based on their color number, once they have been interdicted, will not influence the color number. So, there is no need to reorder the vertices after partially interdicting them, as long as they are interdicted starting with the highest color number first. As it will be shown in our computational study, vertex-color-number provides a very robust measure that delivers excellent heuristic solutions for sparse social networks.

\end{itemize}

\section{Natural formulation}
In this section, we provide an ILP formulation of the problem in the natural space of leader decision variables $w$. In addition, we provide a facial study of the underlying polytope, discussing under which conditions the proposed inequalities are facet-defining.

The following is a valid ILP formulation  for \NICP:
\begin{align}
&&\min \qquad \theta  \label{eq:B1}\\
&& \theta  + \sum_{u \in K} w_u    & \geq |K| & K \in \mathcal{K} \label{eq:B2}\\
   && \sum_{u \in V} w_{u}&\le k \label{eq:B3}\\
  && w_{u}&\in\{0,1\} & u \in V. \label{eq:B4}
\end{align}

To see that the model is valid, observe that for every feasible interdiction policy $\bar{w} \in \mathcal{W}$, the follower's problem boils down to 
$\max_{x \in \mathcal{K}} ~ \sum_{u \in V} x_u (1-\bar{w}_u)$.
Hence, the problem can be restated so that the set of feasible solutions of the follower does not depend on the actions of the leader anymore. Consequently,  one can enumerate all cliques in $G$ and optimize over the set $\mathcal{K}$. That is why the problem can be equivalently restated as
\begin{equation*}
 \min_{w \in\mathcal{W}}   \left\{ \theta ~ : ~ \theta  \geq \sum_{u \in V} \bar{x}_u (1-w_u),   \bar{x} \in \mathcal{K}  \right\},
\end{equation*}
where $\bar x$ represents an arbitrary incident vector of a clique in $G$.  

This single-level ILP formulation contains an exponential number of constraints of type \eqref{eq:B2} that we will refer to as \emph{Clique Interdiction} (CI) cuts. 
These constraints are NP-hard to separate: for each vector $\bar{w}$ and the associated $\bar{\theta}$ given by the current solution of the formulation \eqref{eq:B2} and \eqref{eq:B3} 
, checking if there exists a violated interdiction cut requires finding a maximum weighted clique on $G$ with vertex-weights $c_u=1-\bar{w}_u$ for all $u\in V$. 

In a Branch-and-Cut algorithm applied to the above ILP formulation, it is sufficient to separate integer infeasible points only (fractional points being cut off using standard branching and general cutting plane mechanisms embedded in modern MIP solvers). Whenever $\bar{w}$ is integer, the separation problem consists in 
solving the MCP in the support graph $G[V\setminus V_{\bar{w}}]$. Let $\bar{K}$ be the maximum clique in $G[V\setminus V_{\bar{w}}]$: if $|\bar{K}|>\bar{\theta}$, then a violated CI cut associated with $\bar K$ is found and added to the model.
This separation procedure can be a potential bottleneck for using a Branch-and-Cut algorithm unless an efficient clique solver is used for the separation of CI cuts. We have therefore implemented a tailored separation algorithm based on 
recent state-of-the-art approaches for the MCP.

\subsection{Polyhedral investigation}
In the following, we study the polytope of the single-level \NICP\ formulation \eqref{eq:B1}-\eqref{eq:B4}. 
We provide necessary and sufficient conditions under which the clique interdiction  cuts \eqref{eq:B2} are facet defining and  discuss heuristic lifting procedures designed to strengthen these inequalities.

Given the graph $G$ and the interdiction budget $k$, let $\mathcal{P}(G,k)$ denote the convex hull of feasible solutions of the \NICP\ formulation \eqref{eq:B1}-\eqref{eq:B4}, that is,
  $$ \mathcal{P}(G,k) = \mathrm{conv} \left\{w \in \{0,1\}^{|V|},
  \theta \ge 0 :
 \theta + \sum_{u \in K} w_u \ge |K|, \sum_{u \in V} w_u \le k,  K \in \mathcal{K}\right\}. \label{eq:polyAux}
 $$

Let $(V', q)$ denote a \NICP\ solution where $\theta=q$ and the interdiction policy is defined by $V'$.


\begin{proposition}

The polytope $\mathcal{P}(G,k)$ is full dimensional.
\end{proposition}

\begin{proposition}
Let $u\in V$. 
\begin{enumerate}
\item The trivial inequality $w_u\leq 1$ defines a facet of $\mathcal{P}(G,k)$ if and only if $k\geq 2$.
\item The trivial inequality $w_u\geq 0$ defines a facet of $\mathcal{P}(G,k)$.
\end{enumerate}
\end{proposition}

\begin{lemma}\label{le:facet}
Let $K\in \mathcal{K}$ be an arbitrary clique in $G$. The associated clique interdiction inequality \eqref{eq:B2} does not induce a facet if:
\begin{enumerate}
\item $|K|\le \ell_{\opt}$, or
\item $K$ is not maximal.
\end{enumerate}
\end{lemma}

Even though the value of $\ell_{\opt}$ is not known in advance, the above result is useful for the separation 
of clique interdiction cuts. The result states that the more promising inequalities (in terms of improving the quality of lower bounds) are those with $|K| \ge \ell_{\min}$, considering thereby the value of $\ell_{\min}$ as tight as possible. This is also in line with our preprocessing procedure that removes all vertices from $G$ whose clique number is not greater than $\ell_{\min}$.  
In addition, without loss of generality, in the remainder of this section, we focus on maximal cliques only.

\bigskip 

\begin{lemma}\label{prop:liftZero}
Let $K$ be a  maximal clique and $v\in K$.
If
\begin{equation}
\label{liftinglemma} \omega (G[V\setminus V'])\geq |K|-|V'\cap K|+1 \quad \forall V'\subseteq V \text{ where } v\in V'\text{ and } |V'|\leq k,
\end{equation}
then there exists $\alpha_v \le 0$ such that the associated clique interdiction cut \eqref{eq:B2} can be down-lifted to
\[
\theta +\sum_{u \in K\setminus\{v\}} w_u + \alpha_v w_v\geq |K|.
\]
\end{lemma}

\begin{corollary} \label{cor:facet}
Let $K\subset V$ be a clique. If there exists $v\in K$ satisfying \eqref{liftinglemma}
then the inequality \eqref{eq:B2} cannot define a facet.
\end{corollary}

Finally, the following proposition provides necessary and sufficient conditions under which the CI cuts are facet-defining. This is the major theoretical result of this section, which allows characterizing the strength of the ILP formulation upon which our solution framework is built on. 

\bigskip

 \begin{proposition} \label{prop:facet}
Let $K\in \mathcal{K}$ be a maximal clique. Inequality \eqref{eq:B2}, induced by $K$, defines a facet of $\mathcal{P}(G,k)$  if and only if
\begin{itemize}
\item $|K| \ge \ell_{\opt} +1$,
\item\label{p1:1} 
for all $v\in K$, there exists a subset $V'\subseteq V$ such that $v\in V'$, $|V'|\leq k$ and $\omega(G[V\setminus V'])+|V'\cap K|\leq |K|$.
\end{itemize}
\end{proposition}

\section{Computational results}

We show that the natural model, where the maximality condition of the constraints is considered, allows us to reduce the computational time. Furthermore, using a smart bound, we can reduce the computational time and solve 85\% of considered instances in less than 10 seconds. Without bound and maximality only less than 5\% can be solved in less than 10 seconds.
Another interesting piece of information from the computational results is the climbing of computational time according to the size of the budget ($k$). We also propose an analysis of the strength of the graph by varying the value of $k$.

\subsection{Comparison with a state-of-the-art bilevel solver}

In Table \ref{tab:bilevel}, we compare the results of the state-of-the-art bilevel and interdiction game solver from \cite{FischettiOR} (called \BILEVEL)  with our new approach \MASTERPIECE. Each row corresponds to 44 instances of Set A grouped by the number of vertices $|V| \in \{50,75,100,125,150\}$. For this test we used the same time limit of 600 seconds for each run. For each of the two solvers, we report the following values: the number of solved instances per group (\#solved),
the average computing time in seconds for those instances that were solved to optimality, the average exit gap after the time limit is reached (considering only those instances which were \emph{not} solved to optimality) and the average root gap (over all instances). The exit gap is calculated as $\text{ exit gap } = \text{ UB } - \lceil \text{LB} \rceil$, 
where  UB refers to the global upper bound computed by the corresponding method, and LB refers to the global lower bound of the same method.
In order to measure the quality of lower bounds at the root node of the B\&C tree (denoted by LB$_r$), we compute them with respect to the best known solution (BKS) as $\text{ root gap } = \text{ BKS } - \lceil \text{LB$_r$} \rceil$. 

\begin{table}[h]
\tiny
 \centering
 \setlength{\tabcolsep}{5pt}
 \renewcommand \arraystretch{1.1}
 \begin{tabular}{lrrrrrrrrrrrrrrrr}
  \toprule
	
															 &	&	&	\multicolumn{ 4}{c}{\MASTERPIECE} &&  \multicolumn{ 4}{c}{\BILEVEL}\\																		
\cmidrule(lr){4-7} \cmidrule(lr){8-12}																
$|V|$	&	$\#$	&	&		\# solved				&	time  	&	exit 	gap		&	 root gap	&	&	\# solved			&	time  	&	exit gap	&	 root gap	\\
														\cmidrule(lr){1-2}		\cmidrule(lr){4-7} \cmidrule(lr){8-12}

50 	& 	44 	& 	& 	{\bf44} 	& 	{\bf0.01} 	& 	- 	& 	0.16 	& 	& 	28 	& 	68.58 	& 	6.44 	& 	8.50 	\\
75 	& 	44 	& 	& 	{\bf44} 	& 	{\bf1.45} 	& 	- 	& 	0.41 	& 	& 	14 	& 	120.19 	& 	9.47 	& 	10.91 	\\
100 	& 	44 	& 	& 	{\bf37} 	& 	{\bf9.30} 	& 	1.00 	& 	0.98 	& 	& 	7 	& 	164.42 	& 	12.65 	& 	13.11 	\\
125 	& 	44 	& 	& 	{\bf35} 	& 	{\bf13.43} 	& 	1.33 	& 	1.20 	& 	& 	2 	& 	135.33 	& 	13.88 	& 	14.73 	\\
150 	& 	44 	& 	& 	{\bf33} 	& 	{\bf27.23} 	& 	1.91 	& 	1.43 	& 	& 	1 	& 	397.52 	& 	16.42 	& 	16.39 	\\

    \bottomrule
\end{tabular}
 \caption{Comparison with state-of-the-art  bilevel solver (\BILEVEL) from \cite{FischettiOR} and our approach (\MASTERPIECE). } \label{tab:bilevel}
\end{table}

Table \ref{tab:bilevel} demonstrates that our new approach greatly outperforms the general-purpose bilevel solver of \cite{FischettiOR} by several orders of magnitude. All instances with 50 and 75 vertices are solved to optimality in less than 2 seconds on average by \MASTERPIECE, whereas the bilevel solver manages to solve only 2/3 and  1/3 of the instances, respectively. Similar behavior can be observed for larger instances, where, for example, for graphs with 150 vertices the bilevel solver manages to solve only a single instance in more than 5 minutes, whereas we solve 70\% of them, in less than 1/2 minute on average. These results can be easily explained by the quality of root bounds: whereas the average absolute gaps at the root node for our approach are between 0.16 and 1.4, those of \BILEVEL\ range between 8.5 and 16.4. This has a strong impact on the size of the B\&C tree. For example, for 28 instances of size 50 solved by both approaches, \MASTERPIECE\ requires an average number of 1.5 B\&C nodes, compared to 1960 nodes required by \BILEVEL.

\section{Conclusion}

In this chapter we have studied the interdiction game in a network in which the leader chooses up to $k$ vertices to delete, in order to minimize the clique number determined by the follower in the resulting network. We have studied the problem complexity for special graphs, and derived a single-level ILP formulation with an exponential number of constraints called clique-interdiction cuts. In a polyhedral study of the underlying polytope, we have provided the necessary and sufficient conditions for these cuts to be facet-defining and designed  effective lifting procedures. The separation of these cuts required the development of an efficient exact solver for the maximum clique problem, which we tailored for the clique interdiction game. The deep understanding of the underlying problem allowed us to derive tight combinatorial lower and upper bounds, along with an efficient preprocessing phase for drastically reducing the problem size. 

\newcommand{\IncomingLinks}[1]{\omega^-\left(#1\right)}
\newcommand{\OutgoingLinks}[1]{\omega^+\left(#1\right)}

\newcommand{\NodeSet}{V}
\newcommand{\NodeIdx}{v}
\newcommand{\NodeSource}[1]{s_{#1}}
\newcommand{\NodeDest}[1]{t_{#1}}
\newcommand{\CutIdx}{{\NodeSet'}}

\newcommand{\LinkSet}{E}
\newcommand{\LinkIdx}{e}
\newcommand{\LinkCost}[1]{C_{#1}}
\newcommand{\LinkCR}[1]{\mu_{#1}}
\newcommand{\LinkCapa}[1]{b_{#1}}
\newcommand{\LinkDelay}[1]{\lambda_{#1}}

\newcommand{\DemandSet}{K}
\newcommand{\DemandIdx}{k}
\newcommand{\CRDemandSet}{\DemandSet_\texttt{C}}
\newcommand{\NCRDemandSet}{\DemandSet_\texttt{NC}}
\newcommand{\DemandDelay}[1]{\Lambda_{#1}}

\newcommand{\SlotSet}[1]{S^{#1}}
\newcommand{\Slotset}[1]{s^{#1}}
\newcommand{\SlotIdx}{s}
\newcommand{\SlotSize}[2]{\xi_{#1#2}}
\newcommand{\SlotAllocationFunc}[1]{A\left(#1\right)}
\newcommand{\LinkUsageFunc}[2]{u_{#1}\left(#2\right)}

\newcommand{\PathSet}{P}
\newcommand{\PathIdx}{p}

\newcommand{\Bandwidth}[1]{D_{#1}}
\newcommand{\RatioBandwidth}{D^r}

\def\ProbName{Routing and slot allocation problem}
\def\ProbAc{RSA}
\def\Network{IP-RAN}
\newcommand\FlexECG{\emph{FlexE-CG}\xspace}
\newcommand\allCuts{\emph{allCuts}\xspace}
\newcommand{\OneSlotSetIdx}{S^{\mathbbm{1}}}
\newcommand{\OneSlotSetSet}{\mathcal{S}^{\mathbbm{1}}}

\chapter*{Conclusion}

The main guiding thread, of this document, is the blocker and interdiction notions. Several works were presented to tackle combinatorial optimization problems. For each of these problems, efficient algorithms based on integer linear models were given. These algorithms use powerful methods like Branch-and-Cut or Branch-and-Price algorithms. Theoretical results on complexity and polyhedral analysis were presented. Blocker and interdiction notions provide a new dimension of classical combinatorial optimization problems. This area of research is promising and allows modelling dynamicity of some problems where leader and follower are opposed. This manuscript focus on bilevel notion applied to classical optimization problems like, matching, flow, path, clique, scheduling problems.
Another topic of our research is the telecommunication area where we have provided several main results and proposed several patents on telecommunication networks. 
In the rest of this conclusion, we will present a quick overview of the work done at Huawei in the telecommunication area. 
We will consider several telecommunication protocols and mechanisms inducing some new optimization problems and systems in order to handle these challenges. 
\vspace{-0.5cm}
\paragraph{Slicing}
In order to guarantee data isolation between virtual networks, it is possible to provide either soft or hard slicing. While the former approach, presented in  \cite{slicing3} is easier to deploy, it provides no guarantees on the status of the network when one of the slices experiences a heavy load. The latter fixes this
problem by providing a stricter subdivision of the network.\\

This mechanism introduces some new requirements that can be translated into specific constraints. The first constraint is given by the TDMA resource subdivision implying a division of capacity of each link in slot capacities. For instance, if the first slot is 5Gbs for a link then if a slice needs 3Gbs then we must reserve 5Gbs, and thus pay for these 5Gbs. Considering this constraint needs to change the traditional capacity constraint \cite{slicingOur1} \cite{HUIN202372}. Another requirement is the statistical multiplexing where for a given set of demands, called also services or commodities, the needed capacity to route a subset of demands is less than the sum of their capacities \cite{slicingOur1}. In \cite{MARTIN2022105819}, we propose an efficient algorithm for solving the SPTP. This consists in determining a shortest path going through some node subsets of the graph. Each node subset represents a service that must be done. 
\vspace{-0.7cm}
\paragraph{Deterministic network}
Traditional IP services cannot provide strict QoS guarantees and even if certain service classes can be given preferential treatment, performance is still statistical. Deterministic performance is needed to support applications with requirements of low and worst-case latency guarantee \cite{detnet1}. 
In \cite{detnetOur1,detnetOur2,detnetOur3,detnetOur4,detnetOur5} we propose efficient algorithms to solve this problem and propose algorithms for some variants of this problem based on other protocols ensuring low latency and bounded jitters. 
Based on the CSQF mechanism we have proposed two patents. The first one is to split traffic into several paths to help the load balancing \cite{patent1Det}. This patent introduces a new requirement where the paths associated with a demand must have the same delay or almost the same delay. Furthermore, the capacity constraints change to consider the splitting of demand on several paths. The second patent \cite{patent2Det} introduces the protection mechanism associated with the CSQF mechanism. We extend the 1+1 protection mechanism to  CSQF. 
\vspace{-0.7cm}
\paragraph{Distributed network: Open Shortest Path First}
The arising of Network Virtualization will be a key enabler for the deployment of virtualized components capable of performing efficient path computation on behalf of the routers, thus allowing the optimization of operational IP networks. This perspective change draws again the traffic engineering community's attention to classical problems related to IP network optimization and raises the question of finding effective algorithms allowing to solve those problems for large-scale networks. In particular, traffic routing in IP networks still draws heavily on shortest paths based routing protocols, such as open shortest path first (OSPF), and finding a set of link weights that induce shortest paths while also minimizing the network congestion is one of the key issues for the design of efficient IP networks.

In order to increase routing options, to support different applications' needs or to better load balance traffic in the network, an extension of IGP called Multi-Topology Routing (MTR) \cite{mtr} has been proposed. In this case, multiple IGP instances are running in parallel, each of them working with a different set of link metrics and maintaining a Shortest-Path Tree (SPT).
The problem to find the best OSPF topology, i.e., a set of weights associated with each link, such that the QoS is respected and the maximum link utilization is minimized is an NP-Hard problem. In \cite{nn1} we propose another model based on an exponential number of variables to solve this problem. This model, called Routing-path formulation, is based on a familly of variables representing the whole routing. A \textit{routing} is an assignment of one path for each commodity. 
We propose two patents based on the MTR. The first one \cite{patentVIGP} proposes a way to virtually create a new topology based on the combination of "real" topologies. Looking for a minimum cost path satisfying one or multiple end-to-end constraints such as latency or hop count usually involves algorithms relying on Lagrangian relaxation. They aim at finding feasible Lagrangian multipliers to modify link costs so that a shortest path on the graph with the updated cost gives a feasible solution. Well-known algorithms are LARAC or GEN-LARAC \cite{LARAC}. This invention creates new topologies from existing ones by applying multipliers and combining them linearly. 
The second patent \cite{patentMTR} proposes a mechanism to generate topologies based on almost directed acyclic sub-graphs \cite{GADAG} and the Maximally Redundant Trees (MRT) subgraphs. This mechanism generates an MRT-Blue topology and an MRT-Red forwarding topology. The MRTs are computed locally without exchanging link metrics. In this proposal, silent virtual topologies are tailor-made for fast rerouting and the algorithm is hard-coded into nodes. Our proposal offers more flexibility by building a number of silent virtual topologies, not limited to 2, which can be used for multiple purposes (meet Service Level Agreement Requirements (SLA), load balancing traffic, protect against failures, etc...) and can be updated dynamically.
Other works can be found in \cite{overlay2022} to solve overlay routing and in \cite{coflow, patentCoflow} to consider the co-flow structure of demands. Furthermore, we propose other patents \cite{patent1,patent2,patent3} to help the load balancing, QoS, and protection in networks.

In this manuscript, we have presented achieved works on the blocker and interdiction concepts and also described related achievements in the telecommunication area. For all these works, we have considered mainly methods based on mathematical programming to tackle them. Our experience as a researcher at Huawei allowed us to learn the integration of combinatorial optimization methods into telecommunication systems thanks to patents and how to deliver codes with high performances to solve real instances. This gave me a good overview of combinatorial optimization methods applied to the real world. 
	\appendix

	
\chapter{Definitions} 

\section{Bi-level optimization} 

Some bi-level optimization problems can be seen as a Stackelberg game where two players try to maximize their own profit. The first player, called the leader, moves first and tries to decrease as much as possible the reward of the second player, called the follower, that plays after the leader \cite{dempe2018bilevel}. In the following, we focus on the blocker and interdiction problems that are special cases of the Stackelberg game.
Bi-level problems are divided into two problems: the upper-level (leader) and the lower-level (follower) problem. The choice of the upper-level problem modifies the decision of the lower-level problem. Furthermore, the objective function of the upper-level problem depends on the solution of the lower-level problem. This interaction between the lower-level and upper-level problems induces difficulties since we cannot decouple the two problems. Formally a bi-level problem is written: 
\begin{alignat}{6}
\nonumber  & \min_{x\in X, y \in Y} F(x,y)\\
\label{intro:bilevel_1}   &  \text{subject to: } G_i(x,y)  \leq 0, &\quad  \mbox{ for all } i\in \{1,...,I\} ,\\
\label{intro:bilevel_2}  & y\in \arg\min_{z\in Y}\{f(x,z): g_j(x,z) \leq 0\}, & \quad \mbox{ for all } j\in \{1,...,I\}
\end{alignat}
where:
\begin{itemize}
    \item $x$(resp. $y$) is the vector of upper-level (lower-level) decision variables, variables are in $\mathbb{R}$ or in $\mathbb{Z}$,
    \item $F$(reps. $f$) is the upper-level (resp. lower-level) objective function,
    \item $G$(resp. $g$) are upper-level (resp. lower-level) constraints,
\end{itemize}
In the following, we focus on the blocker and the interdiction variants of bi-level problems. The notion of blocker or interdiction is applied to a property, like maximum flow or perfect matching. Let $\Pi$ be a property and $\alpha(\Pi)$ be the value of this property (for instance $\Pi$ is a maximum flow problem and $\alpha(\Pi)$ is the value of the maximum flow).

\medskip
\textbf{Blocker/Interdiction decision problem}\\ 
\textbf{Input: } Property $\Pi$, max value $\alpha(\Pi)^*$, set of elements $I$, weight  $w_i$ for each element $i\in I$ and a bound $B$.\\
\textbf{Question: } Does there exist a set of elements $I'\subseteq I$ such that the sum of $w_i$, $i\in I'$, is less than or equal to $B$ and $\alpha(\Pi)$ in $I\setminus I'$ is less than or equal to  $\alpha(\Pi)^*$.\\

\vspace{-0.5cm}
\textbf{Blocker problem} searches to minimize $B$.\\

\vspace{-0.5cm}
\textbf{Interdiction problem} searches to minimize $\alpha(\Pi)^*$.\\


\section{Graph definition}

The previous definitions of bi-level/blocker/interdiction notions are introduced in general cases. In the following, we consider properties $\Pi$ based on graph problems.\\
We consider three kinds of graphs:
\begin{itemize}
    \item undirected graph $G=(V,E)$, where $V$ is the set of vertices and $E$ is the set of edges
        \begin{itemize}
        \item $\delta(u)$ is the set of edges incident to the vertex $u$ 
        \item $N(u)$ is the neighbour vertices of the vertex $u$, i.e. $N(u)=\{v|uv\in \delta(u)\}$.   
    \end{itemize}
    \item directed graph $G=(V,A)$, where $V$ is the set of vertices and $A$ is the set of arcs
        \begin{itemize}
        \item $\delta^+(u)$ is the set of arcs outgoing the  vertex $u$ 
        \item $\delta^-(u)$ is the set of arcs incoming the  vertex $u$ 
        \item $N(u)$ is the neighbour vertices of the vertex $u$, i.e. $N(u)=\{v|uv\in \delta^+(u)\}$
    \end{itemize}
    \item bipartite graph $G=(U\cup V,E)$, where the set of vertices is partitioned into two subsets of vertices $U$ and $V$, and $E$ is the set of edges. No edge links two vertices in $U$ or two vertices in $V$ 
    \begin{itemize}
        \item $\delta(u)$ is the set of edges incident to the vertex $u$ 
        \item $N(u)$ the neighbour vertices of the vertex $u$, i.e. $N(u)=\{v|uv\in \delta(u)\}$
    \end{itemize}
    Remark that in a bipartite graph $N(u)\subseteq V$ (resp. $N(u)\subseteq U$) if $u\in U$ (resp. $u\in V$)
\end{itemize}

\section{Complexity}
The complexity of an algorithm is an asymptotic bound of the worst-case running number of operations. We denote by $O(n^k)$ the upper asymptotic bound and $O(n^k)$ is said polynomial if $k$ is a constant. 
A decision or an optimization problem can be solved in polynomial time if and only if there exists a polynomial algorithm to solve this problem.  

\vspace{-0.5cm}
\paragraph{NP-completeness:}
From the theory of complexity, a decision problem is said NP-complete if
\begin{itemize}
\item it is in NP, i.e., for each solution, there exists a polynomial time algorithm to check its feasibility (certificate)
\item there exists a polynomial reduction from another NP-complete problem
\end{itemize}
A polynomial reduction, from a problem $\Pi$ to a problem $\Pi'$, consists in finding polynomial numbers of transformations such that: 
\begin{itemize}
\item applying these transformations on an instance $I$ of $\Pi$ allows obtaining an instance $I'$ of $\Pi'$   
\item and for each instance $I$ the answer of the decision problem $\Pi'$ on $I'$ is the same that the problem $\Pi$ on $I$.
\end{itemize} 

\vspace{-0.5cm}
\paragraph{NP-hardness:}
The NP-completeness can be applied to a decision problem only. Indeed, for an optimization problem where the associated decision problem is NP-complete, checking if a solution is optimal is not in NP since we must prove that there does not exist a better solution (equivalent to the decision problem).

Each decision problem associated with an NP-hard problem is at least NP-complete but it can be harder. Thus every optimization problem, where the associated decision problem is NP-complete, is NP-hard. 

Since the decision problems of interdiction and blocker problems are similar we can easily deduce the following corollary.
\begin{corollary}
The interdiction problem of property $\Pi$ is NP-hard if and only if the blocker problem of property $\Pi$ is NP-hard.  
\end{corollary}

\vspace{-0.5cm}
\paragraph{$\Sigma^P_i$-completeness: }
$\Sigma^P_i$-completeness is a generalization of the NP-completeness notion. Indeed, the $\Sigma^P_1$-completeness is equivalent to the NP-completeness. A decision problem is $\Sigma^P_i$-complete if: 
\begin{itemize}
\item it is in $\Sigma^P_i$, i.e., for each solution there exists a $\Sigma^P_{i-1}$ time algorithm to check its feasibility (certificate)
\item there exists a reduction from another $\Sigma^P_i$-complete problem
\end{itemize}

Since interdiction and blocker notions can be applied on $\Sigma^P_i$-complete problem thus naturally the problem becomes harder and the version of these problems from the interdiction/blocker point of view can become $\Sigma^P_{i+1}$-complete.

\section{Polyhedral analysis}

In this section, we introduce some properties and definitions to understand the results that will be provided in the next chapters. Let $x\in \mathbb{R}^n$ be a vector of size $n$, where $n$ is the number of variables. 
\vspace{-0.5cm}
\paragraph{Linear combination:  } $x$ is a linear combination of $x_1,...,x_k \in \mathbb{R}^n$ if there exists $k$ scalars $\lambda_1,...,\lambda_k\in \mathbb{R}$ such that $x=\sum_{i\in \{1,...,k\}} \lambda_i x_i$. Furthermore if $\sum_{i\in \{1,...,k\}} \lambda_i=1$ then $x$ is an affine combination of   $x_1,...,x_k$. 

\vspace{-0.5cm}
\paragraph{Linearly independent: } A set of vectors $x_1,...,x_k \in \mathbb{R}$ are linearly independent (resp. affine independent) if the following system 

\begin{alignat}{6}
\label{intro:poly1}  & \sum_{i\in \{0,...,k\}} \lambda_i x_i =0 \\
\label{intro:poly2}  & \text{(resp.} \sum_{i\in \{0,...,k\}} \lambda_i x_i =0 \text{ and } \sum_{i\in \{0,...,k\}} \lambda_i =0 \text{)}
\end{alignat}
has a unique solution, $\lambda_i=0$, $\forall i\in \{1,...,k\}$  

Let $S$ be a non-empty set of vectors of $\mathbb{R}^n$. The convex hull of $S$ is the set of vectors obtained by a convex combination of vectors of $S$. This set is denoted by $\text{conv}(S)$

\vspace{-0.5cm}
\paragraph{Polyhedron: } A polyhedron $P$ is a set of solutions from a system of linear inequalities, i.e. $P= \{x\in \mathbb{R}^n| Ax\leq b\}$, where $A$ is a matrix with $m$ lines and $n$ columns, and $b$ a vector of size $m$. The system  $Ax\leq b$ characterizes the polyhedron $P$. A polytope is a bounded polyhedron. Thus a polyhedron is a polytope if and only if there exists $l,u \in \mathbb{R}^n$ such that $l\leq x \leq u$ for each $x\in P$. 

\vspace{-0.5cm}
\paragraph{Dimension: } the dimension $\text{dim}(P)$ of a polyhedron $P$ is the maximum number of solutions affinely independent of $P$ minus 1. A polyhedron is said full dimensional if and only if $\text{dim}(P)=n$. The dimension can be seen as the $n$ minus the number of non-equivalent equality in the system $Ax\leq b$.

\vspace{-0.5cm}
\paragraph{Extreme point:} a point of $P$ is said extreme if it is not a convex combination of other points of $P$.

\vspace{-0.5cm}
\paragraph{Valid inequality: }  an inequality $ax\leq  \alpha$ is valid for the polyhedron $P$ if for all points of $P$ are valid for the inequality $ax\leq \alpha$, i.e., $P\subseteq \{x\in \mathbb{R}^n | ax\leq \alpha \}$

\vspace{-0.5cm}
\paragraph{Facet defining: } an inequality $ax\leq  \alpha$ is facet defining of $P$ if the dimension of $ax\leq  \alpha$ is equal to $\text{dim}(P)-1$.

\section{Exact algorithms: Branch\&Cut and Branch\&Price}

An integer linear program $(P)$ is written in the following way:

\begin{alignat}{6}
& \min \sum_{i\in I} c_i x_i\\
\label{c0:jj}& \sum_{i\in I} a_{ij} x_i \geq b_j & & \forall j\in J,\\
\label{c0:integrity}& x_i\in \{0,1\} & &  \forall i\in I,
\end{alignat}
where $I$ is the set of variables indexes and $J$ is the set of constraints indexes.

The linear relaxation of $(P)$, denoted by $LR(P)$ is obtained by replacing \eqref{c0:integrity} by $0\leq x_i$ and $x_i\leq 1$

\vspace{-0.5cm}
\paragraph{Dual: } The dual of $LR(P)$ is:
\begin{alignat}{6}
& \max \sum_{j\in j} b_j y_j+ \sum_{i\in I} z_i\\
& \sum_{j\in J} a_{ij} y_j + z_i \leq c_i & & \forall i\in I,\\
\label{c0:integrity1}& y_j \geq 0 & &  \forall j\in J,
\end{alignat}
and denoted by $D(P)$, where $z_i$ is the dual variable associated to $x_i\leq 1$ and $y_j$ is the dual variable associated to \eqref{c0:jj}.

\begin{property}
The optimal value of $D(P)$ is equal to the optimal value of $LR(P)$. 
\end{property}
\begin{property}
The value of each feasible solution of $D(P)$ is a lower bound of the value of each solution of $LR(P)$. 
\end{property}

\vspace{-0.5cm}
\paragraph{Totally unimodular: } A matrix is totally unimodular if for every square non-singular submatrix is unimodular.

\begin{property}
If a matrix $A$ is totally unimodular and $b$ is an integer vector then optimal solutions of $LR(P)$ are integers (all values $x$ are equal to 0 or 1).  
\end{property}
An integer linear program (ILP) is said to be compact if the number of variables and constraints are polynomial. A non-compact ILP is a model with a non polynomial number of constraints (cuts) or variables (columns) requiring the use of algorithm methods to generate them dynamically. Note that these methods can be used when a model has an exponential number of constraints or variables but also when the number of constraints or variables is polynomial in $O(n^k)$ where $k>>0$. 
 
 In the following, we call $RLR(P)$ the restricted linear relaxation. Indeed since we add dynamically some constraints or variables at each iteration the model has a subset of constraints or variables.   
It is possible to consider a small subset of variables and/or constraints (the restricted problem) and generates the other one on the fly.
 
In what follows,  we introduce the method to generate dynamically constraints (cuts).

\vspace{-0.5cm}
\paragraph{Separation problem: } Consists in finding the optimization problem that allows finding a violated constraint $ax\geq \alpha$ (from a family of constraints) by the current solution of the $RLR(P)$ denoted by $x^*$, i.e.  $ax^*< \alpha$. If a constraint is violated then adding this constraint will modify the linear relaxation of the $RLR(P)$.

\vspace{-0.5cm}
\paragraph{Cutting plane algorithm: } Cutting plane algorithm is the loop that alternates between all separation problems and the $RLR(P)$. The algorithm stops when separation problems don't add inequality. 

Then we introduce the method to generate dynamically variables (columns).

\vspace{-0.5cm}
\paragraph{Pricing algorithm: }  In this case the pricing problem consists in finding a variable such that the linear relaxation will change. The pricing problem can be seen as a separation problem in the dual. 

\vspace{-0.5cm}
\paragraph{Column generation procedure: } is equivalent to the cutting plane algorithm for the columns. Then the algorithm alternates between all pricing algorithms and the $RLR(P)$.\\

Cutting plane and column generation methods are able to solve the linear relaxation of a model. It is possible to use them simultaneously. To avoid instability, it is better to alternate between a cutting plane method until no constraint is added and a column generation method until no column is added. Indeed, if variables and constraints are generated simultaneously then variables will oppose to the constraints and vice-versa. \\

\vspace{-0.5cm}
\paragraph{Benders decomposition: } 
The Benders decomposition is a procedure to automatically divides into two subsets so that a first-stage master problem is solved over the first set of variables, and the values for the second set of variables are determined in a second-stage subproblem for a given first-stage solution. This new formulation can be solved using the cutting plane algorithm where the second-stage subproblem is equivalent to the separation problem and a new family of constraints, called Benders cuts, is added to the first-stage master problem.

In order, to solve a combinatorial optimization problem if the associated polytope is not integer, it is necessary to plug the cutting plane and column generation method in a branch-and-bound framework to ensure that variables are integers. 

\vspace{-0.5cm}
\paragraph{Branch-and-bound algorithm: } this algorithm is divided into two phases. The first one is the branching phase that aims to partition recursively the search space until a best solution is found. When a search space is divided then nodes in the branching tree are generated. Traditionally, for an integer linear program, the branch-and-bound algorithm consists in partitioning into two search spaces by setting a variable to 0 for the first one and to 1 for the second one. The second is the bounding phase where a search space is implicitly discarded. It is due for instance when the dual bound (given by the linear relaxation) becomes worse than a feasible bound (given by a heuristic solution for instance).
 
The cutting plane and column generation methods can use a branch-and-bound algorithm to find the best solution. 
 
\vspace{-0.5cm}
\paragraph{Branch-and-price: } the goal of this method is to solve the linear relaxation, thanks to the column generation, at each node of the branching tree. Remark that the traditional branching strategy (by setting a variable to 0 in a branch and to 1 in another branch) is not efficient. 

\vspace{-0.5cm}
\paragraph{Branch-and-cut: } this method combines cutting plane algorithm and Branch-and-Bound algorithm. At each node of the branching tree, the cutting plane algorithm is used to determine the linear relaxation.

\section{Operations Research Problems}
In this section, we present some well-known Operations Research problems. 

\vspace{-0.5cm}
\paragraph{Maximum Stable problem} consists in finding a set of vertices $V'\subseteq V$ such that no edge exists between two nodes of $V'$ ( i.e. $u,v\in V'\times V': uv \notin E$). The objective function is to find the set $V'$ of maximum weight.

\vspace{-0.5cm}
\paragraph{Maximum Clique problem} consists in finding a set of vertices $V'\subseteq V$ such that for every two nodes of $V'$ there exists an edge in $E$ ( i.e. $u,v\in V'\times V': uv \in E$). The objective function is to find the set $V'$ of maximum weight.

\vspace{-0.5cm}
\paragraph{Maximum Matching problem} consists in finding a set of edges $E'\subseteq E$ such that two edges of $E'$ do not share the same vertex ( i.e. $uv,u'v' \in E'\times E': u\neq u',u\neq v',v\neq u',v\neq v'$). The objective function is to find the set $E'$ of maximum weight.

\vspace{-0.5cm}
\paragraph{Maximum Flow problem} consists in finding a flow on each arc such that the capacity is respected. Remark that the flow conservation constraint must be respected. The objective function is to find the maximum flow from a source to a destination.

\vspace{-0.5cm}
\paragraph{Minimum Cut problem}  consists in finding a set of arcs such that removing these links disconnects the graph. The minimum $st$-cut consists in ensuring that the source $s$ and the destination $t$ are not in the same connected component. The objective function is to find the cut of minimum weight. 

\vspace{-0.5cm}
\paragraph{Shortest Path problem}  consists in finding a path from a source to a destination. A path is a sequence of consecutive arcs. The objective function is to minimize the length of the path.

\vspace{-0.5cm}
\paragraph{Multi-commodity flow problem} consists in finding a set of paths from different sources to different destinations, such that the capacity constraints on links are respected. The objective function is to minimize the sum of the lengths of selected paths.

\chapter{Appendices - Extended CV} \label{app:appendix_one}
\vspace{-1cm}
\section*{Sébastien Martin}
\paragraph{Principal Research Engineer at Huawei} ~\\
Tel: +33.6.10.71.12.21\\
Mail: sebastien.martin@huawei.com\\
Website: http://martin.site.free.fr/index.php\\
\vspace{-1cm}
\section{Cursus and professional experiences}
\begin{itemize}
    \item 2021 -- now: Principal Research Engineer at Huawei
    \item 2018 -- 2021: Senior Research Engineer at Huawei
    \item 2012 -- 2018: Assistant professor at LCOMS laboratory of University of Lorraine
    \item 2010 -- 2012: Teaching Assistant at MIDO, Université Paris Dauphine, Paris.
    \item 2007 -- 2012: Ph.D. in Computer Science and Combinatorial Optimization at LAMSADE, Université Paris Dauphine, Paris.
    \item 2005 -- 2007: Masters degrees in Computer Science, Artificial Intelligence, Learning and Decision, Université Pierre et Marie Curie, Paris
\end{itemize}

\section{PhD Supervisions}
\paragraph{Guilluame Beraud Sudreau} (2024-)\\
{\it Network optimization with non-linear performance models.}\\
Co-supervised: Hervé KERIVIN (33\%), Walid Ben Ameur (33\%), Sébastien Martin (33\%)

\paragraph{Isma Bentoumi} (2019-2024)\\
{\it Optimization method for the multi-commodity flow blocker problem.}\\
Co-supervised: Fabio Furini (33\%), Ridha Mahjoub (33\%), Sébastien Martin (33\%)

\paragraph{Sarah Minich} (2017-2022)\\
{\it Approximation algorithms for solving some pagination problems.}\\
Co-supervised: Imed Kacem (33\%), Aristide Grange (33\%), Sébastien Martin (33\%)

\paragraph{Youcef Magnouche} (2013-2017)\\
{\it The multi-terminal vertex separator problem: Complexity, Polyhedra and Algorithms.}\\
Co-supervised: Denis Cornaz (10\%), Ridha Mahjoub (40\%), Sébastien Martin (50\%)

\paragraph{Mohammed Albarra Hassan Abdeljabbar Hassan} (2013-2016)\\
{\it Robust scheduling problem for cloud computing}\\
Co-supervised: Imed Kacem(50\%), Sébastien Martin (50\%)

\paragraph{Mohamed Ould} (2011-2014)\\
{\it Combinatorial optimization and knowledge on companies}\\
Co-supervised: Ridha Mahjoub (50\%), Sébastien Martin (50\%)

\section{Master Student supervisions}
Full or partial supervision.
\begin{itemize}
    \item Abdellah Bulaich Mehamdi (2023) {\it Pricing (De)-activation: Multicommodity Flow and Column Generation}
    \item Walid Astaoui (2022) {\it On the $k$-spanning tree problem variants: Complexity and algorithms}
    \item Jiachen Zhang (2022) {\it The Multi-Commodity Flow Problem with disjoint paths}
    \item Miguel Pineda Martin (2022) {\it Quantum computing and combinatorial optimization approach for solving Unsplittable Multi-commodity Flow Problem}
    \item Corentin Clavier (2021) {\it OSPF decomposition area and combinatorial optimization}
    \item Corentin Juvigny(2020) {\it On the shortest path problem with node inclusion} 
    \item Alexandre Schulz(2019) {\it Machine learning applied to shortest path problems} 
    \item Mathieu Kintzinger(2015) {\it Robust nurse assignment problem with priority} 
    \item Karine Laurent(2015) {\it On the knapsack problem with overlapping object} 
    \item Xin Wang(2015) {\it Heuristic and meta-heuristic for the $k$-separator problem} 
    \item Zheng Mengli(2014) {\it Column generation algorithm for the robust nurse assignment problem} 
    \item Xiaotong Jiao(2013) {\it Uni-dimensional cutting problem} 
    \item Ali Taleb(2012) {\it A branch-and-bound algorithm for a scheduling problem} 
\end{itemize}
\section{Responsibilities}
\begin{itemize}
    \item Management of the PRC DataCom patent review board (2022 -- 2024)
    \item Member of the PRC DataCom patent review board (2020 -- )
    \item Member of the LCOMS council (2013 -- 2018)
    \item Management of LCOMS DOP seminar (2013 -- 2018)
    \item Head of computer science department of Metz IUT (2016 -- 2018)
    \item Member of the IUT of Metz council (2012 -- 2018)
\end{itemize}

\section{Projects Involvement}
Management of projects with $^*$ and participation for the other ones. 
\begin{itemize}
\item Huawei project 2022$^*$: MCF solver.
\item Huawei project 2019: Co-flow scheduling algorithm.
\item PEPS 2018 project$^*$: Aide à la décision pour renforcer la robustesse des plannings.
\item Regional 2016 project$^*$: Gestion du personnel paramédical.
\item PHC PAVLE SAVIC 2016: The development of hybrid heuristics for combinatorial optimization.
\item European 2016: Empowering Young Researchers (EYE).
\item PGMO 2015 project: Efficient exact algorithms for Graph Partitioning Problems.
\item PGMO 2014 project: Efficient exact algorithms for Graph Partitioning Problems.
\item ANR 2006 project: Parallel numerical Algorithms for Real time simulation of Algebraic Differential Equations systems (PARADE).
\end{itemize}

\section{Conferences Involvement}

\begin{itemize}
\item CODIT'24: Industry Forum co-Chairs
\item ROADEF 2017: Organizing committee member
\item CIE45: Organizing committee member
\item Sciences Incubator: Organizing committee member
\item CODIT'14: Organizing committee member
\item DASA16: Technical committee member
\item CloudNet16: Technical committee member 
\item CloudNet15: Technical committee member
\end{itemize}

Organization of special sessions in the following conferences:
\begin{itemize}
\item CODIT'24, CODIT'23, CODIT'14
\item CloudNet'14
\item CIE45
\end{itemize}

\section{Publications}
\subsection{International journals}
The authors are given in alphabetic order for the references with a $*$. 
\begin{enumerate}
    \item P. Medagliani, S. Martin, Y. Magnouche, J. Leguay, B. Decraene. \textit{Distributed Tactical TE With Segment Routing}
    IEEE Transactions on Network and Service Management. To appear, 2024.
    \item R. El-Azouzi, F. De Pellegrini, A. Arfaoui, C. Richier, J. Leguay, Q.-T. Luu, Y. Magnouche, S. Martin. \textit{Semi-Distributed Coflow Scheduling in Datacenters}
    IEEE Transactions on Network and Service Management. To appear, 2024.
    \item J. Zhang, Y. Magnouche, P. Bauguion, S. Martin, J. C. Beck. {\it Computing Bi-Path Multi-Commodity Flows with Constraint Programming-based Branch-and-Price-and-Cut}. INFORMS Journal on Computing. To appear, 2024.
    \item* P. Healy, N. Jozefowiez, P. Laroche, F. Marchetti, S. Martin, Z. Róka. {\it A branch-and-cut algorithm for the Connected Max-k-Cut Problem}. European Journal of Operational Research 312 (1), pp. 117-124, 2023. {\it https://doi.org/10.1016/j.ejor.2023.06.015}
    \item* N. Huin, J. Leguay, S. Martin, P. Medagliani. {\it Routing and slot allocation in 5G hard slicing}. Computer Communications 201, pp. 72-90, 2023.
    \item V. Angilella, F. Krasniqi, P. Medagliani, S. Martin, J. Leguay, R. Shoushou, L. Xuan. {\it High capacity and Resilient Large-scale Deterministic IP Networks}. Journal of Network and Systems Management 30 (4), pp. 1-28. 2022.
    \item S. Martin, Y. Magnouche, C. Juvigny, J. Leguay. {\it Constrained shortest path tour problem: Branch-and-Price algorithm}.  Computers and Operations Research 144, pp. 105819.  2022.
    \item* M. Becker, N. Ginoux, S. Martin and Z. Róka. {\it Tire noise optimization problem: a mixed integer linear programming approach}. RAIRO-Operations Research 55 (5), pp. 3073-3085. 2022.
    \item* A. Grange, I. Kacem, S. Martin and S. Minich. {\it Fully Polynomial Time Approximation Scheme for the Pagination Problem with hierarchical structure of tiles}. RAIRO-Operations Research  57, pp. 1–16. 2023. {\it https://doi.org/10.1051/ro/2022022}
    \item J. Krolikowski, S. Martin, P. Medagliani, J. Leguay., S. Chen, X. Chang and X. Geng. {\it Joint Routing and Scheduling for Large-Scale Deterministic IP Networks}.  Computer Communication 165, pp. 33-42. 2020.
    \item* Y. Magnouche, A. R. Mahjoub and S. Martin. {\it The multi-terminal vertex separator problem : Branch-and-Cut-and-Price}. Discrete Applied Mathematics 290, pp. 86-111. 2020
    \item* Y. Magnouche and S. Martin. {\it Most Vital Vertices for the Shortest s-t Path Problem: Complexity and Branch-and-Cut algorithm}. Optimization Letters 14 (8), pp. 2039-2053. 2020.
    \item* P. Laroche, F. Marchetti, S. Martin, A. Nagih, Z. Róka. {\it Multiple Bipartite Complete Matching Vertex Blocker Problem: Complexity, polyhedral analysis and Branch-and-Cut}. Discrete Optimization 35, pp. 100551. 2020. {\it https://doi.org/10.1016/j.disopt.2019.100551}  
    \item* F. Furini, I. Ljubić, S. Martin, P. San Segundo. {\it The Maximum Clique Interdiction Problem}, European Journal of Operational Research. Volume 277, Issue 1, pp. 112-127. 2019.
    \item* D. Cornaz, Y. Magnouche, A. R. Mahjoub and S. Martin. {\it The multi-terminal vertex separator problem: Polyhedral analysis and Branch-and-Cut}. Discrete Applied Mathematics. 256, pp. 11-37. 2019.
    \item* D. Cornaz, F. Furini, M. Lacroix, R. Mahjoub, E. Malaguti, S. Martin. {\it The Vertex $k$-cut Problem}. Discrete Optimization 31, pp. 8-28. 2019.
    \item* M. A. Hassan, I. Kacem, S. Martin, I. M. Osman. {\it On the m-clique free interval subgraphs polytope: polyhedral analysis and applications}. Journal of Combinatorial Optimization 36, pp. 1074–1101. 2018.
    \item* A. Grange, I. Kacem, S. Martin. {\it Algorithms for the Bin Packing Problem with Overlapping Items}. Computers \& Industrial Engineering 115, pp. 331-341. 2018.
    \item* L. Alfandari, T. Davidovi, F. Furini, I. Ljubic, V. Maras, S. Martin. {\it Tighter MIP formulations for Barge Container Ship Routing}. Omega 82, pp. 38-54. 2019.
      \item* M. A. Hassan, I. Kacem, S. Martin, I. M. Osman. {\it Genetic Algorithms for Job Scheduling in Cloud Computing}. Studies in Informatics and Control, ISSN 1220-1766, vol. 24 (4), pp. 387-400, 2015.
    \item* M. Lacroix, A. R. Mahjoub, S. Martin et C. Picouleau. {\it On the NP-Completeness of the Perfect Matching Free Subgraph Problem}. Theoretical Computer Science, Vol. 423, pp. 25-29. 2012.
    \item* M. Lacroix, A. R. Mahjoub et S. Martin. {\it Combinatorial Optimization Model and MIP formulation for the Structural Analysis of Conditional Differential-Algebraic Systems}. Computers \& Industrial Engineering 61, pp. 422-429. 2011.
\end{enumerate}

\subsection{International conferences}
\begin{enumerate}
\item S. Martin, Y. Magnouche, P.
Medagliani, J. Leguay. Alternative Paths Computation for Congestion Mitigation in Segment-Routing Networks. CODIT 24, to appear. 2024.
\item F. Zhang, W. Jiazheng, M. Lacroix, R. Wolfler Calvo, Y. Magnouche, S. Martin. The Multi-Commodity Flow Problem: Double Dantzig-Wolfe Decomposition. CODIT 24, to appear. 2024.
\item Y. Magnouche, S. Martin, J. Leguay, P. Medagliani. In-Band Network Telemetry for Efficient Congestion Mitigation. pp. 34-39, INOC 2024.
\item N. Huin, S. Martin, J. Leguay. Virtual Multi-Topology Routing for QoS Constraints. 37th IEEE/IFIP Network Operations and Management Symposium (NOMS 2024). To appear, 2024.
\item M. Y. Naghmouchi, S. Ren, P. Medagliani, S. Martin, J. Leguay. Optimal Admission Control in Damper-Based
Networks: Branch-And-Price Algorithm. CODIT 23, pp. 488-493. 2023. (This publication won the best paper award.)
\item J. Zhang, Y. Magnouche, S. Martin, A. Fressancourt, C. Beck. The Multi-Commodity Flow Problem with Disjoint
Signaling Paths: A Branch-And-Benders-Cut
Algorithm. CODIT 23, pp. 477-482. 2023. 
\item M. Pineda Martín, S. Martin. Unsplittable Multi-Commodity Flow Problem Via
Quantum Computing. CODIT 23, pp. 385-390. 2023. 
\item A. Benhamiche, M. Chopin, S. Martin. Unsplittable Shortest Path Routing: Extended Model and Matheuristic. CODIT 23, pp. 926-931. 2023. 
\item R. Grappe, M. Lacroix, S. Martin. The Multiple Pairs Shortest Path Problem for
Sparse Graphs: Exact Algorithms. CODIT 23, pp. 956-961. 2023.
\item I. Bentoumi;
F. Furini;
A. R. Mahjoub;
S. Martin. A Branch-and-Benders-Cut Approach to Solve the Maximum Flow Blocker Problem. CODIT 23, pp. 674-677. 2023.
\item Y. Magnouche, S. Martin, J. Leguay. Protected load-balancing problem: Neural-network-based approximation for non-convex optimization. NOMS 2023-2023 IEEE/IFIP Network Operations and Management Symposium (NOMS 2023), pp. 1-9. 2023.
\item M. Y. Naghmouchi, S. Ren, P. Medagliani, S. Martin, J. Leguay. Scalable Damper-based Deterministic Networking. 
2022 18th International Conference on Network and Service Management (CNSM 2022), pp. 367-373. 2022.
\item T. A. Q. Pham, S. Martin, J. Leguay, Xu Gong, Xu Huiying. Intent-Based Routing Policy Optimization in SD-WAN. ICC 2022-IEEE International Conference on Communications, pp. 4914-4919. May 2022.
\item Y. Magnouche, S. Martin, J. Leguay, F. De-Pellegrini, R. El-Azouzi, C. Richier. Branch-and-Benders-Cut Algorithm for the Weighted Coflow Completion Time Minimization Problem. INOC 2022, pp. 1-6. 2022.
\item S. Martin, P. Medagliani, J. Leguay. Network Slicing for Deterministic Latency. 2021 17th International Conference on Network and Service Management (CNSM), pp. 572-577.  2021.
\item  B. Liu, S. Ren, C. Wang, V. Angillela, P. Medagliani, S. Martin, J. Leguay. Towards Large-Scale Deterministic IP Networks. 2021 IFIP Networking Conference (IFIP Networking), pp. 1-9. 2021.
\item* A. Grange, I. Kacem, S. Martin, S. Minich. Approximate solutions for a special pagination problem with 2 symbols per tile. 2021 IEEE International Conference on Networking, Sensing and Control (ICNSC), pp. 1-4. 2021.
\item* S. Chen, J. Leguay, S. Martin and P. Medagliani. Load Balancing for Deterministic Networks. 2020 IFIP Networking Conference (Networking), pp. 785-790. 2020.
\item* N. Huin, J. Leguay, S. Martin, P. Medagliani, S. Cai. Routing and Slot Allocation in 5G Hard Slicing. 9th International Network Optimization Conference. pp. 72-77. 2019.
\item* A. Grange, I. Kacem, S. Martin, S. Minich. Fully polynomial-time approximation scheme for the pagination problem. Proceedings of International Conference on Computers and Industrial Engineering, CIE49, pp. 1-10. 2019
\item* F. Furini, E. Malaguti, S. Martin and I. C. Ternier. ILP Models and Column Generation for the Minimum Sum Coloring Problem.  Electronic Notes in Discrete Mathematics 64, pp. 215-224. 2018.
\item* P. Laroche, F. Marchetti, S. Martin et Z. Roka. " Bipartite Complete Matching Vertex Interdiction Problem with Incompatibility Constraints: Complexity and Heuristics ". 2017 4th International Conference on Control, Decision and Information Technologies (CoDIT), pp. 6-11.  2017.
\item* M. Hassan, I. Kacem, S. Martin and I. M.Osman. " Mathematical Formulation for Open Shop Scheduling Problem ". 2017 4th International Conference on Control, Decision and Information Technologies (CoDIT), pp. 803-808.  2017.
\item* M. Becker, N. Ginoux, S. Martin and Zs. Roka. Optimization of Tire Noise by Solving an Integer Linear Program (ILP) 2016 IEEE International Conference on Systems, Man, and Cybernetics (SMC2016), pp. 1591-1596. 2016.
\item* Y. Magnouche and S. Martin. "The Multi-terminal vertex separator problem: Polytope characterization and TDI-ness". Combinatorial Optimization: 4th International Symposium, ISCO 2016, pp. 320-331. 2016.
\item* M. Hassan, I. Kacem, S. Martin and I. M.Osman. "Unrelated Parallel Machine Scheduling Problem With precedence Constraints: Polyhedral Analysis and Branch-and-Cut". Combinatorial Optimization: 4th International Symposium, ISCO 2016, pp. 308-319. 2016.
\item* Y. Magnouche, A. R. Mahjoub and S. Martin. "The Multi-terminal vertex separator problem: Extended formulations and Branch-and-Cut-and-Price". 2016 International Conference on Control, Decision and Information Technologies (CoDIT), pp. 683-688. 2016.
\item* M. Hassan, I. Kacem, S. Martin and I. M.Osman. "Valid Inequalities for Unrelated Parallel Machines Scheduling with Precedence Constraints". 2016 International Conference on Control, Decision and Information Technologies (CoDIT), pp. 677-682. 2016.
\item* D. Cornaz, Y. Magnouche, A. R. Mahjoub and S. Martin. " The multi-terminal vertex separator problem: polyhedral analysis and branch-and-cut ". Proceedings of 45th International Conference on Computers \& Industrial Engineering (CIE45), ISBN:9781510817456 pp. 857-864. 2015.
\item* M. Hassan Abdel-Jabbar, I. Kacem, S. Martin and I. M. Osman. " Mathematical Formulations for the Unrelated Parallel Machines with Precedence Constraints ". Proceedings of 45th International Conference on Computers \& Industrial Engineering (CIE45), ISBN:9781510817456 pp. 1005-1012. 2015.
\item* P. Laroche, F. Marchetti, S. Martin, Z. Roka and M. Zheng. " Complexity and Heuristics for Multi Bipartite Complete Matching Vertex Interdiction Problem: Application to Robust Nurse Assignment ". Proceedings of 45th International Conference on Computers \& Industrial Engineering (CIE45), ISBN:9781510817456 pp. 1294-1301. 2015.
\item* A. Grange, I. Kacem, K. Laurent, S. Martin. " On the knapsack problem under merging objects' constrants ". Proceedings of 45th International Conference on Computers \& Industrial Engineering (CIE45), ISBN:9781510817456 pp. 1359-1366. 2015.
\item* P. Laroche, F. Marchetti, S. Martin et Z. Roka. " Bipartite Complete Matching Vertex Interdiction Problem: Application to Robust Nurse Assignment ". IEEE International Conference Control, Decision and Information Technologies (CoDIT'14) pp. 182-187. 2014.
\item* D. Cornaz, F. Furini, M. Lacroix, E. Malaguti, A. R. Mahjoub et S. Martin. “Mathematical formulations for the Balanced Vertex k-Separator Problem”. IEEE International Conference Control, Decision and Information Technologies (CoDIT'14) pp. 176-181. 2014.
\item* M. A. Hassan Abdel-Jabbar, I. Kacem et S. Martin. " Unrelated parallel machines with precedence constraints: Application to cloud computing ". IEEE CLOUDNET 2014, pp. 438-442. 2014.
\item* M. Lacroix, A. R. Mahjoub et S. Martin. " Polyhedral Analysis and Branch-and-Cut for the Structural Analysis Problem ". Lecture Notes on Computer Science (International Symposium of Combinatorial Optimization (ISCO) 2012), pp. 117–128. 2012.
\item* M. Lacroix, A. R. Mahjoub et S. Martin. " Structural analysis for Differential-Algebraic Systems: Complexity, Formulation and Facets ". Proceedings ISCO 2010, Electronic Notes in Discrete Mathematics 36, pp. 1073-1080. 2010.
\item* M. Lacroix, A. R. Mahjoub et S. Martin. "Structural analysis in Differential-Algebraic Systems and Combinatorial Optimization". Proceedings of 39th International Conference on Computers \& Industrial Engineering (CIE39), pp. 331-337. 2009. (This publication won the Best Student Paper award)
\item P. Fouilhoux, S. Martin et M. Coupechoux "Combinatorial problems and integer formulations in wireless mesh network design". NCP07 (International Conference on Nonconvex Programming) (2007).
\end{enumerate}

\subsection{Patents}
\begin{enumerate}
\item A system and method for a partial and dynamic distributed sketch assignment (2023). Sébastien Martin, Gabriele Castellano, 	Massimo Gallo,  Isma Bentoumi. To be published.
\item Apparatus and system for Nested Multi-Topology Routing (2023). Sébastien Martin, Youcef Magnouche, Jérémie Leguay,  Zeng Feng. To be published.
\item Apparatus for ensuring QoS requirements using distributed adaptive queue sizing (2022). Sébastien Martin, Antoine Fressancourt,	Paolo Medagliani, Anne Bouillard, Ren Shoushou. To be published.
\item System and Method for Consistent Slices (2022). Sébastien Martin, Antoine Fressancourt,	Paolo Medagliani, Jérémie Leguay. To be published.
\item Apparatus and system to augment path computation elements (2022). Sébastien Martin, Pierre Bauguion, Jérémie Leguay, Zeng Feng, Tang Ziye. To be published.
\item Apparatus and method for distributed load balancing in IP routing (2021). Sébastien Martin, Jérémie Leguay, Youcef Magnouche. To be published.
\item Intent-based smart policy-routing (2021). Pham Tran Anh Quang, Sébastien Martin, Jérémie Leguay, GongXu, Zengfeng. To be published.
\item Apparatus and methods for delay-constrained redundant paths in IP routing (2021). CaiShengming, Jérémie Leguay, Sébastien Martin, Paolo Medagliani. WO2022214164A1.
\item Apparatus and methods for virtual topologies in IP routing (2020). Nicolas Huin, Jérémie Leguay, Sébastien Martin, CaiShengming. WO2022167068A1.
\item Fast Failover for Path Degradation (2020). Jérémie Leguay, Paolo Medagliani, Sébastien Martin, Antoine Fressancourt. WO2022199827A1.
\item Apparatus and Methods for Protected Load Balancing (2020). Sébastien Martin, Jérémie Leguay, Youcef Magnouche, Zhangjie, Liyuechen. WO2022052009A1.
\item Apparatus and Method for coflow tracking to schedule traffic (2020). Jérémie Leguay, Sébastien Martin, Rachid El-Azouzi, Francesco de Pellegrini, Youcef Magnouche, Cedric Richier. WO2022074415A1.
\item A system of policies to improve the resilience of deterministic networks with performance guarantee (2019). Paolo Medagliani, Sébastien Martin, Jérémie Leguay, Chenshuang. WO2021197617A1.
\item Control device, switch device and methods (2019). Paolo Medagliani, Sébastien Martin, Jérémie Leguay, Chenshuang. US20220150159A1/EP3981133A4.
\end{enumerate}
\section{Editorial work} 
I have been solicited to referee papers for the following journals:
\begin{enumerate}
    \item Discrete Applied Mathematics
    \item International Transactions In Operational Research
    \item RAIRO - Operations Research
    \item Discrete Optimization
    \item Computers \& Industrial Engineering
    \item Computers \& Operations Research
    \item Annals of Operations Research
    \item Journal of Combinatorial Optimization
\end{enumerate}
Also, I have been solicited to referee papers for the following conferences:
\begin{enumerate}
    \item CODIT
    \item CIE
    \item CloudNet
\end{enumerate}

\section{Teaching} 

I have taught at University Paris-Dauphine $^1$, University of Lorraine $^2$, Sorbonne University $^3$ and Master Parisien de Recherche Opérationnelle (MPRO) $^4$.

\begin{itemize}
    \item 2019/2024
    \begin{itemize}
        \item Master 2 MPRO $^4$	:	Séminaires optimisation dans les réseaux(16h).
        \item Master 2 TIDE $^3$	:	Optimisation (cours 18h).
    \end{itemize}
        \item 2018/2019
    \begin{itemize}
        \item Projets tutorés IUT L2 $^2$	:	Organisation des projets tutorés (études longues)(10h).
        \item Master 2 ID	$^2$:	Séminaire optimisation dans les réseaux (3h).
    \end{itemize}
        \item 2017/2018
    \begin{itemize}
        \item 2éme Année L2 $^2$	:	Principe des systèmes d'exploitation (Cours 8h, TDs 68h), Complexité (Cours 8h, TDs 8h cours intégré), Projets tutorés (2 groupes 16h)
        \item Année Spéciale L1/2 $^2$	:	Environnement Informatique (TDs 8h), Conception et développement d'applications mobiles (Cours, TDs : 24h cours intégré).
    \end{itemize}
        \item 2016/2017
    \begin{itemize}
    \item 2éme Année L2 $^2$	:	Principe des systèmes d'exploitation (Cours 8h, TDs 38h), Conception et développement d'applications mobiles (Cours, TDs : 24h cours intégré), Complexité (Cours 14h, TDs 14h cours intégré), Projets tutorés (3 groupes  35h)
    \item Année Spéciale L1/2 $^2$	:	Environnement Informatique (TDs 8h), Conception et développement d'applications mobiles (Cours, TDs : 24h cours intégré).
    \end{itemize}
            \item 2015/2016
    \begin{itemize}
 \item Master 1	$^2$ :	Initiation à la recherche (7h).
Optimisation Combinatoire (TPs : 4h).
 \item 2éme Année L2 $^2$	:	Principe des systèmes d'exploitation (Cours 8h, TDs 68h), Conception et développement d'applications mobiles (Cours, TDs : 24h cours intégré), Complexité (Cours 15h, TDs 15h cours intégré), Projets tutorés (1 groupe 20h), suivi de 4 stagiaires (8h)
 \item Année Spéciale L1/2 $^2$	:	Introduction aux systèmes informatiques (Cours 4h, TDs 10h), Conception et développement d'applications mobiles (Cours, TDs : 20h cours intégré), Principe des systèmes d'exploitation (Cours 8h, TDs 22h).
    \end{itemize}
            \item 2014/2015
    \begin{itemize}
\item Master 1 $^2$	:	Initiation à la recherche (7h).
\item 2éme Année L2 $^2$	:	Principe des systèmes d'exploitation (Cours 8h, TDs 72h, TPs 48h), Conception et développement d'applications mobiles (Cours, TDs : 24h cours intégré), Projets tutorés (2 groupes  48h), suivi de 4 stagiaires (8h)
\item Année Spéciale L1/2 $^2$	:	Introduction aux systèmes informatiques (Cours 4h, TDs 10h), Principe des systèmes d'exploitation (Cours 8h, TDs 22h).
\item licence professionnelle métiers du web et commerce éléctronique L3 $^2$	:	4 soutenances de stages  (2h)
    \end{itemize}

            \item 2013/2014
    \begin{itemize}
 \item Master 1 $^2$	:	Initiation à la recherche (2.5h).
 \item 2éme Année L2 $^2$	:	Systèmes (Cours 14h, TDs 63h), Systèmes Approfondis (Cours 7h, TDs 11h), suivi de 8 stagiaires (16h), Projets tutorés (1 groupe  20h).
 \item Année Spéciale L1/2 $^2$	:	Introduction aux systèmes informatiques (Cours 4h, TDs 10h), Principe des systèmes d'exploitation (Cours 8h, TDs 22h).
    \end{itemize}
            \item 2012/2013
    \begin{itemize}
 \item Master 1	$^2$ :	Optimisation en finance (Cours 36h, TPs 12h).
  \item 2éme Année	L2 $^2$ :	Suivi de 6 stagiaires (12h), Projets tutorés (1 groupe  20h).
 \item Année Spéciale	L1/2 $^2$ :	Systèmes (Cours 16h, TDs 14h), Systèmes Approfondis (Cours 16h, TDs 32h).
 \item 1ére Année	L1 $^2$ :	Algorithmique et programmation ADA(TDs 49h, TPs 28h), Langage ADA (TDs 21h, TPs 28h, Projet 12h).
 \item licence professionnelle métiers du web et commerce éléctronique L3 $^2$	: 5 soutenances de Stages  (2,5h).
    \end{itemize}
            \item 2010/2012
    \begin{itemize}
\item DUMIE (L1) $^1$	:	Algorithmique générale (TDs 33h).
\item DUMIE (L1) $^1$	:	Informatique 2 Java (TDs 17h).
\item DUMIE (L2) $^1$	:	Programmation objet en java (TDs 39h, TPs 75h).
\item MIDO (M1) $^1$	:	Optimisation en finance (TPs 20h).
\item MI2E (L3) $^1$	:	Programmation linéaire (TPs 12h).
    \end{itemize}
            \item 2009/2010
    \begin{itemize}
\item  DUMIE (L2) $^1$	:	Informatique 3 (TDs 22h).
    \end{itemize}
            \item 2008/2009
    \begin{itemize}
\item MI2E (L3) $^1$	:	Utilisation des bases de données (TDs 4h, TPs 5h).
\item DUMIE (L2) $^1$	:	Informatique 4 (TDs 32h, TPs 28h).
\item MIDO (M1) $^1$	:	Optimisation en finance (TPs 10h).
\item MI2E (L3) $^1$	:	Java-Objet (TDs 15h).
    \end{itemize}
            \item 2007/2008
    \begin{itemize}
\item DUGEAD (L1) $^1$	:	Initiation à l'informatique (cours, TDs : 39h cours intégré).
\item LSO (L3) $^1$	:	Informatique, Initiation aux SGBD (cours, TDs : 49h cours intégré).
    \end{itemize}
\end{itemize}


  %



 \bibliography{./reference/ref}
\bibliographystyle{StyleThese}


\end{document}